\documentclass[11pt,a4paper]{article}
\pdfoutput=1
\usepackage{graphicx}
\usepackage{aas_macros}
\usepackage{jheppub}
\usepackage[T1]{fontenc} 
\usepackage{amsmath,amssymb,bbm,mathrsfs,nicefrac}
\usepackage{tikz}
\usepackage{amsmath}
\usepackage{amssymb}
\usepackage{textcomp,float,gensymb,wrapfig, enumitem,comment,dsfont,framed,slashed,appendix,wrapfig,xcolor}
\usepackage{ bbold }
\usepackage{braket} 
\usepackage{ mathrsfs }
\usepackage[small]{caption}
\usepackage{subcaption}
\usepackage{etoolbox,hyperref,makecell}
\usepackage{feynmp}
\usetikzlibrary{arrows.meta}
\usepackage{empheq}
\usepackage{braket}

\patchcmd{\abstract}{\null\vfil}{}{}{}
\newcommand{\vev}[1]{\langle #1 \rangle}

\usepackage{hhline}
\usepackage{float}
\usepackage{caption}
\usepackage{soul}

\newcommand{\eref}[1]{Eqn.~(\ref{#1})}
\newcommand{\fref}[1]{Fig.~\ref{#1}}
\newcommand{\sref}[1]{Section~\ref{#1}}

\usepackage[utf8]{inputenc}

\title{Direct Detection of Atomic Dark Matter in White Dwarfs}

\author{David Curtin and Jack Setford}

\affiliation{Department of Physics, University of Toronto, Canada}

\emailAdd{dcurtin@physics.utoronto.ca}
\emailAdd{jsetford@physics.utoronto.ca}

\date{\today}

\abstract{Dark matter could have a dissipative asymmetric subcomponent in the form of  atomic dark matter (aDM). 
This arises in many scenarios of dark complexity, and is a prediction of \emph{neutral naturalness}, such as the Mirror Twin Higgs model. 
We show for the first time how White Dwarf cooling provides strong bounds on aDM.
In the presence of a small kinetic mixing between the dark and SM photon, stars are expected to accumulate atomic dark matter in their cores, which  then radiates away energy in the form of dark photons.
In the case of white dwarfs, this energy loss can have a detectable impact on their cooling rate.
We use measurements of the white dwarf luminosity function to tightly constrain the kinetic mixing parameter between the dark and visible photons, for DM masses in the range $10^{-5}$ -- $10^{5}$ GeV, down to values of $\epsilon \sim 10^{-12}$. 
Using this method we can constrain scenarios in which aDM constitutes fractions as small as $10^{-3}$ of the total dark matter density.
Our methods are highly complementary to other methods of probing aDM, especially in scenarios where the aDM is arranged in a dark disk, which can make direct detection extremely difficult but actually slightly enhances our cooling constraints.
}

\begin{document}

\maketitle

\section{Introduction}

The nature of dark matter (DM) is one of the most important outstanding questions in particle physics, astrophysics and cosmology. Many simple scenarios, including several well-motivated WIMP candidates, have already been ruled out, but a vast range of model space remains to be probed. Constraints on any given scenario are often presented under the assumption that the candidate in question constitutes all of the dark matter in our universe. However, given the richness of Standard Model (SM) matter, it is perhaps begging the question to assume that the dark sector, which accounts for 85\% of the matter of our universe, is comprised of a single species. Less minimal alternatives, in which the dark sector features additional interactions and/or additional states, have been attracting increasing interest~\cite{Zurek:2013wia,Petraki:2013wwa,Krnjaic:2017tio,Tulin:2017ara,Fan:2013yva,Fan:2013tia,Chacko:2018vss,Berlin:2018tvf,Kribs:2018ilo,Renner:2018fhh,Foot:2014uba}. Scenarios of dark complexity are also well-motivated in models that predict a hidden sector related to the visible sector by symmetry~\cite{Berezhiani:2003xm, Foot:2003eq, Chacko:2005pe, Burdman:2006tz, Cai:2008au,Garcia:2015loa,Craig:2015xla,Garcia:2015toa}.

One such scenario is \emph{atomic dark matter} (aDM) \cite{Goldberg:1986nk, Kaplan:2009de, Kaplan:2011yj, Cline:2012is, Fan:2013bea}, in which a subcomponent of dark matter consists, in the simplest realisation, of a dark proton and dark electron bound by dark electromagnetism. In general, constraints on dark matter self-interactions mean that atomic dark matter cannot account for all of dark matter, but mass fractions as high as 5-10\% are much more difficult to rule out (we will discuss existing constraints on this scenario in more detail in Section~\ref{sec:models}). 
Such a model is interesting in its own right, but is particularly compelling within the paradigm of \emph{neutral naturalness} \cite{Chacko:2005pe, Burdman:2006tz, Cai:2008au}. 
Models of neutral naturalness address the Higgs hierarchy problem by introducing a hidden sector related to the SM by some discrete symmetry. 
The new states  are neutral under the Standard Model gauge group, thereby avoiding LHC constraints on colored top partners, but charged under their own dark interactions. 
Examples such as the Mirror Twin Higgs model (MTH)  \cite{Chacko:2005pe, Barbieri:2005ri, Chacko:2005vw, Chacko:2016hvu, Craig:2016lyx} 
are useful benchmark realizations of dark complexity.
As predictive models that give rise to rich dark sectors, they allow for a systematic exploration of the multi-faceted observational and experimental consequences of dark complexity, while also demonstrating that such models can be deeply connected to fundamental questions such as naturalness. 
For instance, the MTH  predicts atomic dark matter in the form of mirror hydrogen and mirror helium, as part of a richer dark sector featuring dark nuclear physics and confinement \cite{Chacko:2018vss}. This raises the fascinating possibility that the hidden sector could even give rise to objects analogous to those found in the visible sector: \emph{mirror stars} and other compact objects that could be detected in optical and X-ray telescope surveys~\cite{Curtin:2019lhm, Curtin:2019ngc}, as well as future microlensing surveys~\cite{Drlica-Wagner:2019xan, Croon:2020wpr} (see also~\cite{Foot:1999hm, Foot:2000vy}).

A dissipative subcomponent of dark matter has interesting phenomenology that is strikingly different from the usual assumptions of cold, thermal dark matter. The fact that aDM can radiate in dark photons means that it has an efficient \emph{cooling} channel, and can undergo collapse from a dark halo to a dark disk, superimposed on the visible baryonic disk of the Milky Way \cite{Fan:2013yva, Fan:2013tia}. This can make direct detection much more difficult, since dark matter contained within a dark disk will be co-rotating with the Earth in a similar orbit around the centre of the galaxy, and as a result will have a reduced velocity relative to us.

In this work we describe a new and powerful means of constraining the local abundance of atomic dark matter. If there is kinetic mixing between the dark and visible photons, which for non-minimal dark sectors is highly motivated even just by gravity~\cite{Gherghetta:2019coi}, then a significant amount of atomic dark matter will be captured by stars. The sheer size and longevity of stars make them excellent dark matter detectors -- even a tiny kinetic mixing can lead to very significant accumulation of aDM in their cores. The same interaction will allow energy transfer from the hot core of the star to this captured \emph{aDM nugget}, which in turn can radiate away energy in the form of dark photons\footnote{We note that the term nugget has been used before in the literature to refer instead to a strongly bound collection of particles held together by a dark confining force, see \cite{Gresham:2017zqi, Gresham:2017cvl, Gresham:2018anj, Bai:2018dxf, Coskuner:2018are}.}.
(This is to be compared to other studies of dark sectors using stellar cooling~\cite{Raffelt:1985nj, Raffelt:1987yb, Raffelt:1987yt, Turner:1987by, vanBibber:1988ge, Burrows:1988ah, Raffelt:1989zt, Blinnikov:1994eoa, Raffelt:1994ry, Gondolo:2008dd, Isern:2008fs, Isern:2008nt, Corsico:2012ki, Melendez:2012iq, Dreiner:2013tja, Redondo:2013wwa, Curtin:2014afa, Bertolami:2014wua, Hurst:2014uda,Vinyoles:2015aba, Chang:2018rso, Dessert:2019sgw}, which rely on particle production in their cores and are therefore limited to probe very light new particles.)
For most stars, such as the Sun, this `dark luminosity' will be only a small fraction of the total energy radiated by the star, and will lead to small deviations from predicted stellar behaviour (see \cite{Michaely:2019xpz} for a related study.)

However, the presence of dark matter in the cores of \emph{white dwarfs} can lead to very dramatic deviations from predicted behaviour. White dwarfs are extremely dense, degenerate remnants of those main sequence stars whose mass is not sufficient to become a neutron star. Since they no longer support nuclear fusion, they do not generate energy, and instead cool slowly over timescales of billions of years. We will show that an accumulation of atomic dark matter in their cores can significantly hasten their cooling. 
White dwarfs have, in general, much lower luminosities than main sequence stars, and as a result the impact of the dark luminosity is correspondingly larger. The high density of white dwarfs also leads to a suppression of the aDM evaporation rate (see Section~\ref{sec:evaporation}), especially for small dark matter masses. 
This effect of the dark photon energy loss becomes more pronounced as the white dwarf ages and gets colder. 
As we shall show, the amount of dark radiation emitted by a white dwarf actually \emph{increases} as it cools -- a highly non-intuitive prediction of \emph{photon portal thermodynamics} between the white dwarf core and the captured aDM. 
Compared to the Standard Model expectation, this leads to a \emph{deficit} in the observed number of white dwarfs at higher magnitudes (lower luminosities). This would be reflected in the white dwarf luminosity function~\cite{Raffelt:1996wa, Raffelt:1999tx}, which describes the distribution of observed white dwarf luminosities. The shape of this distribution is well-understood theoretically, allowing us to place very strong and robust constraints on the total amount of dark matter that can have accumulated in the white dwarfs' cores.

The sensitivity of white dwarf cooling to accumulated aDM enables us to probe aDM scenarios with very small kinetic mixing or local abundance. Since the best observational data comes from relatively nearby white dwarfs (within a few hundred pc), we are effectively probing the \emph{local} dark matter density (on galactic scales). These are the same observables probed by direct detection experiments, and indeed we are highly complimentary to existing/upcoming constraints. As we will show, our constraints actually become \emph{more} sensitive in the disk-like benchmark, where the average dark matter relative velocity is smaller -- unlike conventional direct detection experiments, which are more sensitive at higher relative velocity/higher recoil energy. 
In particular, we find that white dwarf cooling provides the strongest existing constraints to date on direct detection of many aDM scenarios in general, and asymetrically reheated Mirror Twin Higgs models in particular (see also~\cite{MTHastro}).

The remainder of the paper is organised as follows. 
In Section~\ref{sec:models} we define the specific aDM model we study, provide a brief review of atomic dark matter models and their motivation, and discuss existing constraints. 
In Section~\ref{sec:cooling_constraints} we describe how the additional cooling channel affects the white dwarf luminosity function, and how this allows us to derive our constraints by computing anomalous cooling for a single benchmark star.
In Section~\ref{sec:capture} we outline the calculation of the capture rate and the total accumulation of aDM, during three important stages of the star's lifetime: the main sequence (hydrogen burning) phase, the red giant (helium burning) phase, and the white dwarf cooling phase. In Section~\ref{sec:heating_cooling} we describe how we calculate the heating and cooling rate of the aDM nugget, and hence the total emission into dark radiation as a function of $\epsilon$ and the total accumulation. In Section~\ref{sec:results} we present our results, and in Section~\ref{sec:conclusion} we summarise and conclude.

\section{Atomic and mirror dark matter models}
\label{sec:models}

In this section we will outline the model under consideration and how it fits into the wider literature on self-interacting and atomic dark matter. The specific model we will constrain is the simplest realisation of atomic dark matter~\cite{Goldberg:1986nk, Kaplan:2009de, Kaplan:2011yj, Cline:2012is, Fan:2013bea}:
\begin{equation}
    \mathcal{L} = \mathcal{L}_{SM} - \frac{1}{4} \tilde F_{\mu\nu}^2 - \frac{\epsilon}{2} F_{\mu\nu} \tilde F^{\mu\nu} + i \overline{\tilde{p}} ( {\slashed\partial} - i e \tilde {\slashed A} - m_{\tilde p}) \tilde p + i\overline{\tilde e} ({\slashed\partial} + i e \tilde {\slashed A} - m_{\tilde e}) \tilde e,
\end{equation}
where $\tilde p$, $\tilde e$ and $\tilde A_\mu$ are the dark proton, dark electron and dark photon fields respectively. We will always assume that the dark electron is lighter than the dark proton $m_{\tilde e} < m_{\tilde p}$. Note that the dark proton has a positive charge under the dark $\textrm{U(1)}$, while the dark electron has negative charge -- although the sign of the charge assignments will not be important for our constraints. Note the photon kinetic mixing term which acts as a portal interaction between the dark sector and the Standard Model~\cite{Holdom:1985ag}. The kinetic mixing can be removed via the redefinition $A_\mu \to A_\mu - \epsilon \tilde A_\mu$, to leading order in $\epsilon$. This manifests as a millicharge $\pm \epsilon e$ under the visible $\textrm{U(1)}_{em}$ for the dark proton and dark electron respectively. 
In most of our calculations, we assume that the dark electromagnetic coupling is equal to its SM counterpart, $\alpha_D = \alpha_{SM}$, since this is the scenario realized by the Mirror Twin Higgs~\cite{Chacko:2005pe, Barbieri:2005ri, Chacko:2005vw,Chacko:2018vss} and other Mirror Matter models~\cite{Foot:1999hm, Foot:2000vy, Berezhiani:2003xm, Foot:2004pa, Michaely:2019xpz}, but we also present results for smaller ($\alpha_D = 10^{-3}$) and larger ($\alpha_D = 10^{-1}$) couplings to cover more general aDM scenarios.

There are general constraints on $\epsilon$ as a function of the mass of the lightest particle charged under the dark $\textrm{U(1)}$, from stellar cooling \cite{Davidson:1991si,Davidson:1993sj,Raffelt:1996wa,Davidson:2000hf,Vogel:2013raa}, supernova cooling \cite{Davidson:1993sj}, cosmology \cite{Dubovsky:2003yn,Dolgov:2013una,Vogel:2013raa}, and other experiments \cite{Prinz:1998ua,Badertscher:2006fm,Gninenko:2006fi}. For dark electrons with masses similar to the SM electron, $\epsilon$ is constrained to be $\lesssim 10^{-9}$ from supernova constraints. For a marginal operator not forbidden by any symmetries, this may seem an unnaturally small number; indeed, if there are particles in the spectrum charged under both the visible and dark $\textrm{U(1)}$s, they can generate the kinetic mixing $\epsilon \sim 10^{-3} - 10^{-2}$ at 1-loop. However, in specific constructions such as the Mirror Twin Higgs \cite{Chacko:2005pe, Barbieri:2005ri, Chacko:2005vw}, the kinetic mixing is not generated by the IR degrees of freedom even at 3-loop order due to discrete symmetries. On the other hand, it has recently been shown that higher-order loop diagrams involving gravitons can contribute to the kinetic mixing, which if gravity is UV-completed at the Planck scale are expected to contribute at the $\epsilon \sim 10^{-13}$ \cite{Gherghetta:2019coi} level. 
This makes roughly $\epsilon \sim 10^{-13} - 10^{-9}$ our most motivated kinetic mixing range to experimentally probe. 

The idea that the dark sector could feature composite objects analogous to SM atoms has been around for a long time \cite{Goldberg:1986nk, Kaplan:2009de, Kaplan:2011yj, Cline:2012is}. These models generally feature a dark proton and a dark electron, charged with opposite signs under a dark $\textrm{U(1)}$, which we will take to be unbroken \cite{Holdom:1985ag}. Furthermore one generally assumes a dark asymmetry \cite{Kaplan:2009ag}, with the constraint that the universe has net zero dark charge, i.e. the same number of dark protons and dark electrons. The simplest possibility is that the dark atoms are primarily or exclusively ``dark hydrogen'', a bound state of a dark proton and dark electron. This is the simplified scenario that we study.

If aDM makes up all or part of the observed DM abundance (with the  rest being some other relic, in the simplest case conventional non-interacting cold DM)
then this scenario can lead to a subcomponent of dark matter with potentially high self-interaction cross sections. For instance, charged dark ions interact via Rutherford scattering, which has a $1/v^4$ enhancement, while in atomic form the self-interaction cross sections scale with the square of the atomic Bohr radius. Although dark matter self-interactions are generally constrained to be smaller than $\sigma / m \lesssim 1\,\textrm{cm}^2 / \textrm{g}$ \cite{Markevitch:2003at, Kahlhoefer:2013dca,Harvey:2015hha,Robertson:2016xjh, Wittman:2017gxn, Harvey:2018uwf}, for a DM subcomponent of less than 5-10$\%$ of the total DM density these constraints essentially disappear \cite{Fan:2013yva,Cyr-Racine:2013fsa,Chacko:2018vss}.

Thermal \cite{Fan:2013yva, Agrawal:2017rvu} and non-thermal \cite{Reece:2015lch} production mechanisms have been proposed for atomic dark matter scenarios, as well as means of generating an asymmetric aDM component \cite{Reece:2015lch, Agrawal:2017rvu}. In this work we will consider \emph{only} the asymmetric component of atomic dark matter, since the accumulation of a symmetric component in a star's core is generally limited by its annihilation.\footnote{We have checked this explicitly for several benchmark scenarios of the model we study.}  As we will show, we can constrain dark matter density fractions down to $10^{-3}\,\rho_{DM}$ or smaller, so even production mechanisms that generate a small asymmetric aDM relic are relevant \footnote{For a study in which symmetric accumulations of charged dark matter \emph{can} become relevant, see \cite{Fedderke:2019jur}}.

Depending on model parameters, the galactic aDM halo of ionized or partially ionized dark atoms can cool to form a dark disk analogous to the familiar disk structure of baryons in our own galaxy \cite{Fan:2013yva, Fan:2013tia}. Generally speaking, for heavier dark electrons cooling is less efficient, since cooling mechanisms, such as bremsstrahlung and Compton scattering, depend on some fraction of the dark gas being ionized. Thus for dark electrons much heavier than the SM electron, collapse from a halo to a dark disk may not happen on timescales comparable to the age of the universe.

Dark disk dark matter is generally more difficult to constrain with traditional direct detection measurements\footnote{For a study of mirror dark matter in the context of the recent XENON1T excess, see \cite{Zu:2020idx}. Note that this does not take the possibly non-standard morphology of the DM distribution into account.}, since local dark matter is expected to be co-rotating around the centre of the galaxy and hence its velocity relative to Earth is reduced, leading to low recoil in direct detection experiments \cite{Fan:2013yva}. If the dark matter is atomic and ionization fractions are small, prospects for direct detection are especially poor due to the suppression of the atomic form factor \cite{MTHastro}. Indirect observations of stellar kinematics are able to constrain dark disks much thinner than our milky way  due to their purely gravitational effects \cite{Schutz:2017tfp}. However, aDM scenarios with SM-like masses or sources of feedback  like mirror star formation~\cite{Curtin:2019lhm, Curtin:2019ngc, Foot:1999hm, Foot:2000vy} are unlikely to generate such thin disks and likely evade these bounds (see also~\cite{MTHastro}).

Predicting the precise distribution of atomic dark matter in our own galaxy is highly non-trivial, especially if the dark atoms are part of a richer dark sector. For instance the Mirror Twin Higgs model (discussed below) predicts an atomic subcomponent of dark matter, with interactions qualitatively similar to SM nuclear physics. If such a sector can cool efficiently it is highly plausible that mirror matter can collapse into \emph{mirror stars}, fuelled by mirror nuclear fusion and shining in dark radiation \cite{Curtin:2019lhm, Curtin:2019ngc}. This makes the task of predicting the dark disk structure as complex a problem as the formation of the visible sector disk, which is a highly non-linear process depending on radiative and mechanical feedback mechanisms, such as stellar winds and supernova heating. As also discussed in~\cite{MTHastro}, this can make it difficult to interpret a single constraint as a constraint on a particular model. A wide range of direct and indirect probes is therefore needed to truly pin down the nature of the dark sector.

As well as being a useful simplified benchmark model of dark complexity, atomic dark sectors appear in models with independent motivations, such as the Mirror Twin Higgs \cite{Chacko:2005pe, Barbieri:2005ri, Chacko:2005vw}. Twin Higgs models address the hierarchy problem by positing the existence of a new sector related to the SM by a $Z_2$ symmetry. The twin sector generally features a twin $\textrm{SU(3)}^B_c \times \textrm{SU(2)}^B_W \times \textrm{U(1)}^B_Y$ gauge symmetry, a twin Higgs doublet which mixes with the visible Higgs, and mirror copies of the SM fermions which are \emph{neutral} under the SM gauge group. The 125 GeV Higgs boson is interpreted as a pseudo-Nambu-Goldstone boson (pNGB) of an approximate $SU(4)$ global symmetry that emerges at quadratic level of the action from the $SU(2)_L^A \times SU(2)_L^B \times Z_2$ symmetry of the extended Higgs sector, which solves the little hierarchy problem without colored top partners. 
The $Z_2$ twin symmetry (just like supersymmetry in the MSSM~\cite{Martin:1997ns}) must be  broken in the IR to agree with observations, since otherwise the light Higgs is an equal admixture of visible and SM states with a $\sim$50\% decay rate to the hidden sector.  Collider constraints on invisible decays of the Higgs are avoided by assuming that the vacuum expectation value (VEV) of the twin Higgs $v_B$ is a few times higher than the visible Higgs VEV $v_A$, which can be realized in a variety of mechanisms (see e.g.~\cite{
Beauchesne:2015lva, Yu:2016swa, Harnik:2016koz, Yu:2016bku, Jung:2019fsp, Batell:2019ptb, Harigaya:2019shz}).
Values of $v_B / v_A \gtrsim 3$ are sufficient to evade Higgs width constraints \cite{Burdman:2014zta}, implying that the fermions and weak gauge bosons of the mirror sector are heavier than their SM counterparts. There are also cosmological constraints on this scenario, discussed in some detail below.

The simplest MTH realisation in terms of symmetries keeps the couplings of the twin fermions and gauge bosons identical to their SM values, making the properties of the mirror sector relatively easy to predict. In particular, this minimal model implies specific benchmark values of the dark proton and dark electron masses as a function of the ratio $v_B/v_A$. Assuming $Z_2$-symmetric Yukawa couplings (i.e. only soft $Z_2$ breaking), both the dark electron and dark quarks will be a factor of $v_B/v_A$ heavier, and one can then compute the running of the mirror QCD coupling to find the confinement scale of the mirror sector. A numerical fit gives~\cite{Chacko:2018vss}:
\begin{equation}
\frac{\Lambda_{{QCD}_B}}{\Lambda_{{QCD}_A}} \approx 0.68 + 0.41 \log\left(1.32 + \frac{v_B}{v_A}\right).
\end{equation}
Since the nucleon masses are proportional to the confinement scale, $m_{\tilde p}/m_p$ is expected to be in the same ratio. The dependence of $m_{\tilde p}$ on $v_B/v_A$ is rather modest, varying by 30-50\% for $v_B/v_A =$ 3-5.

We have not yet discussed the cosmological implications of the massless dark photon. BBN and CMB measurements provide bounds on the number of relativistic degrees of freedom at their respective times, $N_\mathrm{eff}$, corresponding to the number of effective neutrino species. The latest bounds on $\Delta N_\mathrm{eff}$, which is defined to be zero in the Standard Model, are $\Delta N_\mathrm{eff}\lesssim 0.25$ ($2\sigma$) \cite{Aghanim:2018eyx} or $\lesssim 0.49$ ($2\sigma$) if one uses the $H_0$ measurement from Ref.~\cite{2018ApJ...855..136R}. 
The minimal atomic dark matter model described in \cite{Fan:2013yva,Fan:2013tia} predicts $\Delta N_\mathrm{eff} \approx 0.2$, which will be probed by CMB stage-4 measurements.
However, the prediction for $\Delta N_\mathrm{eff}$ is extremely model-dependent. For instance, the Mirror Twin Higgs model is more severely constrained due to the presence (in the minimal setup) of a massless dark photon and three generations of mirror neutrinos. The Higgs portal keeps the two sectors in thermal equilibrium down to temperatures $\mathcal O(\textrm{GeV})$ \cite{Barbieri:2005ri}, which leads to a very large contribution to the energy density of the universe during CMB times and $\Delta N_\mathrm{eff} = 5.7$ \cite{Chacko:2016hvu,Craig:2016lyx}, which is robustly excluded. The bounds can be evaded by admitting hard $Z_2$ breaking in the Yukawa sector \cite{Farina:2015uea,Barbieri:2016zxn,Csaki:2017spo,Barbieri:2017opf}, for instance in the Fraternal Twin Higgs \cite{Craig:2015phac} and vector-like Twin Higgs \cite{Craig:2016kue} constructions. An alternative approach that does not require $Z_2$ breaking is an asymmetric reheating process (due to an asymmetric late-time decay) that preferentially heats the visible sector \cite{Chacko:2016hvu,Craig:2016lyx}. This has the effect of diluting the hidden sector contribution to the energy density and reducing $\Delta N_\mathrm{eff}$ to levels that are consistent with current bounds but will also be probed by CMB stage-4.

This discussion serves to illustrate the broader landscape of models in which atomic dark matter lives, as well as the various constraints on these scenarios. Existing direct detection constraints on DM self-interactions are limited, especially for DM fractions of 10\% or smaller.
If the local aDM is ionized, future electron-recoil experiments like SENSEI~\cite{Abramoff:2019dfb} could detect a 1\% aDM fraction for kinetic mixings much smaller than $10^{-9}$ for halo and disk-like distributions (see~\cite{MTHastro}), but aDM direct detection is extremely challenging if it lives in a non-ionized disk-like distribution.
Cosmological constraints are a powerful tool but are model dependent, while constraints from indirect measurements of stellar kinematics only apply to particular DM distributions and also suffer from our inability to predict, in a given model, the DM distribution in our galaxy.

The constraints we present in this work are robust, probing the local aDM density \emph{directly}, and do not greatly rely on any assumptions about the distribution of the dark matter. The only assumption that affects our constraints is the average velocity and ionization of the aDM incident on the white dwarf. We will present our constraints for two limiting benchmarks that bracket a reasonable range of velocities, discussed in Section~\ref{sec:velocity}. We  emphasize that our constraints become \emph{stronger} in the low relative velocity regime, in contrast to conventional direct detection experiments, which suffer from dramatically reduced sensitivity in exactly this regime.

\section{Experimental setup: white dwarfs as aDM detectors}
\label{sec:cooling_constraints}

White dwarfs  are relatively simple physical objects: fusion no longer occurs in their cores, and they cool down as their internal energy is lost to radiation. 
This makes them ideal laboratories for constraining or detecting BSM processes that change the white dwarf cooling rate, either by providing new channels for energy loss or by providing new heat sources.

The most well-known application of this method relies on the fact that light BSM particles with small SM couplings can be produced in the cores of stars or supernovae, carrying away energy. 
This provides very stringent bounds on axions and other dark forces~\cite{Raffelt:1985nj, Raffelt:1987yb, Raffelt:1987yt, Turner:1987by, vanBibber:1988ge, Burrows:1988ah, Raffelt:1989zt, Blinnikov:1994eoa, Raffelt:1994ry, Gondolo:2008dd, Isern:2008fs, Isern:2008nt, Corsico:2012ki, Melendez:2012iq, Dreiner:2013tja, Redondo:2013wwa, Curtin:2014afa, Bertolami:2014wua, Hurst:2014uda,Vinyoles:2015aba, Chang:2018rso, Dessert:2019sgw}, but is naturally limited by the core temperature to BSM particle masses below $\sim$ 100 MeV / 100 keV for supernovae \cite{Chang:2018rso} / horizontal branch, red giant and white dwarf stars \cite{Davidson:1991si,Davidson:1993sj,Raffelt:1996wa,Davidson:2000hf,Vogel:2013raa}.
On the other hand, depending on the DM model, it is also possible for ambient DM to be captured by the star and disrupt it in some way \cite{Press:1985ug, Gould:1987ju, Goldman:1989nd, Raffelt:1996wa, Spolyar:2007qv, Bertone:2007ae, McCullough:2010ai, Kouvaris:2010jy, deLavallaz:2010wp, Kouvaris:2011fi, McDermott:2011jp, Iocco:2012wk, An:2013yfc, Leung:2013pra, Bramante:2013hn, Bell:2013xk, Dreiner:2013tja, Bertoni:2013bsa, Vincent:2014jia, Bramante:2014zca, Graham:2015apa, Bramante:2015cua, Raj:2017wrv, Baryakhtar:2017dbj, Graham:2018efk, Bell:2018pkk, Ellis:2018bkr, Garani:2018kkd, Acevedo:2019gre, Acevedo:2019agu, Michaely:2019xpz, Dasgupta:2019juq, Garani:2019fpa, Dasgupta:2020dik, Garani:2020wge}.
Such stellar probes of ambient DM effectively measure the same observable as DM direct detection experiments on Earth, and are not restricted to light DM masses since they do not rely on producing the BSM particles.

In our study, we combine features of both approaches by noticing that aDM with a small SM coupling~--~due to its asymmetric, dissipative, multi-component nature~--~leads to DM being captured in the star, which accumulates and collapses to a small and relatively cold aDM nugget in the core. The aDM nugget saps away heat and radiates the energy away in the form of dark photons. 
Accumulation of aDM therefore acts as an invisible heat sink for stars. 
This is exactly analogous to the capture of SM matter in mirror stars that makes them visible in optical and X-ray telescopes~\cite{Curtin:2019lhm,Curtin:2019ngc}.
White dwarfs, due to their high core density which allows for more heat to be transferred to the nugget, are much more efficient radiators of dark photons compared to main sequence stars, and cooling constraints can provide the strongest bounds on the kinetic mixing parameter $\epsilon$ and the local abundance $n_{aDM}$ for dark electron masses above 10 keV, where most conventional stellar cooling constraints lose sensitivity. 

In this section we review how observational data of white dwarfs constrains new cooling channels and define a simulated benchmark white dwarf star we use for our calculations (Section ~\ref{s.WDLF}). 
We then discuss the assumptions about the local aDM velocity distribution and ionization that we make to derive our bounds (Section~\ref{sec:velocity}). The actual contribution to white dwarf cooling from aDM accumulation will then be discussed in Section~\ref{sec:capture} and~\ref{sec:heating_cooling}.

\subsection{The White Dwarf Luminosity Function}
\label{s.WDLF}

\begin{figure}
\centering
\includegraphics[scale=0.4]{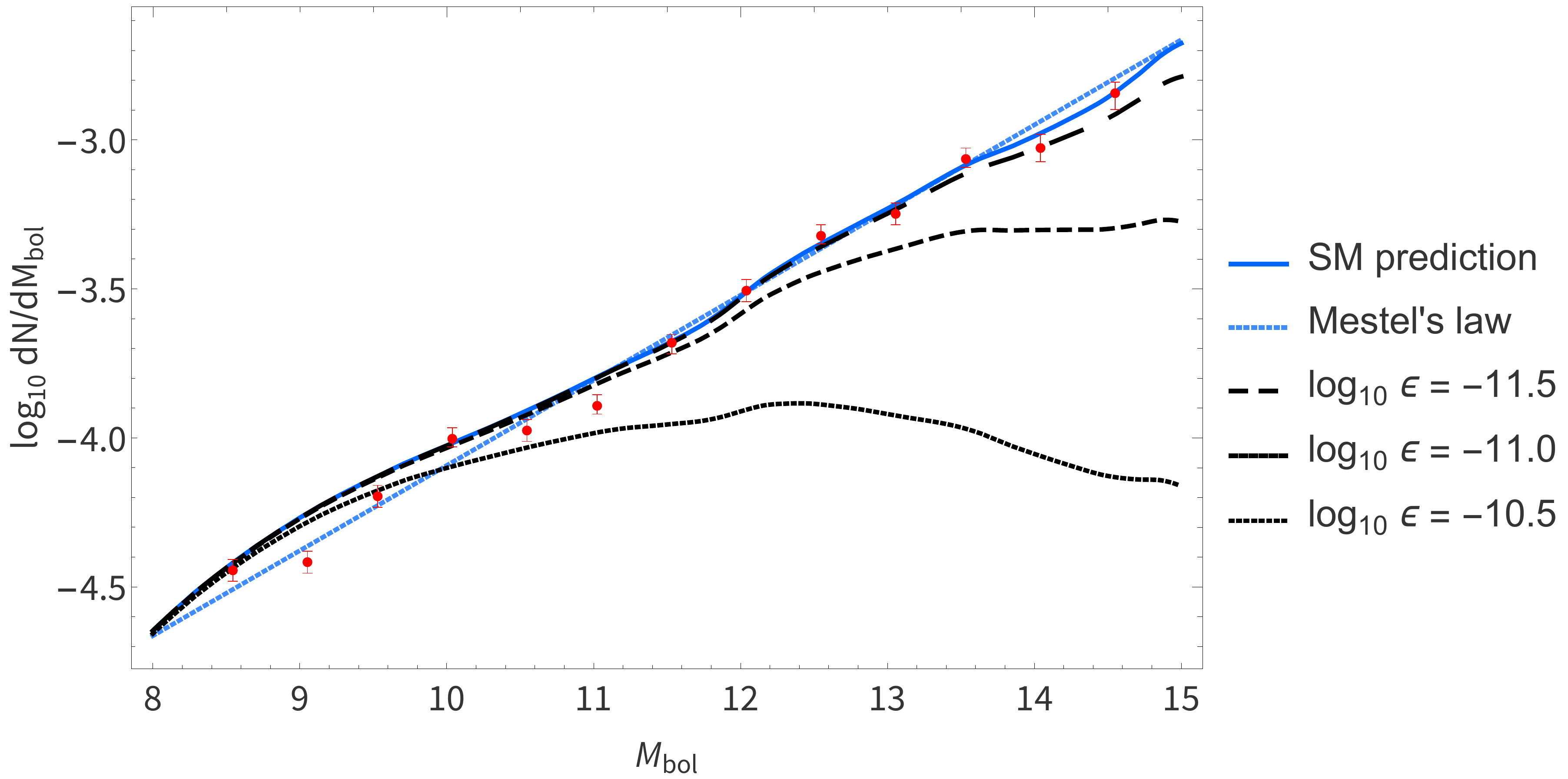}
\caption{
\emph{Red data points:}
The measured white dwarf luminosity function for white dwarfs showing only hydrogen features in their atmospheres, assuming a constant formation rate~\cite{Harris:2005gd}. 
\emph{Blue curve:}  Theoretical prediction resulting from detailed modeling of the WDLF~\cite{Isern:2008fs}. 
\emph{Blue dotted line:} The simple approximate prediction of Mestel's law.
\emph{Black  curves}: 
Predicted luminosity functions with accumulated aDM for different values of $\epsilon$, with dark atom masses $\{ m_{\tilde p}, m_{\tilde e}\} = \{ m_p , m_e \}$ forming 10\% of the local dark matter density. 
The deviations from Standard Model predictions are more pronounced for older, cooler white dwarf stars, and become more significant for higher values of $\epsilon$. 
As is conventional, the SM WLDF predictions are normalized to pass through the point with the smallest error bars. The aDM curves are normalized to the SM one at small magnitudes, to emphasise that deviations from predicted behaviour are most dramatic at larger magnitudes.
}
\label{fig:luminosity_function}
\end{figure}

The WDLF, defined as the number density of white dwarfs of a given bolometric magnitude ${dN}/{dM_\textrm{bol}}$, 
offers a means to test our understanding of white dwarf cooling. (See e.g.~\cite{Raffelt:1996wa, Raffelt:1999tx} for reviews.) 
The observed WDLF~\cite{Harris:2005gd} for white dwarfs showing only hydrogen features in their atmospheres is shown as the red data points in Fig.~\ref{fig:luminosity_function}.
The detailed physics of white dwarf cooling is complex and beyond the scope of this work, but as we show in the following, these details are not important to the robustness of our BSM constraints. 

Assuming a constant white dwarf production rate, and assuming all white dwarf stars of a given magnitude cool at the same rate,\footnote{As we explain below, this is a reasonable approximation for our purposes, but a more complete treatment would involve modifying the RHS of Eqn.~(\ref{eq:LF}) to integrate over different cooling times for a distribution of white dwarf masses and possibly compositions.} 
the number density of white dwarfs in a given magnitude interval $\{M_\textrm{bol}, M_\textrm{bol} + dM_\textrm{bol} \}$ is proportional to the time a white dwarf takes to cool through that interval:
\begin{equation}
    \label{eq:LF}
    \frac{dN}{dM_\textrm{bol}} \propto \frac{dt}{dM_\textrm{bol}}.
\end{equation}
The bolometric magnitude $M_\textrm{bol}$ is defined via $M_\textrm{bol} - M_{\textrm{bol},\odot} = -2.5 \log_{10} (L_\gamma / L_\odot)$, where $L_\gamma$ is the visible luminosity (into Standard Model photons) of the white dwarf and $M_{\textrm{bol},\odot}$ is taken to be 4.75, following convention. 
We can rewrite \eqref{eq:LF} as the following:
\begin{equation}
    \frac{dN}{dM_\textrm{bol}} \propto \frac{dt}{dU} \frac{dU}{dL_\gamma} \frac{dL_\gamma}{dM_\textrm{bol}},
\end{equation}
where $U$ is the total internal energy of the white dwarf. Both the visible luminosity and the energy loss to dark radiation contribute to the cooling:\footnote{Here we are ignoring cooling by neutrino emission, since this is only a relevant effect for very hot white dwarfs, with magnitudes around $M_\textrm{bol}\approx$ 6-7 \cite{Dreiner:2013tja}.}
\begin{equation}
    -\frac{dU}{dt} = L_\gamma + L_{dark},
\end{equation}
and from the definition of bolometric magnitude we have that
\begin{equation}
    \frac{dL_\gamma}{dM_\textrm{bol}} \approx -0.921 \, L_\gamma.
\end{equation}
Thus we find
\begin{equation}
    \frac{dN}{dM_\textrm{bol}} \propto \frac{L_\gamma}{L_\gamma + L_{dark}} \frac{dU}{dL_\gamma}.
\end{equation}
If we assume that the effect of the aDM on the structure of the white dwarf is minimal (besides the increased rate of cooling) then we can argue that the value of $dU/dL_\gamma$ is unaffected by the new physics. Thus the logarithm of the luminosity function is given by
\begin{equation}
\label{e.dNdMbol}
    \log \left( \frac{dN}{dM_\textrm{bol}} \right) = \log\left( \frac{dN}{dM_\textrm{bol}} \right)_{SM} + \log \left( \frac{L_\gamma}{L_\gamma + L_{dark}} \right),
\end{equation}
where the first term on the RHS is the WDLF predicted by the Standard Model, and the second term is the correction due to new physics. Note that this correction neatly separates from the complex white dwarf physics needed to accurately calculate the luminosity function. This form of the correction is robust even if one does not assume a constant white dwarf production rate.

Using white dwarfs as probes of new physics requires predictions for both terms in 
\eref{e.dNdMbol}.
For the SM contribution, a simplified analysis in which the luminosity of the white dwarf scales with $L_\gamma \propto T^{7/2}$ (Mestel's approximation), predicts \cite{1952MNRAS.112..583M}
\begin{equation}
    \log_{10}\left( \frac{dN}{dM_\textrm{bol}} \right)_{SM} = C + \frac{2}{7} M_\textrm{bol}.
\end{equation}
This already gives quite a reasonable fit to the data, but more sophisticated white dwarf stellar modeling \cite{Isern:2008fs} allows the observed WDLF to be very closely matched, see \fref{fig:luminosity_function} (blue curve). 
Note that the overall normalization of the WDLF is conventionally taken from data, since it depends on the total white dwarf formation rate. On the other hand, the shape of $dN/dM_\textrm{bol}$ is a robust prediction that is quite sensitive to BSM effects.

\begin{figure}
 \centering
    (a)\hspace{3mm}\begin{subfigure}[]{0.4\textwidth}
        \includegraphics[scale=0.5]{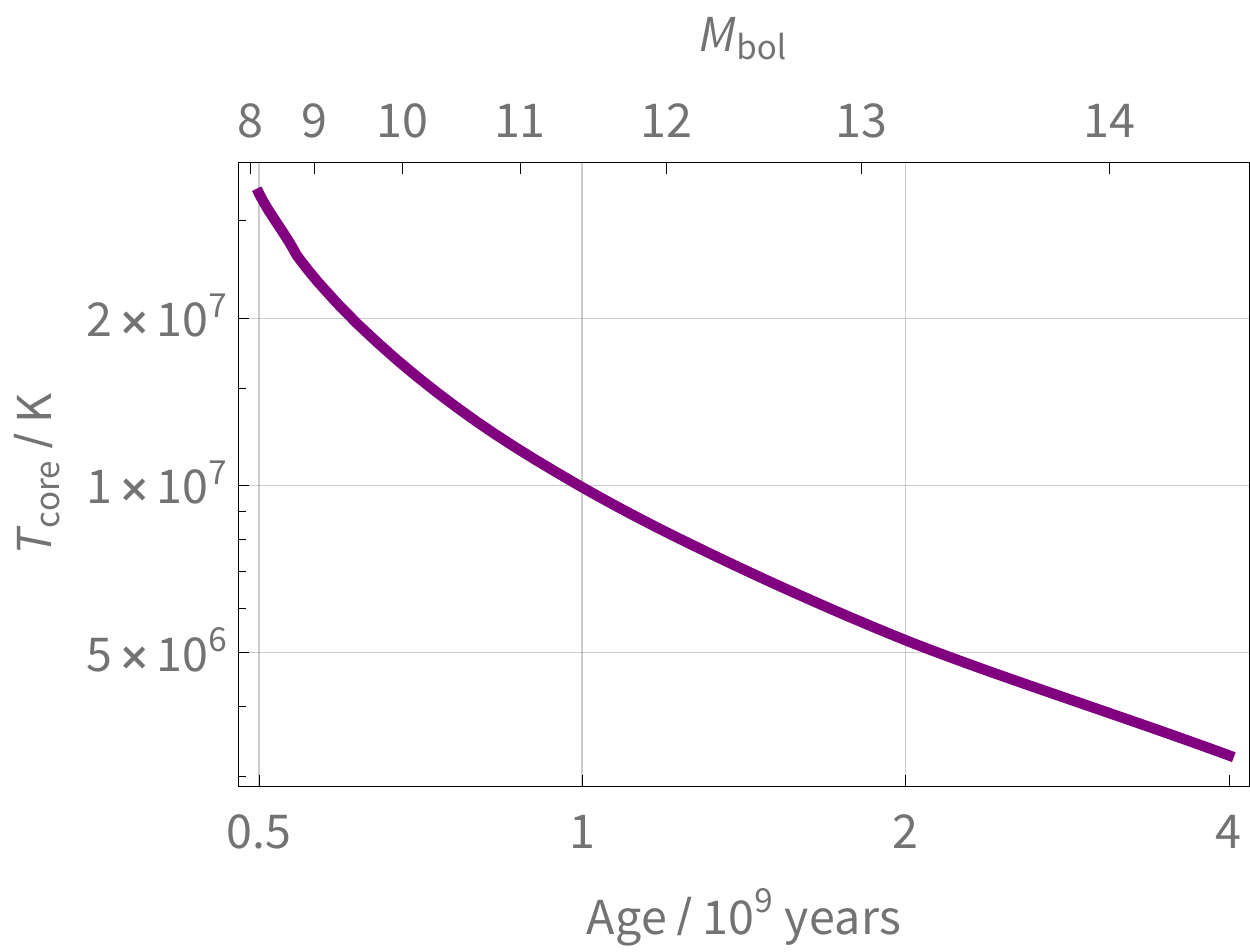}
    \end{subfigure}\\
    (b)\hspace{3mm}\begin{subfigure}[]{0.4\textwidth}
        \includegraphics[scale=0.5]{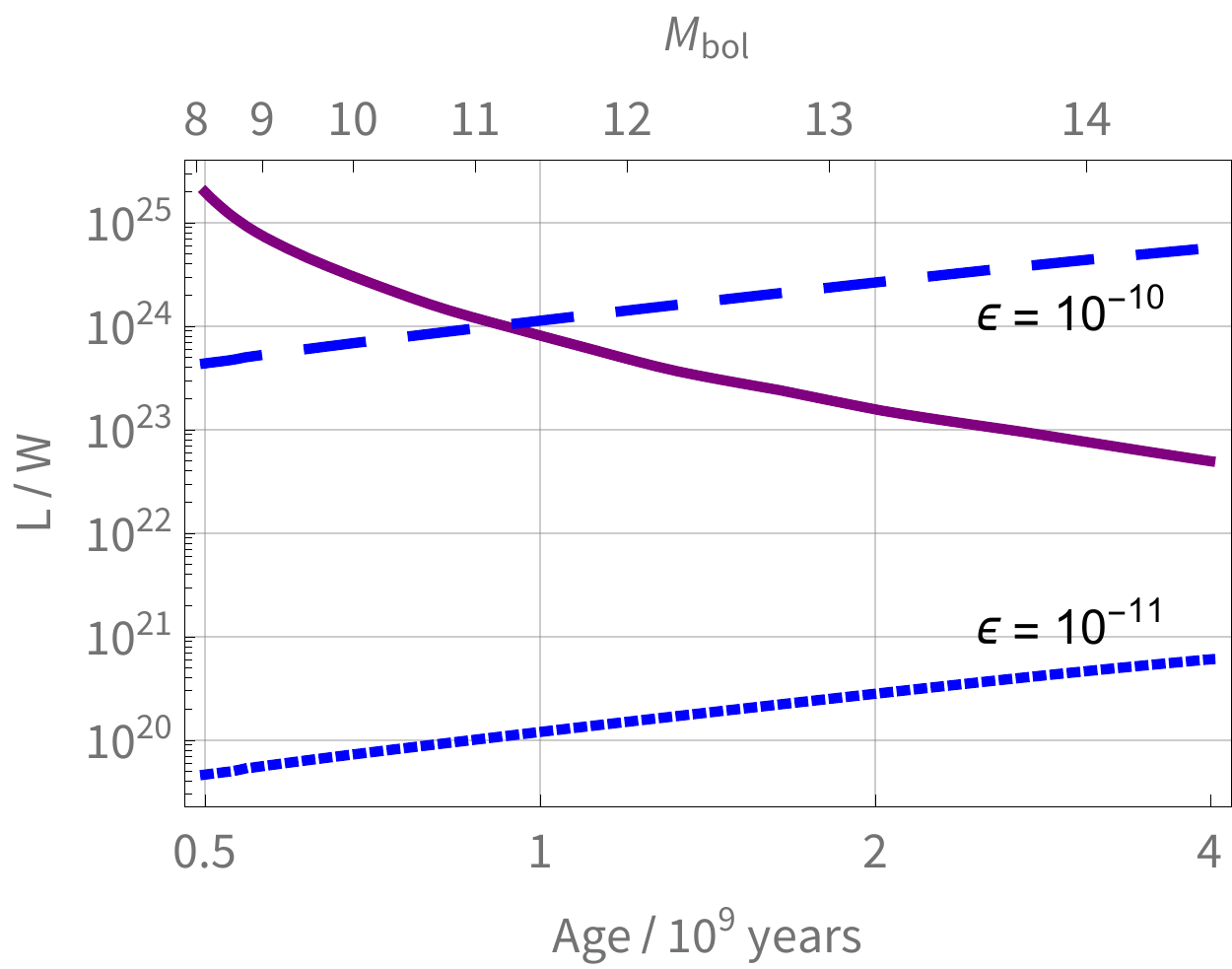}
    \end{subfigure}
    \hspace{0.5cm}
    (c)\hspace{3mm}\begin{subfigure}[]{0.4\textwidth}
        \includegraphics[scale=0.5]{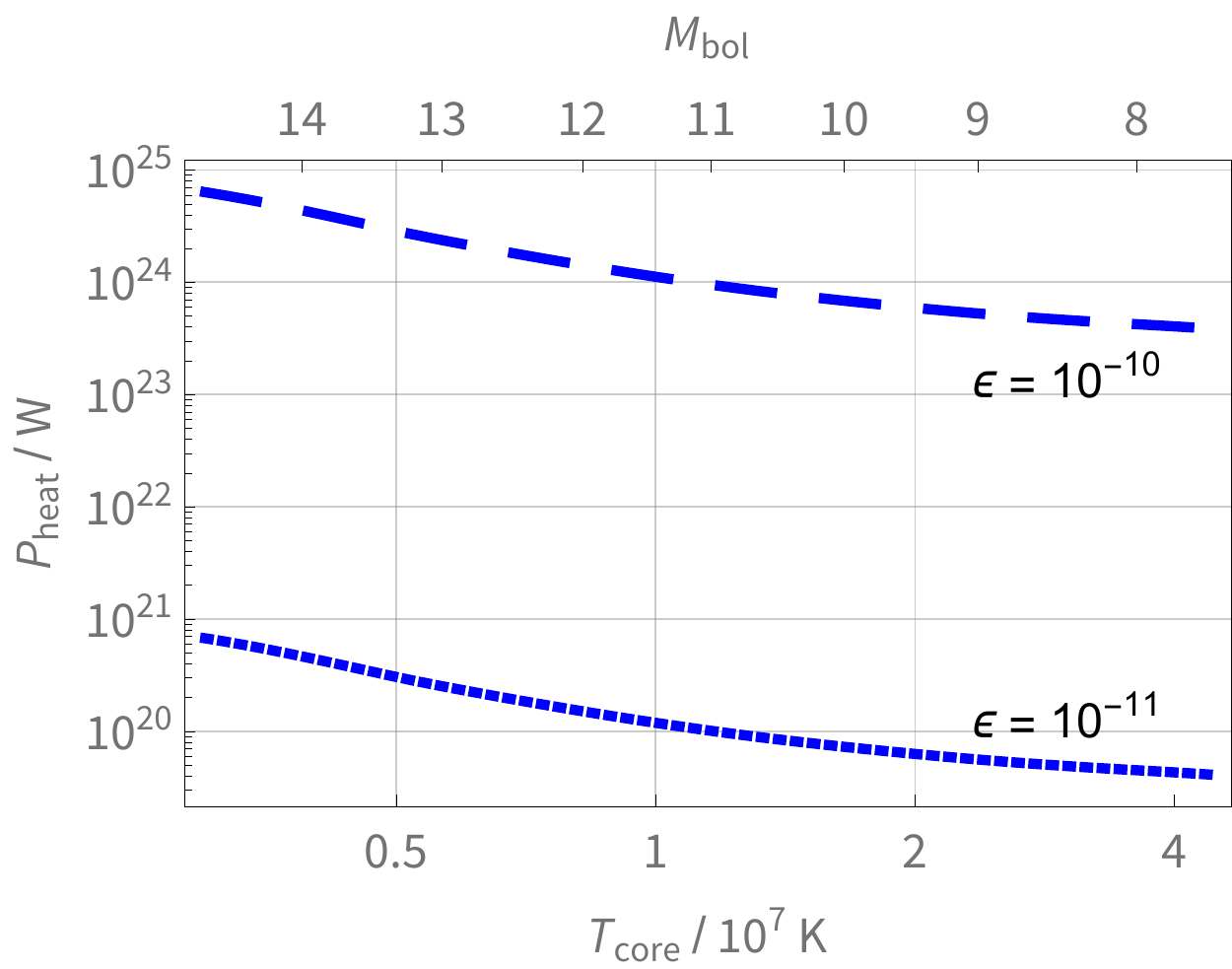}
    \end{subfigure}
\caption{
(a) Core temperature of our benchmark star once it enters the white dwarf phase, as a function of its age or bolometric magnitude.
(b) shows the white dwarf energy loss from normal SM photon radiation (purple curve), as well as radiation of dark photons due to the accumulated aDM nugget (under the assumption that the extra cooling does not affect stellar structure).
(c) shows the heating rate of the DM nugget as a function of the WD core temperature, to illustrate the surprising result that cooler white dwarfs can actually be \emph{more} efficient at heating (see Section~\ref{sec:heating_cooling}).
For the sake of this example we have assumed the ambient atomic dark matter density to be 1\% of the total dark matter density, and assumed the fast $\vev{v} = 220$ km/s DM velocity distribution.
}\
\label{f.WDevolution}
\end{figure}

The second term in Eqn.~(\ref{e.dNdMbol}) is determined by computing the energy lost to BSM processes $L_{dark}$ as a function of  the instantaneous white dwarf luminosity $M_\textrm{bol}$ (as well as the BSM parameters).
In keeping with the approximation of \eref{eq:LF}, we choose a single \emph{benchmark star} to represent the entire white dwarf population. This benchmark star has 
a mass of $3\,M_\odot$ during its main sequence phase before moving to the Asymptotic Giant Branch (AGB) as it commences its He-burning phase, and finally 
becoming a $0.6\,M_\odot$ white dwarf. 
These stellar parameters were chosen since the white dwarf mass function is strongly peaked around $0.6\,M_\odot$ \cite{2016MNRAS.461.2100T}.
We simulate the entire life cycle of the star using MESA \cite{Paxton2011,Paxton2013,Paxton2015,Paxton2018,Paxton2019}, starting from its pre-main sequence phase and ending the simulation when the white dwarf has cooled to a luminosity of $10^{-4} L_\odot$. See \fref{f.WDevolution} for the properties of the benchmark star during its white dwarf cooling phase.

Note that we can use the simulation of this single benchmark star to directly compute $\log({dN}/{dM_\textrm{bol}})_{SM}$. Our result closely matches the more sophisticated modeling of the WDLF shown as the blue curve in \fref{fig:luminosity_function}. This verifies our approach of using a single benchmark star to derive cooling bounds.

At each time during the benchmark star's life, we can compute aDM capture and evaporation rates to track the amount of accumulated aDM and compute the energy lost to dark photon radiation. This allows us to compute $L_{dark}(M_\mathrm{bol})$ during the white dwarf phase, and hence the deviation from the SM WDLF for a given aDM scenario.\footnote{As part of our calculation we also compute energy loss during the main sequence phase of our benchmark star (and for other stellar masses), and verify that bounds on white dwarf cooling supply the strongest bounds on aDM.}

\begin{table}
    \begin{center}
    \hspace*{-8mm}
    \begin{tabular}{| c | c | c | c | c | c | c | c | c |}
    \hline
     & {\Large$\frac{R}{R_\odot}$} & {\Large$\frac{T_\mathit{core}}{10^7 \,\textrm{K}}$} & {\Large$\frac{\rho_\mathit{core}}{\textrm{g\,cm}^{-3}}$} & {\Large$\frac{L_\gamma}{L_\odot}$} & {\Large$\frac{\tau}{\textrm{years}}$}  \\
    \hline
    Main sequence (MS) & 2.49 & 2.43 & 37.6 &  92.3 & $3 \times 10^8$ \\
    Asymptotic giant (AGB) & 15.1 & 11.3 & $2.97\times 10^4$ & 112 & $1.4\times 10^8$ \\
    White dwarf (WD) & 0.0121 & 0.288 & $3.95 \times 10^6$ & $10^{-4}$ & $4 \times 10^9$ \\
    \hline
    \end{tabular}
    \end{center}
    \caption{Properties of the benchmark star we simulate at various points throughout its lifetime. We note that $L_\gamma$ is the visible luminosity of the star into SM photons (as opposed to dark photons). The last column $\tau$ is the time the star spends in each phase. For the white dwarf phase, this marks the age at which it reaches $M_{bol} = 14.7$, populating the faintest bins in the measured WDLF, see Fig.~\ref{fig:luminosity_function}.}
    \label{table:benchmarks}
\end{table}

In practice, we use three snapshots of the star's evolution to compute aDM capture and evaporation rates during the corresponding parts of the stars life cycle: 1) halfway through its main sequence lifetime, 2) halfway through its helium burning lifetime, and 3) at the end of its white dwarf cooling period. Specifically, this means that the properties of the star (its temperature, density and composition profiles) are held constant during each phase, although the total dark matter accumulation (and the capture and evaporation rates, which depend on the accumulation) are computed as a continuous function throughout the star's lifetime.
Using the white dwarf model at the \emph{end} of its cooling to estimate the capture rate throughout the entire cooling period is strictly an underestimate, since the white dwarf gets smaller and denser as it cools, and so its geometric cross section and that of the aDM nugget will be underestimated.
%\footnote{Note that while the capture and evaporation rate are `updated' only at these snapshots along the star's life, the resulting white dwarf cooling rate is computed continuously to track the effect on the WDLF.}
%
In Table~\ref{table:benchmarks} we list the properties of our benchmark star at these three snapshots.

For the purpose of setting aDM constraints, we assume that we can exclude scenarios that result in the benchmark star having a dark luminosity \emph{equal} to its visible luminosity, $L_{dark}~=~L_\gamma$ at any observed magnitude, which corresponds to a deviation in $\log_{10} dN/dM_\mathrm{bol}$ of
\begin{equation}
\label{e.WDLFchange}
    \log_{10} \left( \frac{L_\gamma}{L_\gamma+L_{dark}} \right) = \log_{10} \left( 1/2 \right) \approx 0.30
\end{equation}
compared to the SM WDLF. 
This is well outside the error bars of Fig.~\ref{fig:luminosity_function} and numerous other studies \cite{refId0,Harris:2005gd}, making our choice conservative\footnote{We note that in Ref. \cite{Isern:2008fs} the authors model an alternative WDLF under the additional assumption that the white dwarf formation rate drops exponentially on galactic timescales. This can have the effect of producing an excess of white dwarfs in the cool, old end of the luminosity function, around $M_\mathrm{bol} \approx$ 15. This raises the possibility that for a particular value of $\epsilon$ this excess could cancel the deficit induced by aDM cooling effects. We do not expect this to affect our analysis significantly, since even if this cancellation obscures BSM effects at the coldest end of the WDLF, we have verified that the fit to the data at smaller $M_\mathrm{bol} \approx$ 13 would be worse, and this discrepancy would be revealed by precision studies.}.

\subsection{Local aDM distribution}
\label{sec:velocity}

The white dwarfs which constitute the WDLF dataset \cite{Harris:2005gd} are in our galactic neighborhood with distances below a kiloparsec. Therefore we assume that their aDM capture rate reflects the local aDM density, ionization and velocity distribution. 

We will derive constraints on the local density  $\rho_{aDM}$ by assuming that the ambient aDM is fully atomic, i.e. has negligible ionization. This is largely conservative, since dark-atomic form factors would slightly suppress aDM capture. It also corresponds to the scenario that is most difficult to probe in direct detection experiments, since it precludes the possibility of detecting the dark-electron population in sensitive electron-recoil experiments~\cite{MTHastro}. 
However, across most of the parameter space we consider (and in the case of white dwarf capture, across \emph{all} of the parameter space we consider) the stellar escape velocity is large enough that the kinetic energy of incoming aDM exceeds the binding energy of the dark atom, so we not expect ionization of aDM to play a major role in its accumulation. We therefore expect our results to apply to scenarios with locally ionized aDM as well.\footnote{This assumes that dark plasma screening effects can be neglected, see~\cite{MTHastro} for the first examination of this effect.}

The velocity dependence in the SM-aDM interaction cross section makes our results somewhat dependent on the local aDM velocity distribution. As we already discussed, this distribution is extremely difficult to predict, and we therefore adopt two benchmarks which should bracket the range of possibilities, also used in~\cite{MTHastro}. These benchmarks correspond to aDM that is arranged in a CDM-like halo-configuration, or aDM that has efficiently cooled and collapsed into a dark disk that is perfectly aligned with our Milky Way:

\begin{itemize}
    \item \emph{Halo benchmark} -- if the atomic dark matter is halo-like, then it is expected to have the usual cold dark matter velocity distribution, taken to be a Maxwell-Boltzmann distribution with $\langle v \rangle = 220$~km/s. 
    \item \emph{Disk benchmark} -- in the case of a dark disk which is closely aligned with the visible galactic disk, the dark matter could be co-rotating with the Earth around the centre of the galaxy. In this case we take the relative velocity distribution to be a Maxwell-Boltzmann distribution with $\langle v \rangle = 20$ km/s. 
    This is the approximate velocity dispersion of stars and gas in our local stellar neighborhood~\cite{Dehnen:1997cq,LopezSantiago:2006xv} and therefore represents a lower bound on the velocity dispersion of DM relative to the white dwarfs we study. 
\end{itemize}
We will see that our constraints are somewhat stronger for lower dark matter velocities, due to the enhanced capture rate. This is notably in contrast to ordinary direct detection experiments, which generally have worse sensitivity at lower velocities.

Note that in the halo benchmark, we neglect the relative velocity between the white dwarf and the aDM halo. This would vary from star to star, is likely not known for each star over its lifetime, and at any rate is unlikely to have a major effect on our sensitivities, since we find that the sensitivity for such widely different assumptions as the halo and disk benchmark only differs by a factor of few in kinetic mixing. Any decrease in sensitivity from taking this relative velocity into account (which would slightly decrease capture) would therefore be more than compensated by performing a more realistic fit to the WDLF than our very conservative criterion in \eref{e.WDLFchange}. This simplification will therefore not significantly change our results.

\section{Accumulation of aDM in stars}
\label{sec:capture}

In this section we describe the calculation of the total accumulation of the aDM nugget in the white dwarf. 
We focus on aDM scenarios where the dark electron and hence also the dark proton mass is above 10 keV, where stellar cooling bounds lose sensitivity.
This includes dark matter that is captured during the main sequence lifetime of the parent star, as well as dark matter captured by the white dwarf during its cooling period. The capture of dark matter by stars has been covered extensively in the literature \cite{Gould:1987ir, Zentner:2009is, Petraki:2013wwa, Catena:2015uha, Garani:2017jcj}. In particular, we treated the case of capture of SM baryons by Mirror Stars in \cite{Curtin:2019lhm, Curtin:2019ngc}, which is entirely analogous to the capture of dissipative, asymmetric dark atoms by ordinary stars, except that here we also have to keep detailed track of evaporation due to the wide mass range of aDM constituents we consider.

We note in advance that we will assume that the densities and temperatures of the captured aDM are never high enough that \emph{dark nuclear fusion} can occur. Of course, the details of dark fusion rates will depend on the detailed properties of the dark sector, but for the sake of generality we will neglect this possibility.

\subsection{Capture}

We note that it has been suggested \cite{Foot:2014mia,Michaely:2019xpz} that a significant amount of aDM could have accumulated during the star's formation period. However the formation of stars from a dense molecular cloud is a complex and highly non-linear process, and we are not aware of any robust estimates of the resulting aDM accumulation. Dissipative dark matter in overdense regions can presumably also collapse into compact objects, but since the dark matter will be at a different temperature and density, its associated Jeans' scale will be different, and there is no \emph{a priori} reason to suppose that the resulting compact objects will overlap significantly with each other. In any case, accounting only for the capture of dark matter during the remainder of the star's lifetime will be an underestimate, and our bounds will remain robust.

There are two main capture channels, capture of dark matter via its interactions with SM matter in the star (SM-dark capture) and capture via interaction of the dark matter with already captured dark matter (dark-dark self capture).

The total capture rate is given by 
\begin{equation}
\frac{d N_{aDM}}{dt}  = n_{aDM} \sum_i C_i ,
\end{equation}
where the sum is over different target nuclei in the star with which the incoming dark matter can scatter, and $n_{aDM}$ is the ambient aDM density. The capture coefficient for scattering with species $i$ is given by \cite{Gould:1987ir}
\begin{equation}
    C_i = \int dV \int du \,\frac{f(u)}{u} w(r)\, \Omega_+^i(w, r),
\end{equation}
where the integrals are over the volume of the star and the distribution of incoming dark matter velocities $u$. The local capture rate $\Omega_+$ is defined as 
\begin{equation}
    \Omega_+^i(w, r) = n_i(r) w(r) \,\Theta(E_R^{max} - E_R^{min}) \int_{E_R^{min}}^{E_R^{max}} d E_R \frac{d\sigma_i}{dE_R},
\end{equation}
where $n_i$ is the number density of species $i$, $w$ is the incoming scattering velocity, related to the velocity at infinity via $w(r) = \sqrt{u^2 + v_{esc}(r)^2}$. The limits $E_R^{min}$ and $E_R^{max}$ are the minimal target recoil energy for capture, and the maximum possible recoil energy, respectively.

The capture cross section for scattering via the vector portal is closely related to the Rutherford cross section:
\begin{equation}
    \label{eq:rutherford}
    \frac{d\sigma}{d\Omega} = \frac{4 \epsilon^2 \alpha \alpha_D Z_1^2 Z_2^2 \mu^2}{q^4} = \frac{\epsilon^2 \alpha \alpha_D Z_1^2 Z_2^2}{4 \mu^2 v^4 \sin^4(\theta/2)},
\end{equation}
where $Z_i$ are the charges under the respective $\textrm{U(1)}$s, $\mu$ is the reduced mass of the DM-target system, $v$ is the relative velocity, and $\theta$ is the deflection angle. The momentum transfer $q$ for two body elastic scattering is given by $2 \mu v \sin(\theta/2)$. Note that we assume that the target species are at rest, rather than taking into account the velocity distribution of the hot stellar material, which is a reasonable and common assumption for dark matter capture. For the case where the incoming dark matter is atomic, rather than ionized, the finite size of the atom screens the dark charge at distances greater than the Bohr radius \cite{Cline:2012is}, and the cross section is regulated as follows:
\begin{equation}
    \label{eq:rutherford_screened}
    \frac{d\sigma}{d\Omega} = \frac{4 \epsilon^2 \alpha \alpha_D Z_1^2 Z_2^2 \mu^2}{(\Lambda^2 + q^2)^2} = \frac{4\epsilon^2 \alpha \alpha_D Z_1^2 Z_2^2 \mu^2}{(\Lambda^2 + 4 \mu^2 v^2 \sin^2(\theta/2))^2},
    \end{equation}
where $\Lambda = 1/\tilde a_0$ is the reciprocal of the dark atomic Bohr radius and $Z_2$ is the charge of the dark nucleus, taken to be 1 for dark hydrogen. This can be expressed in terms of the recoil energy of the target nucleus:
\begin{equation}
    \label{eq:capture_recoil}
    \frac{d\sigma}{dE_R} = \frac{8\pi\epsilon^2 \alpha \alpha_D Z_1^2 Z_2^2 m_T}{v^2 (\Lambda^2 + 2m_T E_R)^2},
\end{equation}
where $m_T$ is the mass of the target. As discussed above, note that we always assume the incoming dark matter is atomic, rather than ionized.

For the purposes of estimating dark-dark self capture, the density profile of the aDM nugget can be estimated by assuming it is isothermal at some temperature $T_{nugget}$. In that case its density profile is well approximated by a Gaussian with width obtained from the virial theorem:
\begin{equation}
\label{e.rvirial}
    r_{virial} = \left( \frac{9 k T_{nugget}}{4\pi G \rho_{core} \overline m_{DM}} \right)^{1/2},
\end{equation}
where $\rho_{core}$ is the density at the core of the parent star, and $\overline m_{DM}$ is the mean mass of the aDM nugget constituents.

There are two upper limits worth noting: dark-dark capture cannot exceed the \emph{geometric} limit set by the total size of the aDM nugget, and the total capture cannot exceed the geometric limit set by the size of the star. The geometric limit of the capture rate arises because the total capture cross section cannot exceed the area of the target presented to the incoming dark matter flux, namely $\sigma_{capture}^{max} \sim \pi R^2$, with $R$ the approximate radius of the nugget or the star. Taking into account gravitational focussing, the geometric rate scales with
\begin{equation}
    C_{geo} \sim \frac{\overline v_{esc}^2}{u} \pi R^2,
\end{equation}
where $\overline v_{esc}$ is the escape velocity averaged over the density distribution of the target material, and $u$ is the incoming velocity of the dark matter \cite{Petraki:2013wwa, Curtin:2019ngc}.

The total capture coefficient is then:
\begin{equation}
\label{e.Ctot}
    C_{tot} = \min( C_{SM} + \min(C_{DM}, C_{geo}^{nugget}), \,C_{geo}^{star}),
\end{equation}
so that the total capture rate is given by $n_{aDM} C_{tot}$.

Note that $C_{DM}$ is proportional to the amount of dark matter captured, so that before self-capture reaches its geometric limit, the accumulation grows exponentially. The geometric limit for dark-dark capture is usually reached relatively quickly on the scale of the star's lifetime. When calculating the total dark matter abundance we compute the SM-dark capture using the three snapshots of the star's evolution in Table~\ref{table:benchmarks} and take into account the variation of the total capture rate due to accumulation throughout the star's lifetime, so that the total accumulation is proportional to $\int_0^{\tau_{star}} dt \,C_{tot}(t)$.

For low values of $\epsilon \lesssim 10^{-10}$, geometric self capture from the aDM nugget often dominates the total capture rate. This means an estimate of the nugget temperature is necessary for a reliable calculation. In Section~\ref{sec:heating_cooling} we describe our calculation of the heating and cooling rate of the nugget, which allows us to self-consistently solve for the temperature of the nugget as a function of the accumulation. A colder nugget will be smaller, and therefore have a smaller geometric capture rate, which scales with $T_{nugget}$.

The benchmark white dwarfs that we study are cool, 0.6 solar mass white dwarfs coming from 3 solar mass parent stars. The main sequence lifetime of a 3 solar mass star is approximately 300 million years, while we will be considering white dwarfs that have had up to around 4 billion years to cool (in order to populate the entire WDLF, see Section~\ref{sec:cooling_constraints}).
 This means that, although the white dwarf's dark matter capture rate is generally lower than the main sequence capture rate, the accumulation during the white dwarf lifetime can be comparable to the main sequence accumulation, especially when evaporation is taken into account (see below).
  Thus we will account for the dark matter accumulation over the entire lifetime of the star.

\subsection{Evaporation}
\label{sec:evaporation}

Captured dark matter can be lost from both the parent star and the white dwarf via evaporation, which occurs when captured particles gain enough thermal energy that their velocity exceeds the escape velocity of the star and they are ejected \cite{Griest:1986yu}.

\subsubsection*{Rough estimate}

Without making use of detailed stellar simulations, one can estimate the dark matter mass at which evaporation starts to become important by comparing the average thermal energy at the core of the star, $\frac{3}{2} k T$, with the kinetic energy required to escape, $\frac{1}{2} m v_{esc}^2$. The latter is given by the work that must be done to eject a particle of mass $m_{DM}$ from the core of the star,
\begin{equation}
    \frac{1}{2}m v_{esc}^2 = \int_0^R \frac{G M(r) m}{r^2} dr + \frac{G M(R) m}{R},
\end{equation}
where $M(r)$ is the mass enclosed at radius $r$. 
Numerically integrating the mass profiles for our benchmark star we find:
\begin{equation}
    v_{esc} \approx \begin{cases}
        1700~\text{km/s} & \text{Main sequence,} \\
        2600~\text{km/s} & \text{Asymptotic giant,} \\
        7000~\textrm{km/s} & \text{White dwarf}.
    \end{cases}
\end{equation}
The dark matter masses for which $\frac{3}{2}k T_{core} = \frac{1}{2} m_{DM} v_{esc}^2$ are, for each phase of our benchmark star:
\begin{equation}
    \label{eq:critical_masses}
    m_{DM} \approx \begin{cases}
        200~\text{MeV} & \text{Main sequence,} \\
        380~\text{MeV} & \text{Asymptotic giant,} \\
        1.4~\textrm{MeV} & \text{White dwarf}.
    \end{cases}
\end{equation}
Naively we expect dark matter masses around or below these values to be efficiently evaporated during the relevant part of the star's lifetime. This estimate is somewhat conservative in that it does not take into account details such as the collisional cross section and energy transfer. For dark matter particles much lighter than the mean nuclear mass in the core of the white dwarf, energy transfer will be less efficient and evaporation will be further suppressed. This in fact will allow us to probe to much lighter masses than the 1.4 MeV suggested by the above estimate. The evaporation rate tends to be very sensitive to the mass, because of the exponential dependence $\sim e^{-\frac{m v^2}{2 k T}}$ in the velocity distribution. Given this sensitivity and the fact that the values in \eqref{eq:critical_masses} are well within the range we wish to probe, a more detailed calculation of the evaporation rate is warranted.

\subsubsection*{Detailed calculation}

The rate of evaporation of dark matter due to collisions with the SM material in the core is given by \cite{Griest:1986yu}
\begin{multline}
\Omega^-(v_{esc}, r) = \sum_i n_i(r) \int_0^{v_{esc}(r)} d^3v_{DM} 
\int_0^\infty d^3v_i \\ f(v_{DM}, T_{DM}(r)) \, f(v_i,T(r)) \,\sigma_{evap}(m_{DM}, m_i, v_{DM}, v_i, v_{esc})\, |\vec v_{DM} - \vec v_i|,
\end{multline}
where $f(v,T)$ is the Maxwell-Boltzmann velocity distribution\footnote{We note that using a Maxwell-Boltzmann distribution for the velocities of the captured material is a conservative assumption. Since faster particles will have a higher evaporation rate, the equilibrium distribution may feature tapering in the high velocity tail, which would \emph{decrease} the total evaporation rate. We do not attmpet to model this effect, noting only that our approach is conservative.} at temperature $T$, and $m_i$, $v_i$, $n_i$ are the mass, velocity and number density of the different SM constituents of the star. The evaporation cross section is the cross section to scatter up from velocity $v_{DM}$ to the escape velocity:
\begin{equation}
    \label{eq:evap_cross_section}
\sigma_{evap}(v_{esc}) = \int_{E_R^{min}(v_{esc})}^{E_R^{max}} dE_R \frac{d\sigma}{dE_R},
\end{equation}
where $E_R$ is the recoil energy of the DM particle, $E_R^{min}(v_{esc})$ is the minimum recoil required to eject the DM (a function of the local escape velocity), and $E_R^{max}$ is the maximum possible energy transfer in a two body collision.

The total evaporation rate is then
\begin{equation}
-\left(\frac{dN}{dt}\right)_{evap} = \int_0^{R_{star}} dr\, 4\pi r^2 \, n_{aDM}(r) \, \Omega^-(v_{esc}(r ), r),
\end{equation}
where we assume the density profile is Gaussian with width set by the virial theorem (see above).

However, due to the strong self interactions of the dark matter, in some cases this is a drastic overestimate of the evaporation rate, since it does not account for further scatters of the dark matter in which it may become ``recaptured''. In fact the aDM nugget is very efficient at recapture, especially if the evaporated particles are only moving slightly faster than the escape velocity. We estimate the true evaporation rate by accounting for a probability of escape, approximately given by
\begin{equation}
P_{escape}(v_{evap}, r) = \begin{cases}
\exp\left( - \frac{R_{nugget} - r}{\lambda_{recap}} \right) & 0 \leq r < R_{nugget}, \\
1 & r \geq R_{nugget},
\end{cases}
\end{equation}
with
\begin{equation}
\exp\left( - \frac{R_{nugget} - r}{\lambda_{recap}} \right) = \exp\big({-(R_{nugget} - r) \,n_{aDM}(r)\, \sigma_{recap}(v_{evap})}\big),
\end{equation}
where we approximate the distance the DM must travel to escape the nugget as $R_{nugget} - r$, with $R_{nugget}$ the nugget's virial radius. The recapture cross section is defined analogously to \eqref{eq:capture_recoil} and \eqref{eq:evap_cross_section}, and is a function of the velocity the dark matter has acquired after its initial collision with the hot stellar material. This refined evaporation rate is then integrated over $v_{evap}$ and the volume of the aDM nugget:
\begin{equation}
-\left(\frac{dN}{dt}\right)_{evap} \approx \int dV \int_{v_{esc}(r)}^\infty dv \, n_{aDM}(r) \, P_{escape}(v,r) \, \Omega^-(v,r).
\end{equation}

\begin{figure}
    \centering
    (a)\hspace{3mm}\begin{subfigure}[]{0.33\textwidth}
        \includegraphics[scale=0.15]{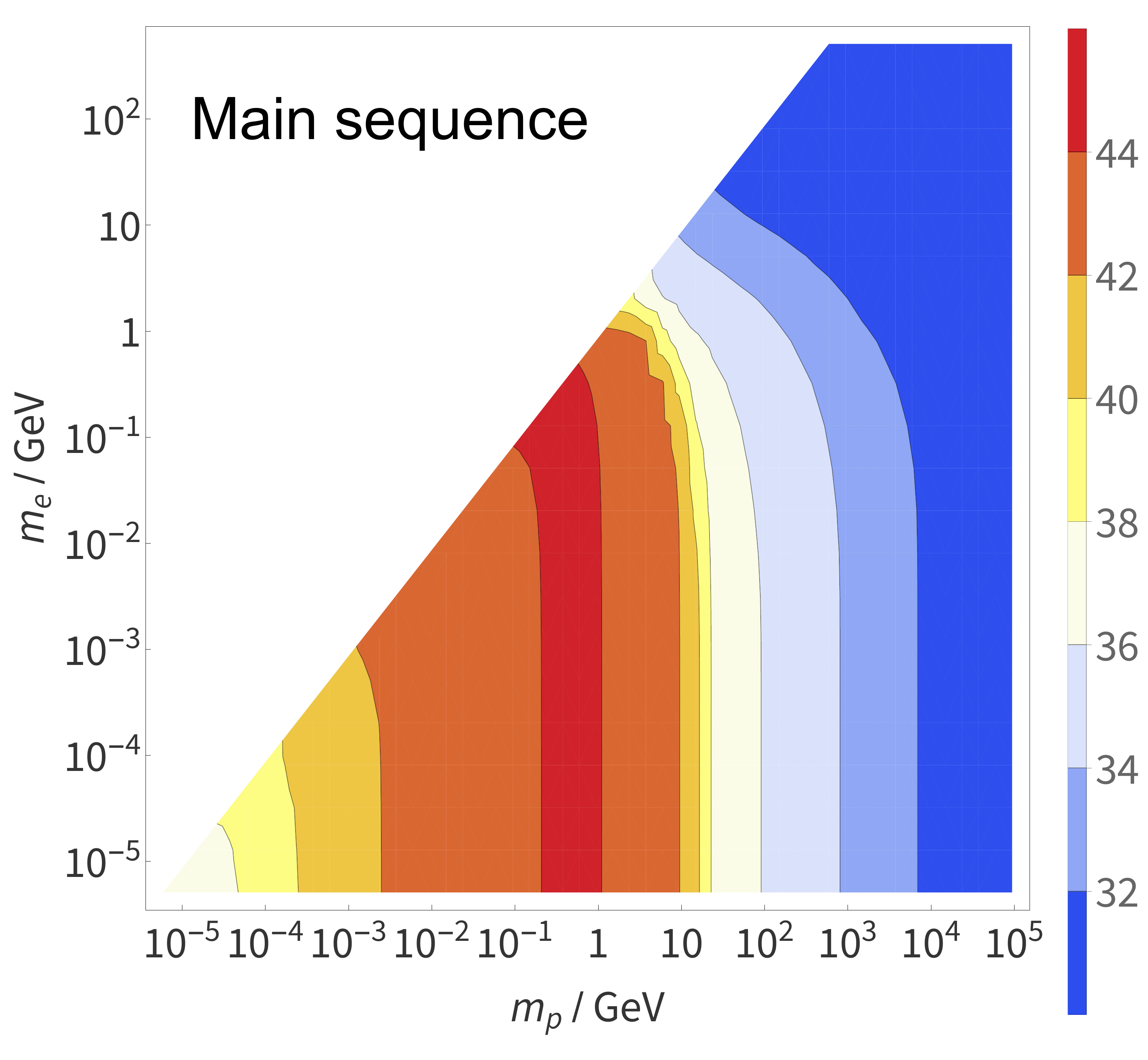}
    \end{subfigure}
    \hspace{0.5cm}
    (b)\hspace{3mm}\begin{subfigure}[]{0.33\textwidth}
        \includegraphics[scale=0.15]{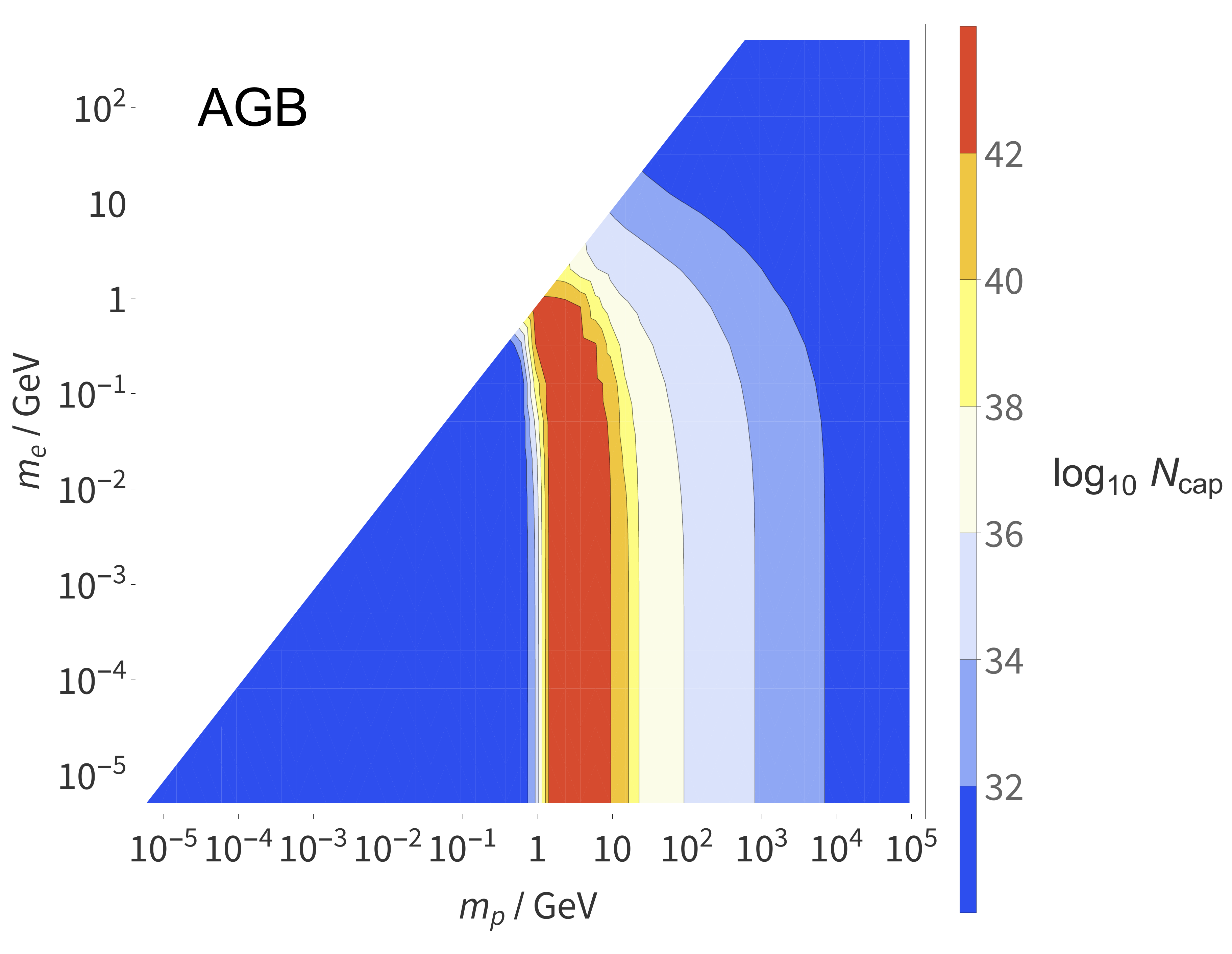}
    \end{subfigure}\\
    (c)\hspace{3mm}\begin{subfigure}[]{0.33\textwidth}
        \includegraphics[scale=0.15]{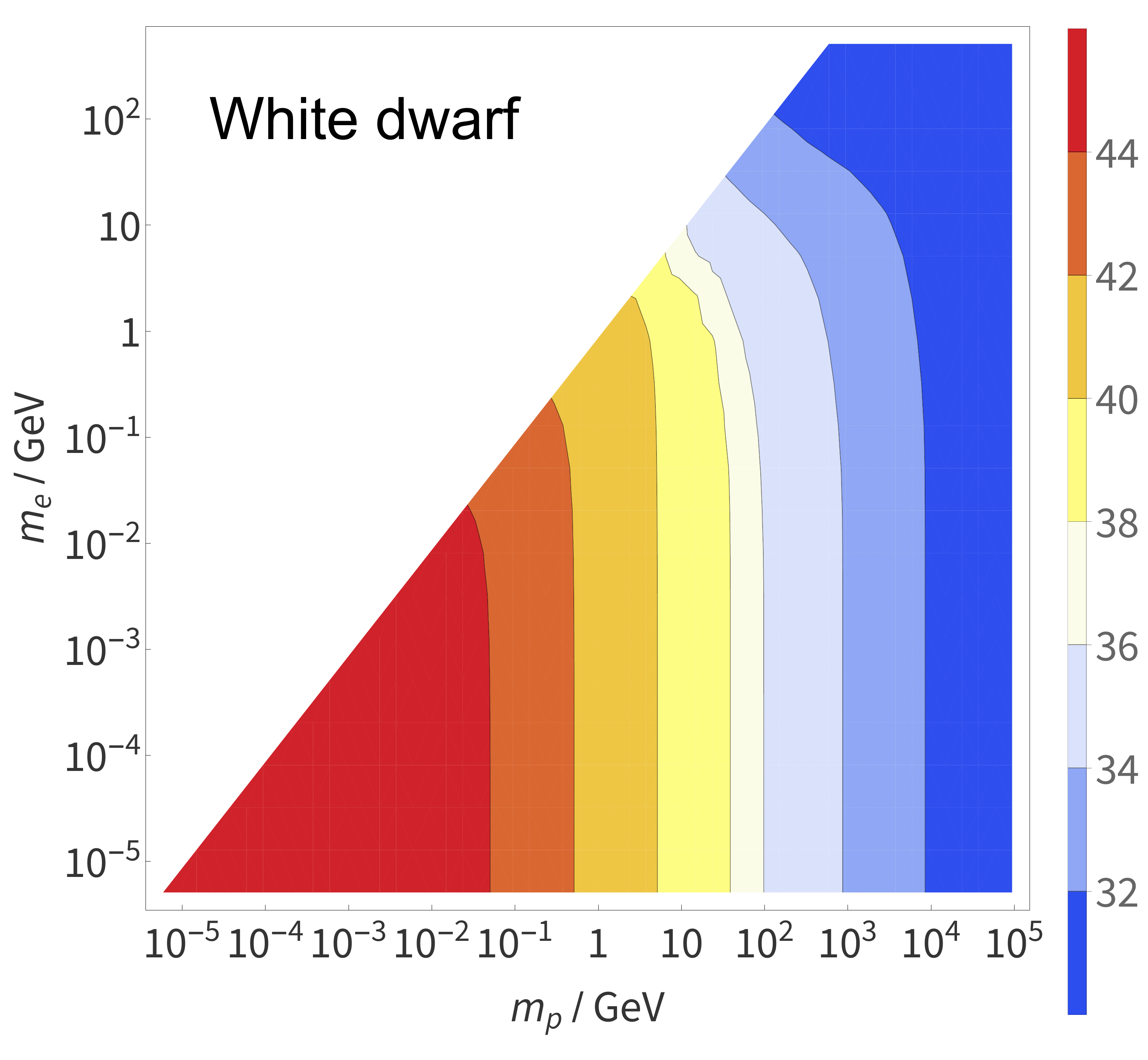}
    \end{subfigure}
    \hspace{0.5cm}
    (d)\hspace{3mm}\begin{subfigure}[]{0.33\textwidth}
        \includegraphics[scale=0.15]{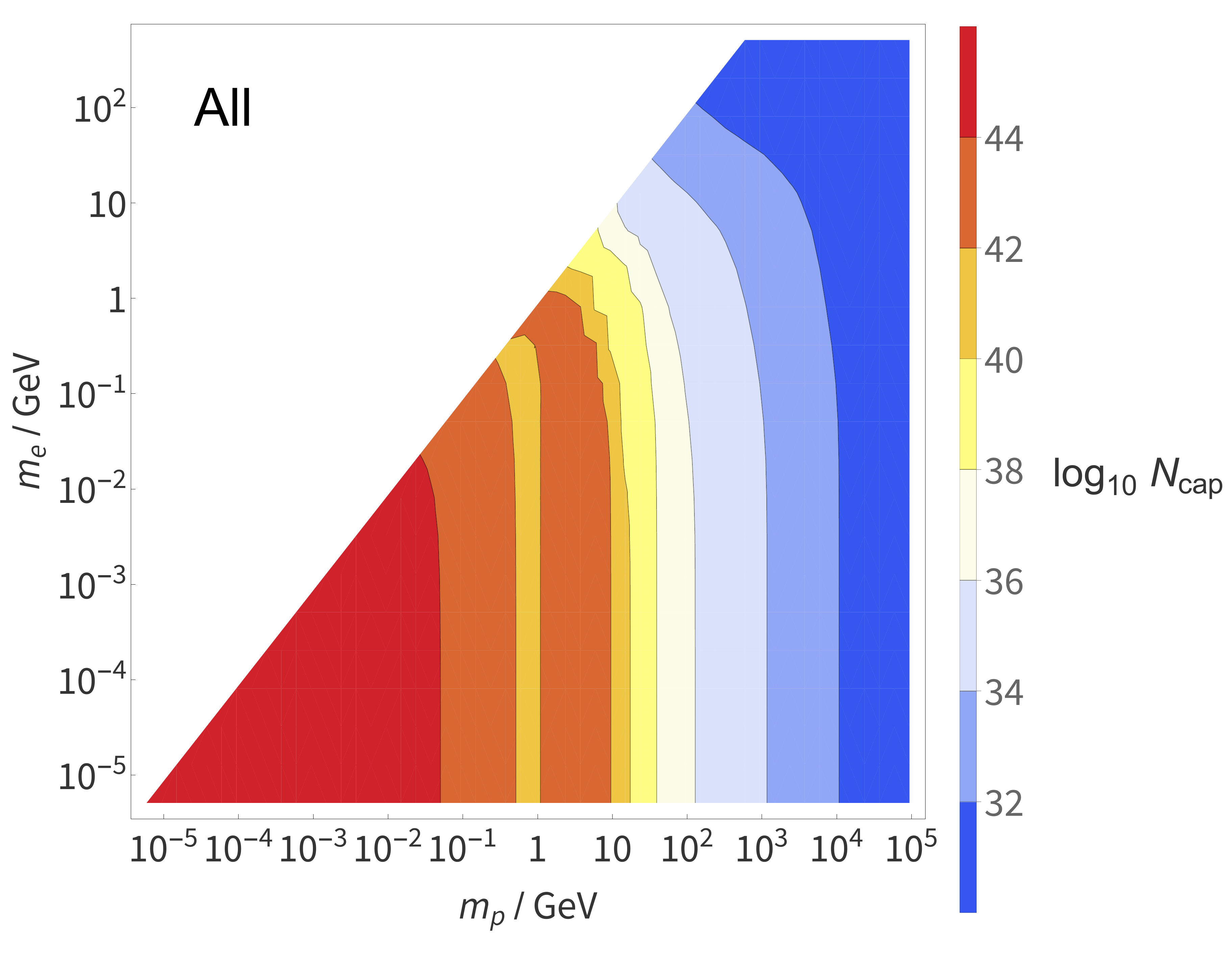}
    \end{subfigure}
    \caption{aDM accumulation in our benchmark star (described in Section~\ref{s.WDLF}) for $\epsilon = 10^{-11}$, assuming $\rho_{aDM} = 0.1 \rho_{DM}$ and using the halo-like velocity assumption $\langle v \rangle = 220$ km/s: (a) accumulation at the end of the star's main sequence lifetime, (b) accumulation at the end of the star's AGB phase, shortly before it becomes a white dwarf. DM lighter than around 0.5 GeV is very efficiently evaporated away during the 100 million years the star spends burning helium. (c) DM accumulated during the star's lifetime as a white dwarf only; (d) total accumulation, taking into account accumulation during the main sequence, AGB and white dwarf phases. Nugget density can be estimated from total accumulation for a given nugget temperature using the virial theorem, equation~\eqref{e.rvirial}.}
    \label{fig:capture}
\end{figure}

Taking all of this into account, we find that, for our 3 solar mass benchmark star, 
evaporation starts to limit the main sequence accumulation for DM masses below around $0.1 \,\textrm{GeV}$. 
The evaporation rate is significantly \emph{enhanced} during the helium burning phase, in line with our previous estimation, primarily due to the much hotter temperature at the core. Below masses of about $0.5$ GeV the dark matter is efficiently evaporated away.
On the other hand, evaporation for the white dwarf is all but negligible even for the lowest 10 keV masses we consider. This is primarily due to the much higher escape velocity from the extremely dense white dwarf core, and also the inefficient energy transfer from the heavy SM nuclei to the light dark matter, which is suppressed by $m_{DM}/m_{nucleus}$.
Furthermore, at low masses the accumulation is enhanced by the fact that the ambient DM number density increases with decreasing mass.
The captured number of dark hydrogen atoms for a representative aDM scenario is shown in \fref{fig:capture}. This clearly shows the effect of evaporation during the different phases of the benchmark star's life cycle, and explains the non-trivial dependence of aDM accumulation on dark proton mass by the time the white dwarf is 4 billion years old.

At any given point in time we therefore know the number of aDM atoms captured by the star, $N_{aDM}$. To compute heating and cooling rates we must also know the distribution of the aDM in the star. We assume the aDM nugget has a Gaussian temperature profile of width equal to the virial radius \eref{e.rvirial} corresponding to the temperature at the center, and solve for the density profile to be consistent with the conditions of hydrostatic equilibrium. This is a reasonable approximation for our purposes~\cite{Curtin:2019lhm, Curtin:2019ngc}, although we have checked that other profile assumptions (i.e. isothermal, and linearly decreasing with radius) have less than a percent-level effect on the equilibrium temperature that we obtain.

\section{Heating and cooling of the captured aDM nugget}
\label{sec:heating_cooling}

Our task in this section is to calculate the total rate of energy loss into dark photons. The emission rate of dark photons depends on the temperature of the aDM nugget $T_{nugget}$, which is usually \emph{different} to the temperature of the white dwarf core $T_{core}$ even at equilibrium.
Therefore we solve for the equilibrium value of $T_{nugget}$ at which the heating rate and cooling rate (both in general a function of $T_{nugget}$) are equal. We do not attempt to calculate the precise spectrum of the dark photon emission, since our goal is only to calculate the total cooling rate of the white dwarf.

\subsection{Heat transfer calculation: Photon Portal Thermodynamics}
\label{s.heating}

The kinetic-mixing suppressed interaction between the white dwarf interior at temperature $T_{core}$ and the captured aDM nugget at equilibrium temperature $T_{nugget}$ presents a rather unique physical setup: 
two gases/plasmas, one of them very hot (the white dwarf interior),
occupying the same physical location, yet coupling to each other only via extremely ``rare'' but otherwise electromagnetism-like interactions.
This ``photon portal thermodynamics'' gives rise to several unfamiliar and seemingly counter-intuitive phenomena:
\begin{enumerate}
\item Since the cooling of the aDM nugget arises from unsuppressed self-interactions that cause emission of dark photons, it is generally very efficient compared to the $\epsilon^2$-suppressed heating rate.  The two gases might therefore never reach thermal equilibrium, and the captured aDM with $T_{nugget} \ll T_{core}$ acts as a constant heat sink for the white dwarf core. 

\item The high temperature of the white dwarf core, far above the ionization threshold for SM atoms as well as dark atoms (for most of the relevant parameter space), means those rare $p \tilde p \to p \tilde p$ interactions that do transfer heat to the aDM nugget are \emph{high energy processes} sensitive to the non-trivial velocity dependence of unscreened massless photon exchange, unlike the ``billiard-ball-like'' interactions of classical thermodynamics. 

Photon portal thermodynamics therefore leads to the extremely surprising conclusion \emph{that older and colder white dwarfs lose more heat to the aDM nugget than younger and brighter white dwarfs!} This has important consequences for the detectability of aDM-catalyzed white dwarf cooling.

\item There are situations, especially of interest to us since they are at the boundaries of the aDM parameter space that can be probed by white dwarf cooling, where so little aDM is accumulated that cooling becomes inefficient and the nugget comes closer to reaching equilibrium with the core, $T_{nugget} \to T_{core}$.  In that case, we must understand the dependence of the heating rate on the temperature difference $\Delta T = T_{core} - T_{nugget}$ between the two gases. Unlike in classical thermodynamics, the non-trivial velocity dependence of the interaction means the heating rate is no longer proportional to $\Delta T$.
\end{enumerate}
To correctly account for all these effects, we must derive the heat transfer rate between the white dwarf core and the aDM nugget starting from microphysical first principles. That is the goal of this section, and the full calculation bears out all the phenomena described above.

The cross section for a particle with SM charge scattering off a particle with dark charge is the same cross section used in the capture calculation \eqref{eq:rutherford} (essentially $\epsilon^2$-suppressed Rutherford scattering), and scales with $1/v^4$. The IR singularity of the cross section means that the thermally averaged energy transfer rate diverges, but in reality the integral is cut off by screening effects. If the dark matter is not ionized, then the physical size of the atom, given by the generalized Bohr radius, screens the charge of the dark proton \cite{Cline:2012is}. In general the equivalent of the Bohr radius for a dark atom is
\begin{equation}
\tilde a_0 = \frac{1}{\alpha_D}\left(\frac{1}{m_{\tilde e}} + \frac{1}{m_{\tilde p}} \right).
\end{equation}

There is also plasma screening: in a non-degenerate plasma, charges are screened at distances greater than the Debye length. 
For a gas of degenerate electrons such as is found in the core of a white dwarf, the appropriate screening length is the Thomas-Fermi length~\cite{1957ApJ...126..213S}
\begin{equation}
a_{TF}^2 = 16\pi \alpha\, m_e \left( \frac{3 n_e}{\pi} \right)^{1/3},
\end{equation}
where $n_e$ is the number density of electrons. The Thomas-Fermi length is larger than the Debye length; the degenerate electrons are less efficient at screening the charges in the plasma.
We take the screening length $\lambda_{screen}$ to be whichever is smaller of the Thomas-Fermi length $a_{TF}$ and the Bohr radius of the dark atom $\tilde a_0$ unless the dark atoms become ionized, in which case we use $a_{TF}$. (The dark Debye screening of the ionized dark plasma is always less efficient at screening than the Fermi screening of the white dwarf electrons, across the range of parameters we consider.)
For a screening cutoff $\Lambda = 1/\lambda_{screen}$, the modified cross section is identical to \eref{eq:rutherford_screened}:
\begin{equation}
    \label{eq:rutherford_again}
\frac{d\sigma}{d\Omega} = \frac{4 \epsilon^2 \alpha\alpha_D Z_1^2 Z_2^2 \mu^2}{(\Lambda^2 + q^2)^2} = \frac{4\epsilon^2 \alpha\alpha_D Z_1^2 Z_2^2 \mu^2}{(\Lambda^2 + 4 \mu^2 v^2 \sin^2(\theta/2))^2},
\end{equation}
which for small velocities no longer diverges and approaches a constant contact interaction (i.e. the billiard-ball limit of classical thermodynamics).

The rate of heating per unit volume from gas 1 at temperature $T_1 = T_{core}$ to coincident gas 2 at temperature $T_2 = T_{nugget}$, both assumed to be comprised of a single species of masses $m_1$ and $m_2$ respectively, is in full generality:
\begin{multline}
    \label{eq:heating_full}
	\frac{dP}{dV}(T_1,\,T_2) = n_1 n_2 \int_0^\infty dv_1 \;f(v_1, T_1) \int_0^\infty dv_2 \;f(v_2, T_2)\frac{1}{\pi} \int_0^\pi d\theta_i\sin\theta_i \int_0^{\pi} d\theta_f \sin\theta_f \\
	|\vec{v}_1 - \vec{v}_2| \; 2\pi \; \frac{d\sigma}{d\theta_f}(v_1, v_2,\theta_i,\theta_f) \; \Delta E(v_1,v_2,\theta_i,\theta_f),
\end{multline}
where $f(v,T)$ is the Maxwell-Boltzmann velocity distribution at temperature $T$, $\theta_i$ is the initial collision angle in the star frame, $\theta_f$ is the scattering angle in the centre of mass frame, and $\Delta E$ is the energy gained by particle 2 during the collision in the star frame. The relative velocity is given by $|\vec{v}_1 - \vec{v}_2| = \sqrt{v_1^2 + v_2^2 - 2 v_1 v_2 \cos{\theta_i}}$. The energy transfer $\Delta E$ can be positive or negative depending on the velocities and the scattering angles. The heating rate is positive (negative) for $T_1 > T_2$ ($T_1 < T_2$), goes to zero when $T_1 = T_2$, and is maximised for $T_2/T_1 \to 0$.

Here we are implicitly neglecting any effects of evaporation on the distribution of velocities. For instance if the temperature of the dark matter approached that of the core, thermal velocities may approach the escape velocity, leading to a truncation of the velocity distribution. This approximation is, however, valid for our purposes, since evaporation from the white dwarf core becomes important only for masses $\lesssim 10^{-3}$ GeV, and for such small masses cooling is extremely efficient (see the following section), such that it is impossible for the dark matter to thermalize with the core unless the accumulation is very small.

We make a few brief comments on the evaluation of equation \eqref{eq:heating_full}. The heating rate is calculated separately for each species of nuclei in the white dwarf core\footnote{The species present in the white dwarf core are, in descending order in mass fraction: ${}^{16}$O~(76.1\%), ${}^{12}$C~(21.2\%), ${}^{22}$Ne (1.95\%), ${}^{24}$Mg (0.40\%) and ${}^{20}$Ne (0.21\%).}. It is important to evaluate the integrand, including the cross-section, in the star frame rather than the zero-momentum frame of the collision, since it is the energy gained/lost in that frame which determines the rate of heating. We analytically perform the $\theta_f$ integral and obtain an analytic expression for the remaining integrand, which is then integrated numerically. It is useful to first rewrite this integrand in terms of a function of the dimensionless variables:
\begin{equation}
    \frac{m_2}{m_1}, \;\;\;\; \frac{T_2}{T_1}, \;\;\;\; \frac{\sqrt{\mu T_1}}{\Lambda},
\end{equation}
such that
\begin{equation}
\label{e.dPheatingdV}
    \frac{dP}{dV} = n_1 n_2 \, \frac{\epsilon^2 \alpha \alpha_D Z_1^2 Z_2^2}{\sqrt{\mu\,T_1}} \; G\left( \frac{m_2}{m_1}, \frac{T_2}{T_1}, \frac{\sqrt{\mu T_1}}{\Lambda} \right).
\end{equation}
Note that the effect of screening depends on the dimensionless parameter  $\sqrt{\mu T_1}/\Lambda$, which can be thought of as the ratio between the average collisional energy and the energy required to penetrate the screening.
The master function $G(x,y,z)$ is then a numerical integral over the velocities and angles, for which a lookup table need only be computed once, over a suitable range of values, to allow fast evaluation for our studies of the aDM parameter space.

\begin{figure}
    \centering
    \hspace{6mm}\includegraphics[scale=0.4]{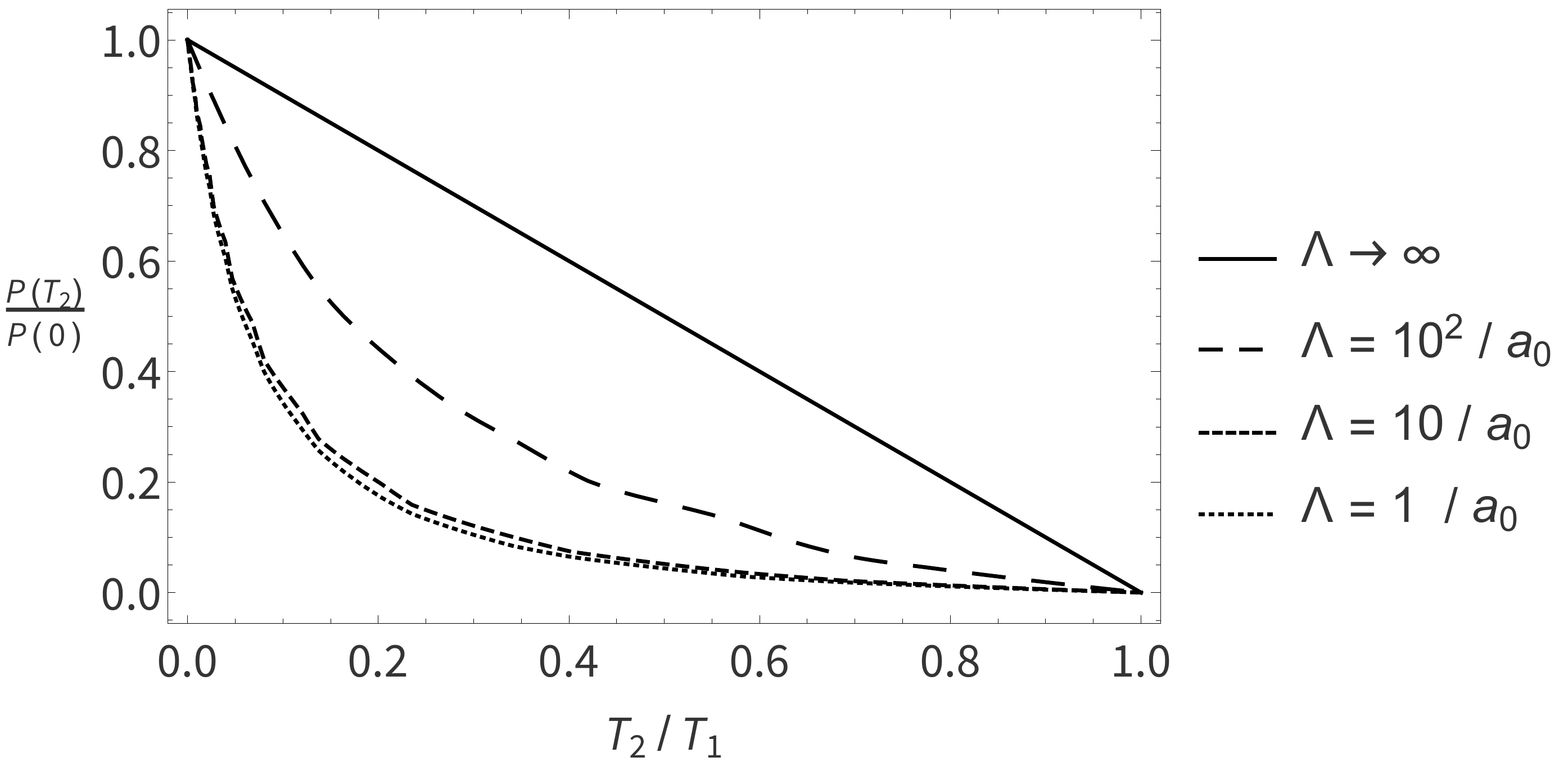} \\
    \hspace{-7mm}\includegraphics[scale=0.4]{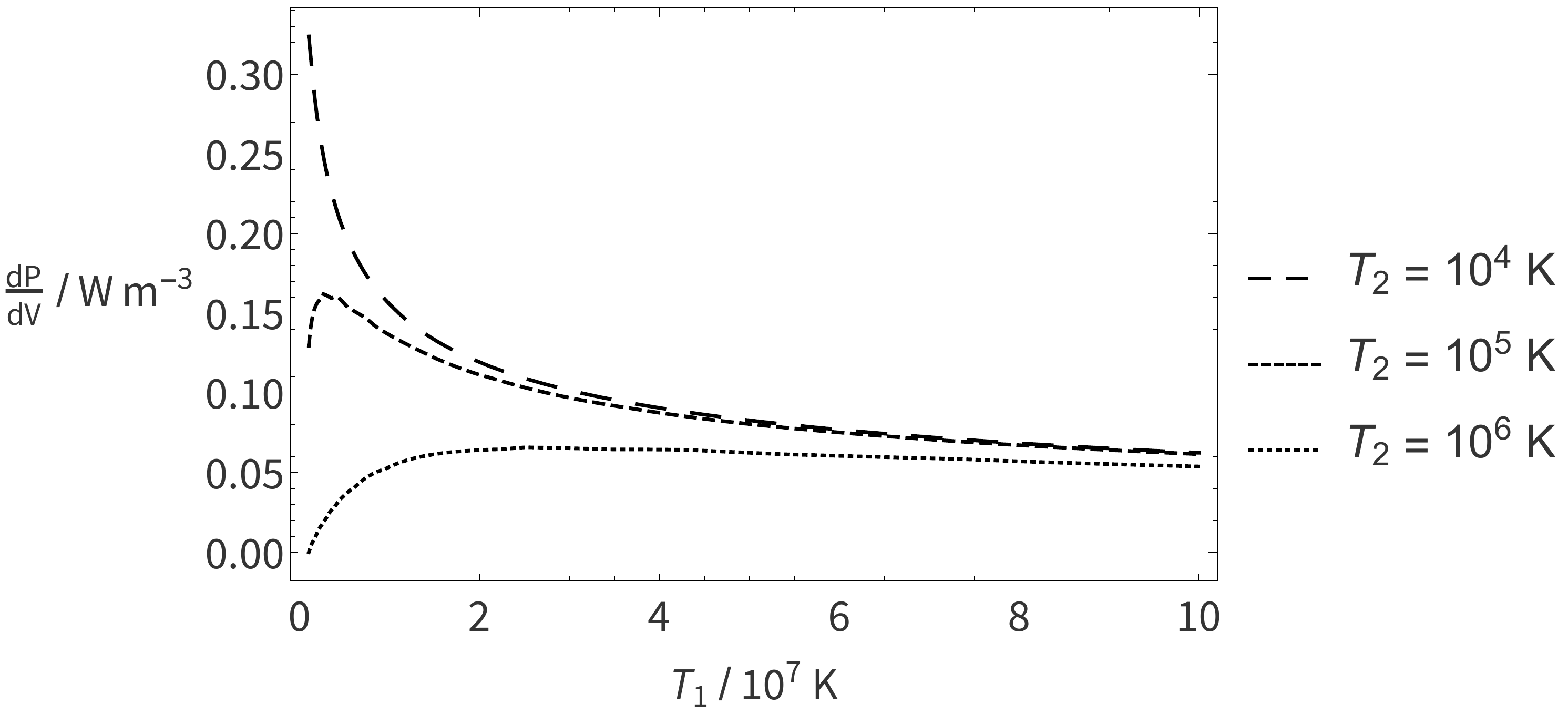}
    \caption{
    \emph{Top:}~Heating rate from hot gas 1 to colder gas 2 via $\epsilon^2$-suppressed photon exchange, as a function of $T_2/T_1$, normalised to unity at $T_2 = 0$. We show the dependence for different values of the inverse screening length $\Lambda$, which appears in the scattering cross section \eqref{eq:rutherford_again}. The solid line is the familiar contact interaction limit in which the heating rate is proportional $\Delta T$.  The amount by which the heating rate departs from this behaviour is characterised by the dimensionless parameter $\sqrt{\mu T_1}/\Lambda$, where $\mu$ is the reduced mass. For this plot we chose $T_1 = 2.9\times10^6$ K, $m_1 = 10$ GeV and $m_2 = 1$ GeV, and $a_0$ is the usual Bohr radius of the hydrogen atom, so that $\sqrt{\mu T_1}/\Lambda = 130,\,13,\,1.3$ for the three curves from bottom to top.
    \emph{Bottom:}~Heating rate per unit volume as a function of $T_1$, the temperature of the hotter gas. This is to illustrate the unfamiliar behaviour that a hotter gas can actually be \emph{less} efficient at heating than a cooler gas, if the interaction is mediated by $\epsilon^2$-suppressed massless vector exchange. For this plot we chose $m_1 = 10$ GeV, $m_2 = 1$ GeV, $\Lambda = 1/a_0$, and took the density of both species to be $10^{26}\,\textrm{m}^{-3}$ (the overall heating rate is just proportional to $n_1 n_2$).}
    \label{fig:heating}
\end{figure}

We illustrate the importance of this careful calculation in Fig.~\ref{fig:heating} (top), where we show the normalised heating rate for different values of the screening scale $\Lambda$, compared to the naive expectation from classical thermodynamics that the heating rate should be proportional to $\Delta T = T_1 - T_2$. 
For higher screening scales, i.e. $\sqrt{\mu T_1}/\Lambda \to 0$, we approach the contact interaction limit, in which collisions are not sensitive to the velocity dependence of the cross section. In this case we do indeed recover the classical expectation. 
However for smaller screening scales and/or higher temperatures,  $\sqrt{\mu T_1}/\Lambda \gg 1$, the screening is penetrated and the velocity dependence has a dramatic effect of the functional form of the heating rate. 
It is especially important to get the heating rate correct in the limit where the two gases are very close in temperature $T_1 \approx T_2$, since we can see from Fig.~\ref{fig:heating} that the heating rate can be significantly suppressed compared to the linear dependence on $\Delta T$ of the contact interaction limit.

In the limit where $T_2 \ll T_1$, the calculation can be simplified by assuming the particles of the cooler gas are at rest.
Furthermore, in the limit where $\mu \langle v \rangle \gg \Lambda$, in which the particles are fast enough that they penetrate the screening and see the velocity dependence of the cross section, 
\eref{e.dPheatingdV} can be analytically evaluated to recover the $T_2$-independent heating rate in the cold-nugget limit derived by \cite{Curtin:2019ngc}:
\begin{equation}
\label{e.dPheatingdVsmallT2}
\frac{dP}{dV} \approx n_1 n_2 \frac{2\pi\epsilon^2\alpha\alpha_D Z_1^2 Z_2^2}{m_2 \langle v \rangle} \left( \log\frac{8 \mu^2 \langle v\rangle^2}{\Lambda^2} - 1\right),
\end{equation}
where $\langle v\rangle = \sqrt{3 k_B T_1 / m_1}$ is the average thermal velocity of SM particle 1. 
We see that in this limit the dependence on the IR cutoff of the cross section is logarithmic. Note the unfamiliar $T^{-1/2}$ scaling of the heating rate; this makes hotter gases \emph{less} efficient at heating a cooler gas due to the velocity suppressed scattering cross section. 
We illustrate this behaviour in Figure~\ref{fig:heating} (bottom). 
This realizes the third effect discussed above, that cooler white dwarfs actually emit a \emph{higher} luminosity in dark photons, and that the energy loss effect only grows more pronounced as the white dwarf cools (so long as we are in the limit where screening is ineffective).

To obtain the full heating rate we sum the result of \eqref{eq:heating_full} (obtained numerically) over all nuclei in the white dwarf core. We account for scattering off dark protons, dark electrons and dark atoms, using the appropriate level of ionization at a given DM temperature and density derived using Saha's equation. For $m_{\tilde{e}}$ lighter than around $10\,\textrm{MeV}$, heating of dark electrons is suppressed by the plasma screening, since the momentum transfer is always significantly smaller than the inverse Thomas-Fermi length. However, we include this source of heating in our calculations, since for heavier dark electrons, and in cases where the dark electron and proton have similar masses, the effect can become relevant.

We note that collisions involving SM electrons are never relevant, partly due to the plasma screening, and also because electron transitions in the degenerate plasma are degeneracy suppressed. For this reason we do not account for these in our calculations.

\subsection{Cooling rate and optical depth}

In general, to obtain the anomalous cooling rate of the white dwarf, we must find the equilibrium temperature $T_{nugget}$ at the center of the aDM nugget such that the total heating and cooling rates are equal. This not only determines the rate of energy loss (which, as shown by \eref{e.dPheatingdVsmallT2}, is actually \emph{independent} of $T_{nugget}$ if it is much smaller than $T_{core}$), but also the size of the aDM nugget, which plays a role in the aDM capture rate, see \eref{e.Ctot}.

The physics of the various aDM nugget cooling mechanisms can be very complex~\cite{Rosenberg:2017qia}. Bremsstrahlung emission, ionization, recombination and atomic transitions play significant roles in different regimes of temperature and density.
The dark atomic binding energy is given by 
\begin{equation}
\tilde E_0  \  = \  \frac{1}{2} \alpha_D^2 \frac{ m_{\tilde p}m_{\tilde e}}{m_{\tilde p} + m_{\tilde e}} 
\ \approx \ \frac{1}{2} \alpha_D^2 m_{\tilde e} \ \  \ \mathrm{if} \ \ \  \ m_{\tilde e} \ll m_{\tilde p}.
\end{equation}
In our cooling calculation, we make the simplification of \emph{only} considering bremsstrahlung cooling.
The degree of  ionization within the aDM nugget is determined by the local temperature and density via Saha's equation. 
This simplification essentially restricts our attention to scenarios where the white dwarf core temperature is high enough to cause a significant degree of ionization in the nugget. 
Therefore, will not obtain aDM limits for $m_{\tilde e} \gg 2 T_{core}/\alpha_D^2$, leaving the required more complex cooling calculations for future work. 

We have to take into account the optical depth of the nugget to the emitted dark radiation, since in the optically thick regime radiated dark photons are more likely to be reabsorbed before escaping, which can occur  via photoionization or through free-free transitions (inverse bremsstrahlung).
We therefore need to first understand the circumstances that determine whether the aDM nugget is optically thin or thick.
Dark photons can scatter from dark protons and dark electrons via Thomson scattering, with the frequency-independent Thomson cross section:
\begin{equation}
\label{e.thomson}
    \sigma_{thoms} = \frac{8\pi}{3} \frac{\alpha_D^2}{m_i^2}.
\end{equation}
Furthermore they can be absorbed by inverse bremsstrahlung; the attenuation coefficient (the reciprocal of the photon mean free path $\lambda_{ff}$) for this process is \cite{2011piim.book.....D}
\begin{equation}
\label{e.kappaff}
    \kappa_{ff} = \frac{16\pi^2}{3} \left(\frac{2\pi}{3}\right)^{1/2} \frac{\alpha_D^3}{m_i^{3/2} T^{1/2} \omega^3} \left( 1 - e^{-\omega/T} \right) n_i^2 \, g_{ff},
\end{equation}
where $g_{ff}$ is the free-free Gaunt factor and $\omega$ is the frequency of the dark photon. Dark photons from the nugget interior will random-walk their way to the surface of the nugget, traveling a total distance $D \sim R_{nugget}^2 / \lambda_{thoms}$, where $\lambda_{thoms}$ is the mean free path for Thomson scattering. Therefore if $D < \lambda_{ff}$, most photons are absorbed before escaping, and the aDM nugget is optically thick.
Recall we assume the aDM nugget has a Gaussian temperature profile with width given by the virial radius, which also sets the density profile from the conditions of hydrostatic equilibrium. In that case, for a given set of aDM parameters and for our specific benchmark star, there is a certain amount of aDM accumulation $N_{aDM}^{thick}(T_{nugget})$ below which the nugget is optically thin, and above which it is optically thick. (In reality the transition is not so sharp but this suffices for our purposes.)

For purposes of illustration, we can derive an estimate for $N_{aDM}^{thick}$ by treating the DM nugget as a sphere of uniform density and radius given by the virial radius at temperature $T$. Solving for the accumulation such that the path length to escape $D$ is equal to the path length $\lambda_{ff}$ (and assuming both bremsstrahlung and absorption are dominated by the dark electron, which is valid if $m_{\tilde e} \ll m_{\tilde p}$, we find:
\begin{eqnarray}
N_{aDM}^{thick} &\sim& 5\times10^{44} \left( \frac{T_{nugget}}{T_{core}} \right)^{7/3} \left(\frac{\alpha_{D}}{\alpha_{SM}}\right)^{-5/3} \left( \frac{m_{\tilde e}}{m_{DM}} \right)^{7/6}
\\ \nonumber
&\approx& 1\times10^{35}  \left(\frac{T_{nugget}}{0.05 \tilde E_0}\right)^{7/3}  \left(\frac{\alpha_D}{\alpha_{SM}}\right)^3 \left( \frac{m_{\tilde e}}{m_e} \right)^{7/2} \left( \frac{m_{DM}}{m_H} \right)^{-7/6},
\end{eqnarray}
where $m_{DM} = m_{\tilde p} + m_{\tilde e}$ is the total mass of the dark matter atom, which appears in the expression for the virial radius. In the second line we have again assumed $m_{\tilde e} \ll m_{DM}$ in the expression for $\tilde E_0$. We have expressed the result in terms of the ratio $(T_{nugget}/0.05 \tilde E_0)$ because, in the efficient cooling regime, the nugget equilibrium temperature tends to be around $\sim 0.05 \tilde E_0$ (this is true across the vast majority of our parameter space, and is relatively insensitive to the heating rate, for reasons discussed below). We can now discuss cooling in the optically thin and thick regime separately.

We first consider the optically thin regime.
The cooling rate per unit volume due to bremsstrahlung is given by \cite{2011piim.book.....D}
\begin{equation}
    \label{eq:bremsstrahlung}
    \frac{dP_{cool}}{dV} \approx \sum_i \frac{16}{3} \left(\frac{2\pi}{3}\right)^{1/2} \frac{\alpha_D^3}{m_i^2} (m_i T_{nugget})^{1/2} n_i^2 \ \  ,
\end{equation}
where $m_i$ is either the dark electron or dark proton mass, and $n_i = n_{\tilde e} = n_{\tilde p}$ is the number density of \emph{free} dark electrons/dark protons, determined by Saha's equation from the temperature and total density. 
 For $T_{nugget} \lesssim \tilde E_0$, the ionization fraction is small but rises very sharply with temperature. 
This means a small increase in $T_{nugget}$ leads to a huge increase in $n_{i}$ and hence the bremsstrahlung cooling rate, resulting in $T_{nugget} \lesssim  E_0 \ll T_{core}$ for most of aDM parameter space. 
This makes the heating (and hence cooling) rate independent of $T_{nugget}$.
However, if the aDM nugget density is low enough, the nugget can become \emph{fully} ionized.
In this situation, which in practice only occurs when very little aDM has accumulated in the star, the emission rate can only increase with $\sqrt{T_{nugget}}$, and it is possible for bremsstrahlung cooling to become inefficient in the optically thin regime. 
This edge case, in which $T_{nugget}$ could approach $T_{core}$, is actually of great interest to us, since we want to be able to study the smallest possible amount of aDM accumulation that can still lead to detectably anomalous white dwarf cooling. 
This is why we needed the full heat transfer calculation of 
\eref{e.dPheatingdV}, which captures the correct non-linear dependence on $\Delta T$ between the core and the nugget.

We now turn to the optically thick regime. As discussed for the analogous problem of SM matter capture in mirror stars in Ref.~\cite{Curtin:2019lhm,Curtin:2019ngc}, the cooling rate is now determined by black body emission from the nugget surface:
\begin{equation}
P_{blackbody} = 4 \pi R_{nugget}^2 \sigma_B T_{surface}^4
\end{equation}
The difficulty lies in relating the \emph{interior} temperature of the aDM nugget, which determines its size via the virial theorem $R_{nugget}^2 \sim T_{nugget}$, to the \emph{surface} temperature $T_{surface}$. 
The temperature profile of the nugget will depend on radiative and/or convective heat transport, as well as a detailed understanding of the aDM opacity, which depends on the detailed composition (e.g. possible small amounts of heavier ``dark elements'') and would be analogous to solving the structure equations of a star~\cite{1983psen.book.....C}. This is beyond our scope and reliant on unknown details of the dark sector.
Fortunately, it is also not necessary. 

It is natural to expect that cooling, which generally relies on aDM self-interactions, is very efficient in the optically thick / large accumulation regime. As long as the white dwarf core temperature is hot enough to ionize the aDM, we therefore always expect that $T_{nugget} \ll T_{core}$ in the optically thick regime. In that case, the heating and hence cooling rate does not depend on $T_{nugget}$, while its size for purposes of computing aDM self-capture  can be approximated by setting $T_{nugget} \sim 0.05\tilde E_0$.

However, this argument has one loophole that must be checked. One might worry that in the optically thick regime, the outer layers of the nugget could act as an ``insulating blanket'', leading to a low $T_{surface}$ while $T_{nugget} \to T_{core}$, which would lead to a much smaller cooling rate. 
To confirm that this cannot happen, we performed the following check: 
\begin{enumerate}
\item Assume some aDM nugget  temperature profile $T(r)$. We checked many possibilities: linear, constant, Gaussian, as well as unphysical blanket-like scenarios with an interior region at a high constant temperature and an exterior region with a lower constant temperature, transitioning at some $r < R_{nugget}$. 

\item For this temperature profile, derive the density profile consistent with hydrostatic equilibrium. 

\item Compute the probability of bremsstrahlung dark photons emitted at varying radii $r$ inside the nugget to escape to the surface via diffusion. This determines the cooling rate. 
\end{enumerate}

We perform the third step numerically by estimating the chance of photons at different depths to diffuse to the surface before being absorbed (see discussion below \eref{e.kappaff}), taking into account also the possibility of absorption via photoionisation, and photon energy loss for each Thomson scattering. We find that $T_{nugget}$ does not vary significantly for all the temperature profiles we tried, and is always $\ll T_{core}$. %
Therefore, we can use a Gaussian temperature profile for our final aDM constraint calculations, regardless of whether the nugget is optically thin or thick. We include the escape probability of bremsstrahlung photons as described above, which consistently takes into account the change as we transition from the optically thin to the optically thick regime. In the latter, the heating rate will always be independent of $T_{nugget}$ (considering only ionization cooling).

Evidently, it is not possible to realize a ``blanket effect'' for the aDM nugget. This is reasonable, since any cold outer layer would be transparent to dark photons from the hot interior, while a hot outer layer would be opaque but radiate energy away efficiently. 
We can further illustrate why $T_{nugget} \ll T_{core}$ is always true in the optically thick regime by crudely analytically estimating the cooling rate for the special case of an aDM nugget with constant density and total ionization. 
Photons that start out distance $r$ from the nugget center and diffuse to the surface are assumed to travel a total distance less than $\lambda_{ff}$ in order to avoid being absorbed.  
The total distance traveled through the nugget must then satisfy
$D = (R_{nugget} - r)^2/\lambda_{thoms} < \lambda_{ff} $.
The cooling rate can therefore be roughly estimated by multiplying the bremsstrahlung rate in \eref{eq:bremsstrahlung} by the volume of a shell near the surface of thickness $\sqrt{\lambda_{ff} \lambda_{thoms}}$.
We can combine these expressions to obtain a crude estimate for the scaling of the cooling rate in the optically thick regime. Letting the frequency of the dark photons be characteristic of the temperature $\omega \sim T_{nugget}$, we find
\begin{equation}
\label{e.Pcoolestimate}
    P_{cool} \approx
    1.2 \times 10^9 \, L_{WD} \left( \frac{N_{aDM}}{N_{aDM}^{thick}} \right)^{1/2} \left( \frac{T_{nugget}}{T_{core}} \right)^{11/3} \left( \frac{\alpha_D}{\alpha_{SM}} \right)^{-1/3} \left( \frac{m_{\tilde e}}{m_{DM}} \right)^{5/6}
\end{equation}
where $N_{aDM}$ is the total amount of captured dark matter. %
We have expressed the result with respect to  values for the particular white dwarf benchmark we have chosen: $T_{core} = 2.9\times10^6\,\textrm{K}$, and $L_{WD} = 10^{-4} L_\odot = 3.8\times10^{22}\,\textrm{W}$. We have assumed that the bremsstrahlung is dominated by the dark electron, which is true if the dark electron is lighter than the dark proton.

This demonstrates that, for the range $10^{-10} < m_{\tilde e} / m_{DM} < 1$ we consider, an optically thick nugget would radiate away orders of magnitude more energy in dark photons than the whole white dwarf radiates in SM photons if the nugget temperature approached the temperature of the white dwarf core. Since we set aDM constraints by requiring that the anomalous heat loss never exceeds the SM photon luminosity,  this shows that $T_{nugget}$ is always much smaller than $T_{core}$ for our parameters of interest.
The exception to this is if the dark electron is heavy enough such that $T_{core}$ is no longer sufficient to ionize the dark atoms, in which case the estimate \eref{e.Pcoolestimate} breaks down since it assumed full ionization. However, our numerical calculation of the cooling rate in the optically thick regime consistently takes into account the lower bremsstrahlung emission with lower ionization. Therefore, our limits in the high $\tilde E_0$ regime are conservative, since they do not include atomic cooling effects.

\section{Results}
\label{sec:results}

We now present the numerical results of our study, outlining the deviations in white dwarf cooling behavior we predict for aDM, and presenting bounds on aDM parameter space from the WDLF.

\subsection{The White Dwarf Luminosity Function in aDM}

Our aDM scenario is fully described by the choice of 
$\epsilon, m_{\tilde p}, m_{\tilde e}, \alpha_D$, as well as the local aDM fraction $\rho_{aDM}/\rho_{DM}$ and assuming either the halo- or disk-like aDM velocity distribution. 
For each choice of these parameters we compute aDM accumulation in our benchmark star from the start of its life until it reaches an age of 4 billion years as a white dwarf.
At each point in the star's evolution we find the properties of the aDM nugget by solving for its equilibrium temperature and hence finding the anomalous cooling rate.
Assuming the evolution of the star is unaltered, we then set limits on $\epsilon$ for fixed $m_{\tilde p}, m_{\tilde e}, \alpha_D, \rho_{aDM}/\rho_{DM}$ by adjusting the kinetic mixing until the deviation from the SM WDLF is equal to the criterion in \eref{e.WDLFchange}.

\fref{f.WDevolution} (c) shows the aDM-catalyzed cooling rate for a variety of aDM scenarios.
The unusual phenomenon we discuss in \sref{s.heating} is clearly illustrated: as the white dwarf core temperature drops with age, the absolute energy lost to the aDM nugget increases due to the velocity dependence in the scattering cross section which sets the rate of heat transfer. 
The effect on the WDLF is illustrated in 
Fig.~\ref{fig:luminosity_function}, assuming a mirror-matter-like aDM scenario constituting 10\% of the local DM density for varying kinetic mixing $\epsilon$.
As expected, the effect of the energy loss from the aDM nugget on the WDLF is most pronounced for cooler, older white dwarfs, both because the anomalous energy loss increases in absolute terms, and because the visible luminosity drops significantly with decreasing core temperature (scaling approximately with $T^{7/2}$), so that as the star cools, the dark photon energy loss becomes more important relative to the Standard Model energy loss. 
(Note that we normalize the aDM WDLF curves to agree with the SM WDLF at high white dwarf luminosities, since this truthfully represents the fact that deviations become more apparent later in the white dwarf's evolution. A more detailed WDLF analysis would normalize all WDLF curves to the data point with the smallest error bar, but  our criterion for excluding BSM effects to the WDLF \eref{e.WDLFchange} is so conservative that this does not affect our results.)

This is to be compared to constraints on axions and other light degrees of freedom from white dwarf cooling \cite{Isern:2008fs, Dreiner:2013tja, Curtin:2014afa}, in which the effects are usually the most pronounced for younger white dwarfs, since the production of the new degree of freedom is enhanced at higher temperatures. In our case, the energy loss mechanism does not rely on the \emph{production} of new degrees of freedom, but rather radiation from the already captured dark matter.
It also shifts the deviations into the best-measured part of the WDLF. As Fig.~\ref{fig:luminosity_function} demonstrates, this means our $\epsilon$ bounds could be improved by a factor of at least a few by performing a more careful fit to the WDLF (though at that level of precision one should also take into account the distribution of stellar velocities relative to the aDM halo).

\subsection{Constraints on aDM parameter space}

\begin{figure}[t!]
    \centering
    \begin{subfigure}[]{0.33\textwidth}
        \includegraphics[scale=0.15]{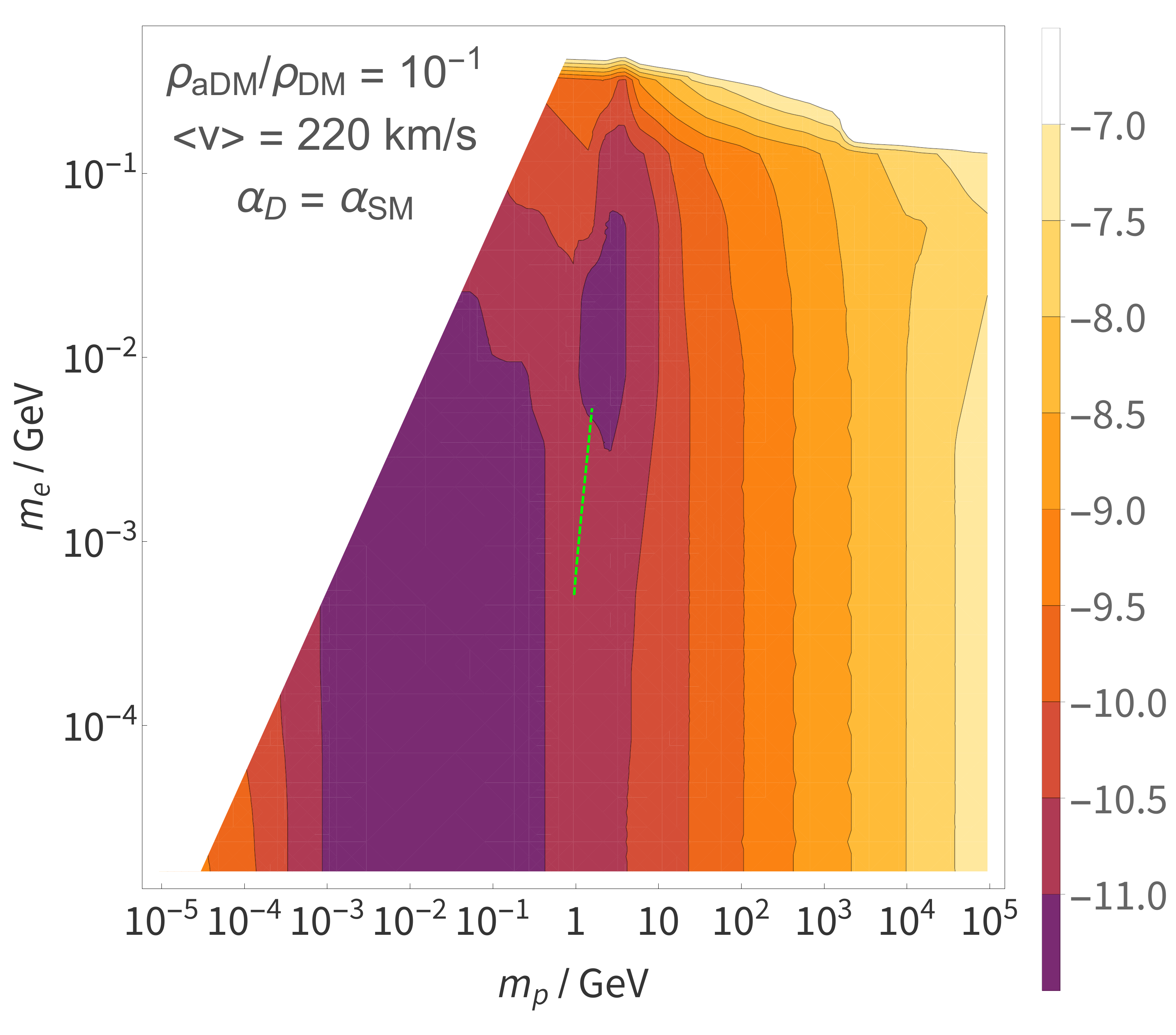}
    \end{subfigure}
    \hspace{-0.2cm}
    \begin{subfigure}[]{0.33\textwidth}
        \includegraphics[scale=0.15]{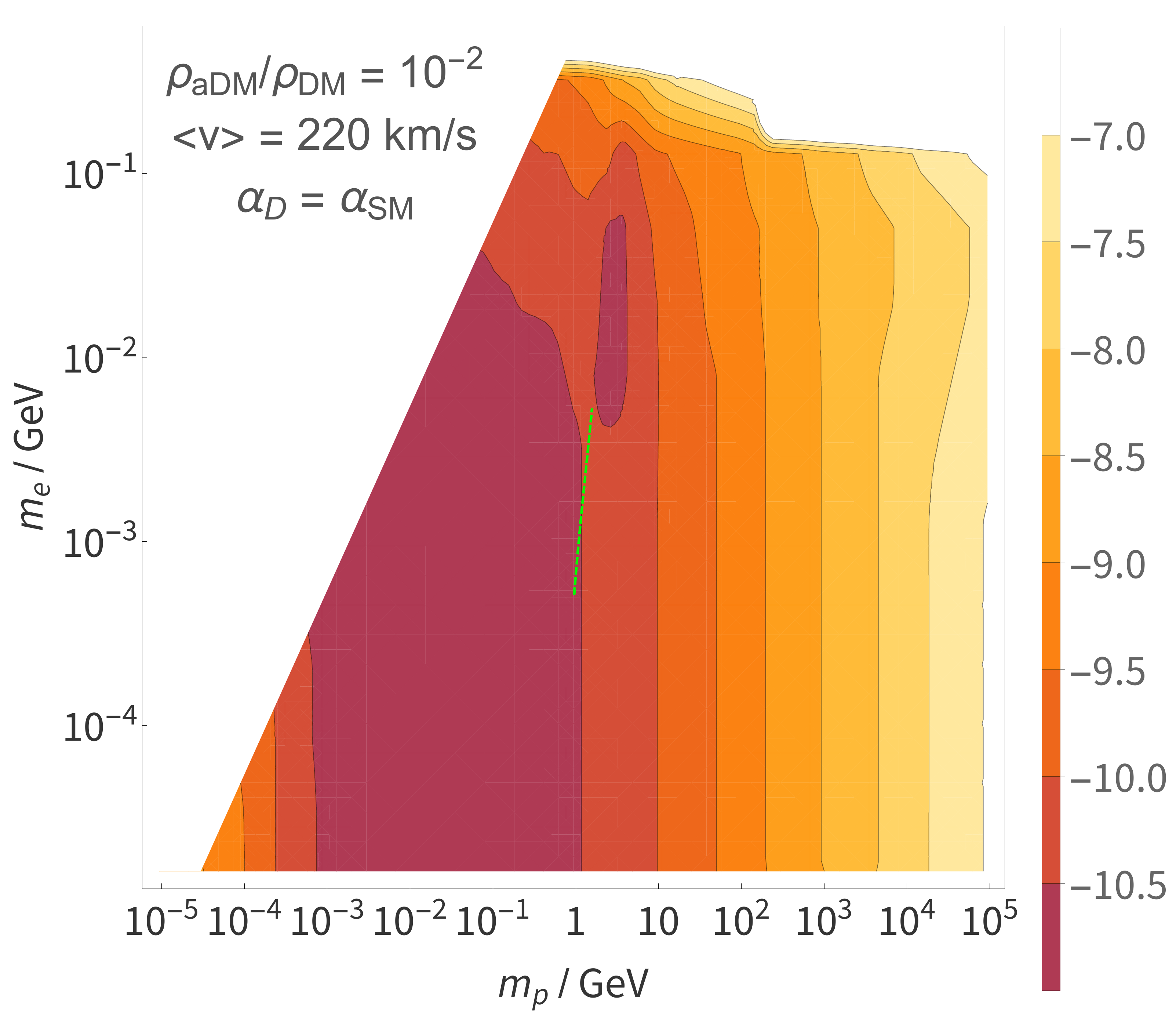}
    \end{subfigure}
    \hspace{-0.2cm}
    \begin{subfigure}[]{0.33\textwidth}
        \includegraphics[scale=0.15]{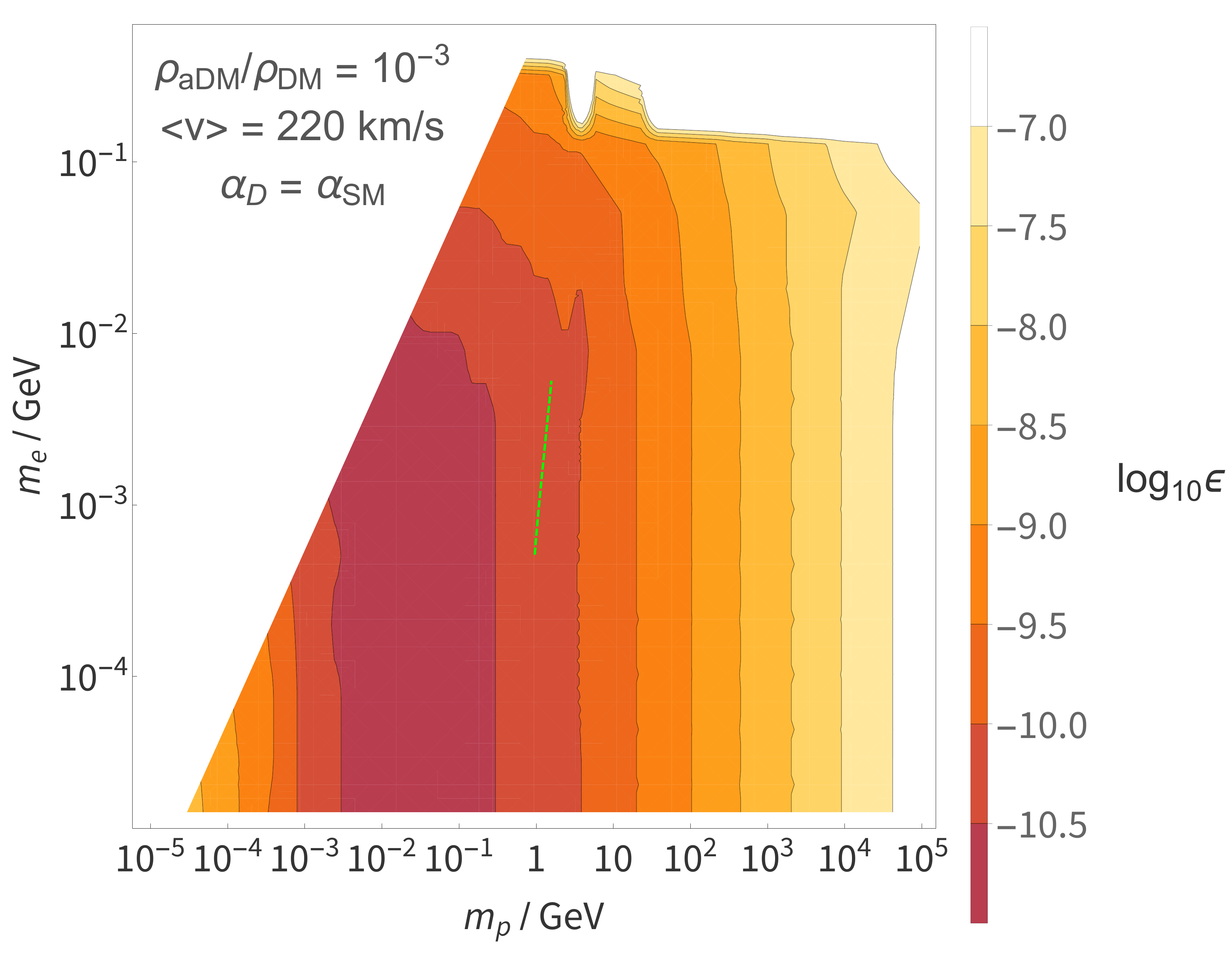}
    \end{subfigure}
    \begin{subfigure}[]{0.33\textwidth}
        \includegraphics[scale=0.15]{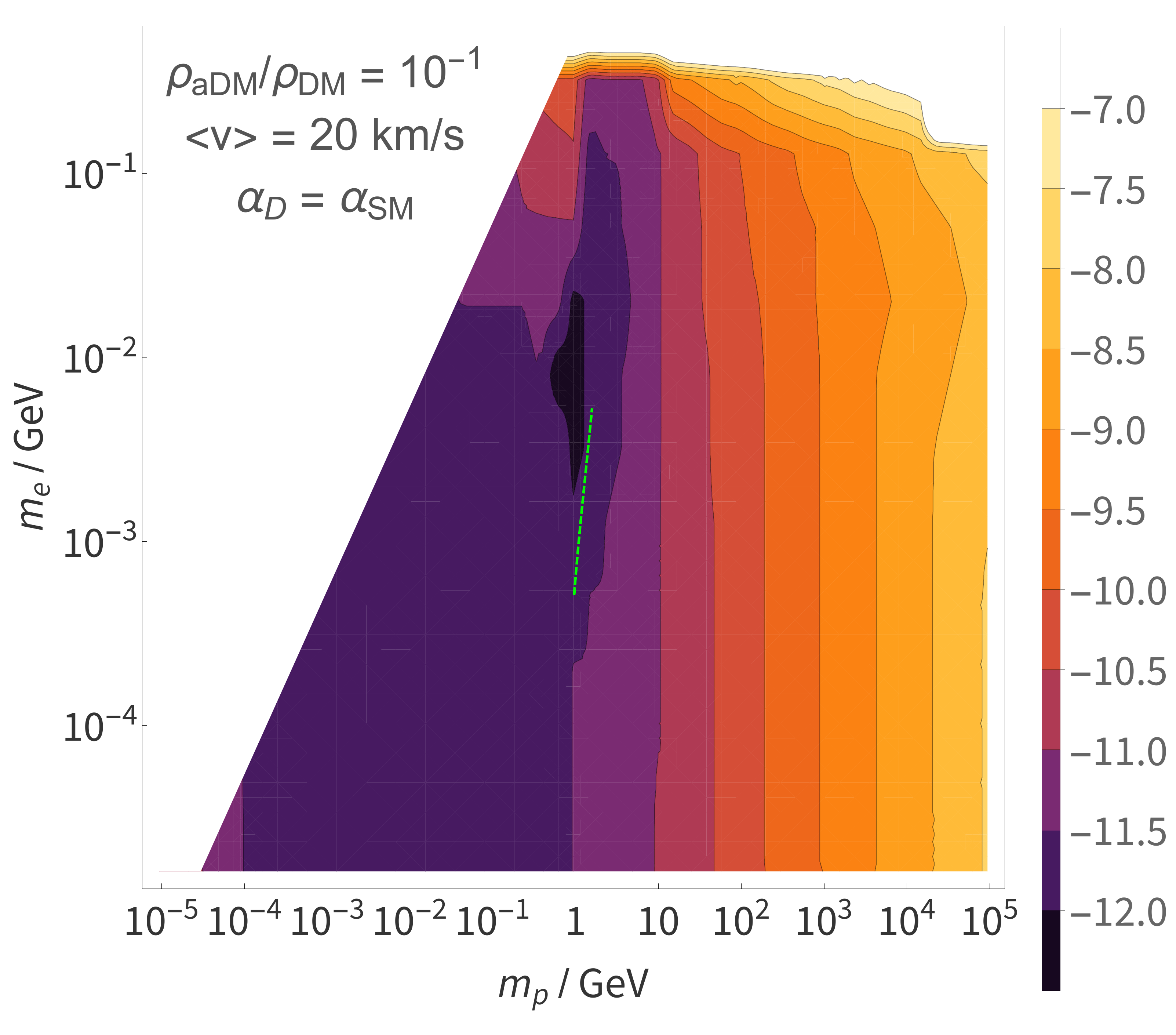}
    \end{subfigure}
    \hspace{-0.2cm}
    \begin{subfigure}[]{0.33\textwidth}
        \includegraphics[scale=0.15]{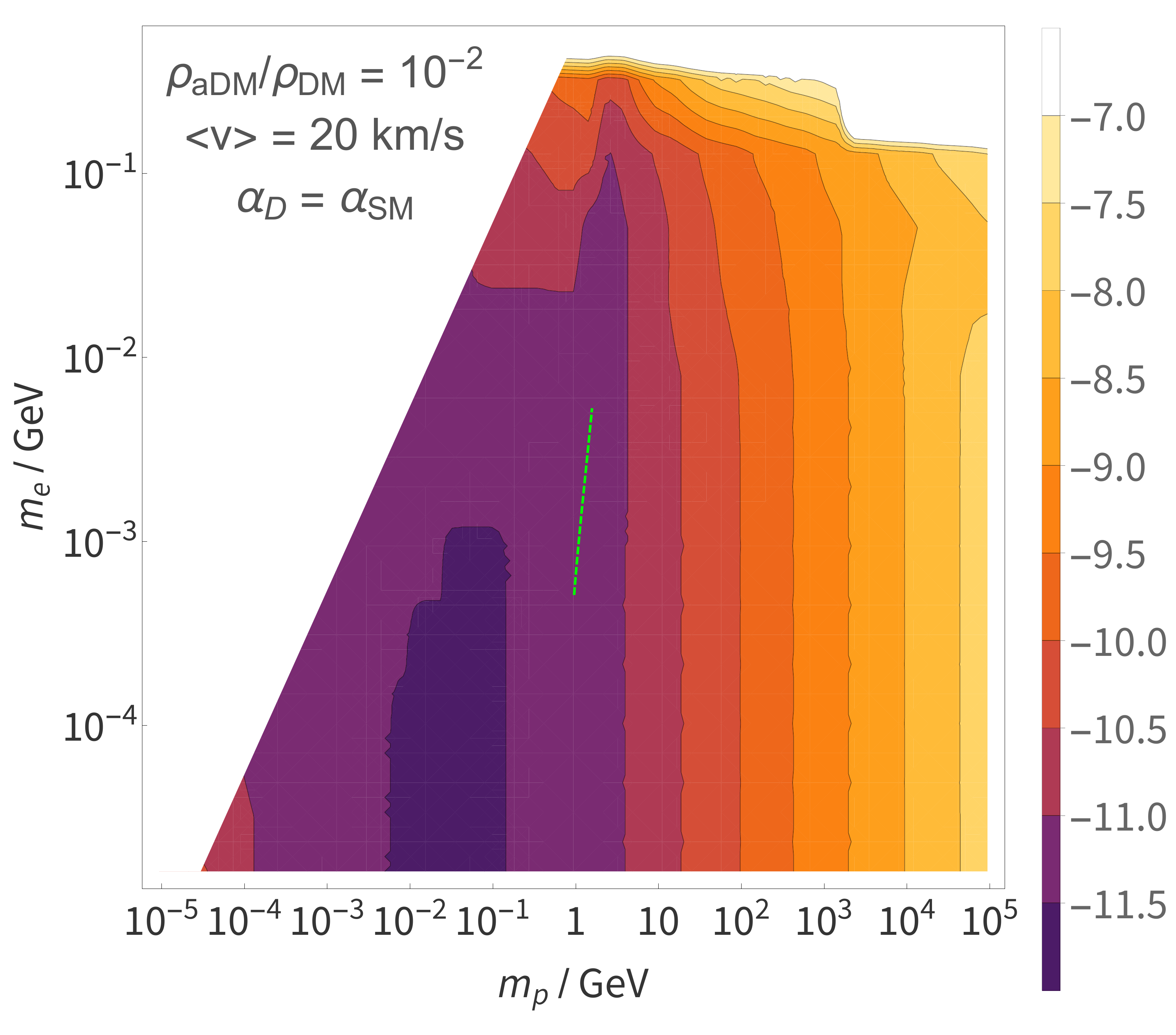}
    \end{subfigure}
    \hspace{-0.2cm}
    \begin{subfigure}[]{0.33\textwidth}
        \includegraphics[scale=0.15]{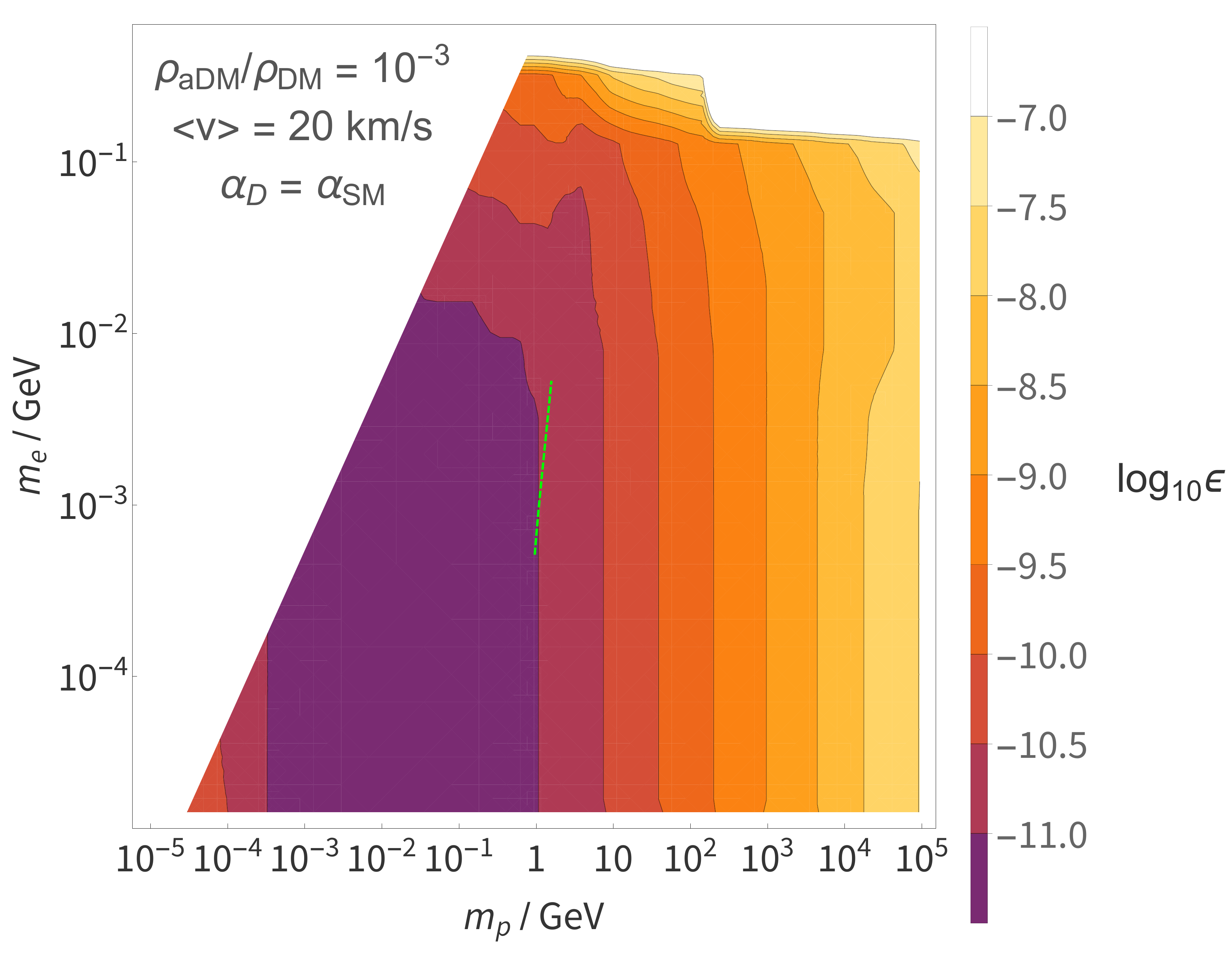}
    \end{subfigure}
    \caption{Colored contours show constraints on the dark photon mixing $\epsilon$ in aDM for $\alpha_D = \alpha_{em} \approx 1/137$ as a function of dark proton and dark electron mass. 
    The top row shows constraints assuming $\langle v \rangle = 220$ km/s (halo-like), while the bottom row shows constraints for $\langle v \rangle = 20$ km/s (disk-like). From left to right the columns correspond to dark matter density fractions of $\rho_{aDM}/\rho_{DM} = 10^{-1}$, $10^{-2}$, and $10^{-3}$.
    The dashed green line shows how the  dark electron and proton masses are correlated in the Mirror Twin Higgs scenario for $v_B/v_A = 1 - 10$.
    }
    \label{fig:gridplots}
\end{figure}

We present our aDM constraints from white dwarf cooling as exclusion limits on the value of the kinetic mixing $\epsilon$ in the plane of dark proton and dark electron mass, for different assumptions about the aDM density fraction and choice of velocity distribution (see Section~\ref{sec:velocity}).

In Figure~\ref{fig:gridplots} we show the bound on $\epsilon$ when $\alpha_D = \alpha_{em} \approx 1/137$ as a function of $m_{\tilde p}$ and $m_{\tilde e}$, for three different dark matter density fractions and for the halo-like and disk-like velocity distributions.
Comparing to Figure~\ref{fig:capture} we see that, as expected, the constraint is correlated to the amount of dark matter that has accumulated in the white dwarf. In particular, the fact that sensitivity peaks around $m_{\tilde p} \approx 1$ GeV is due to the fact that this is the mass region in which the dark matter survives evaporation during the helium burning period. In other regions of the parameter space, the accumulation is dominated by the white dwarf period, which increases for lower mass. Note however that the constraints get weaker at the low-mass corner of the plot, despite the large accumulation. This is because the heat transfer from the white dwarf to the aDM nugget becomes less efficient for small masses -- the average energy transfer per collision is kinematically suppressed by $m_{DM} / m_{N}$ in the limit of small dark matter mass, where $m_N$ is the mass of the colliding SM nucleus.
Our constraints die off around $m_{\tilde e} \sim 10^{-1}$~GeV, corresponding to a dark atomic binding energy of $\tilde E_0 \sim 10$~keV. This is because the limits are dominated by cold and old white dwarfs with a core temperature of $3\times 10^6$ K (see \fref{f.WDevolution} (c)). For larger dark electron masses, atomic processes in the aDM nugget would dominate cooling. We leave the corresponding analysis for future studies.

\begin{figure}
    \centering
    \hspace{-7mm}
    \begin{tabular}{ccc}
        \includegraphics[height=4.6cm]{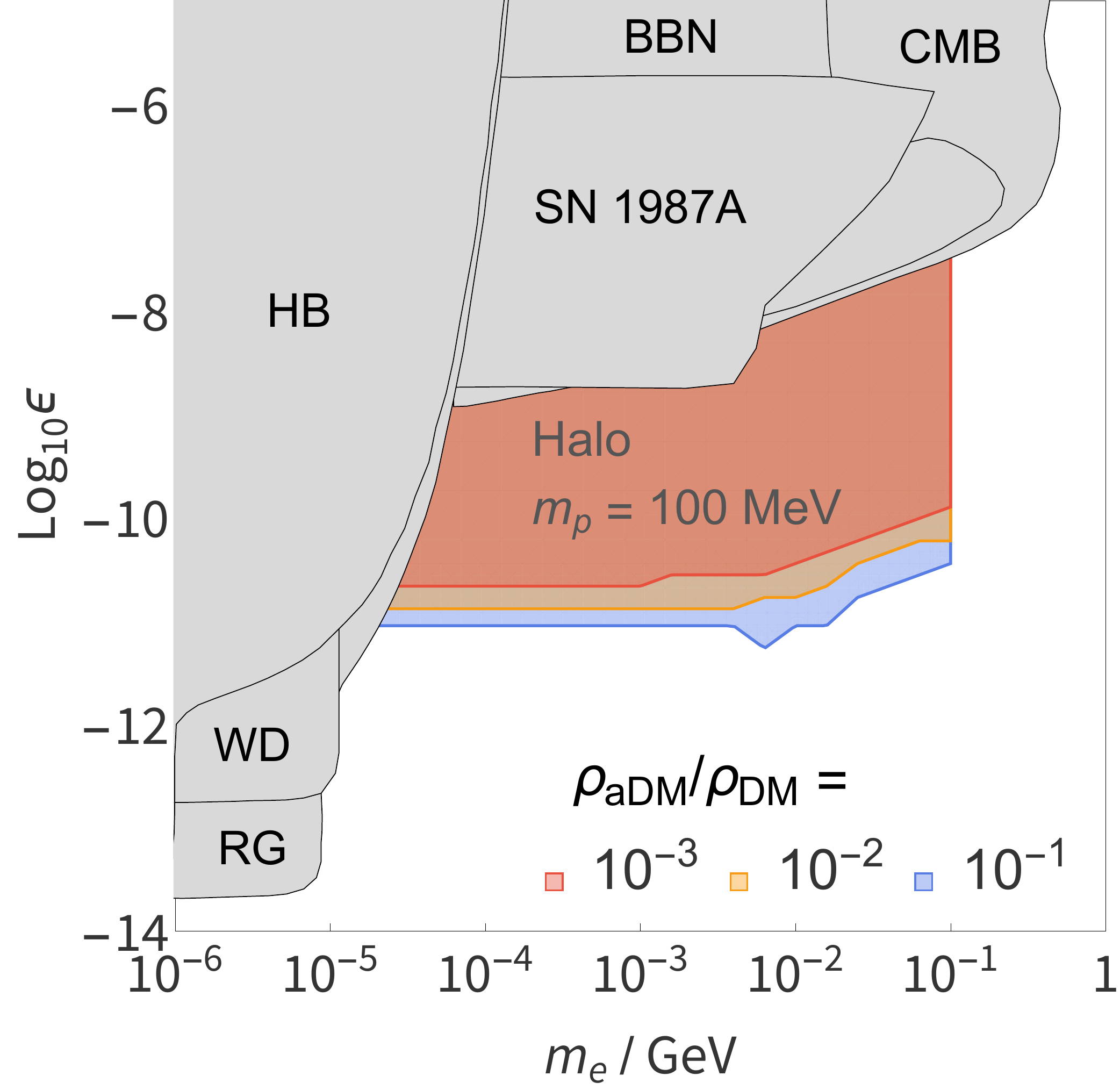}
        &
        \includegraphics[height=4.6cm]{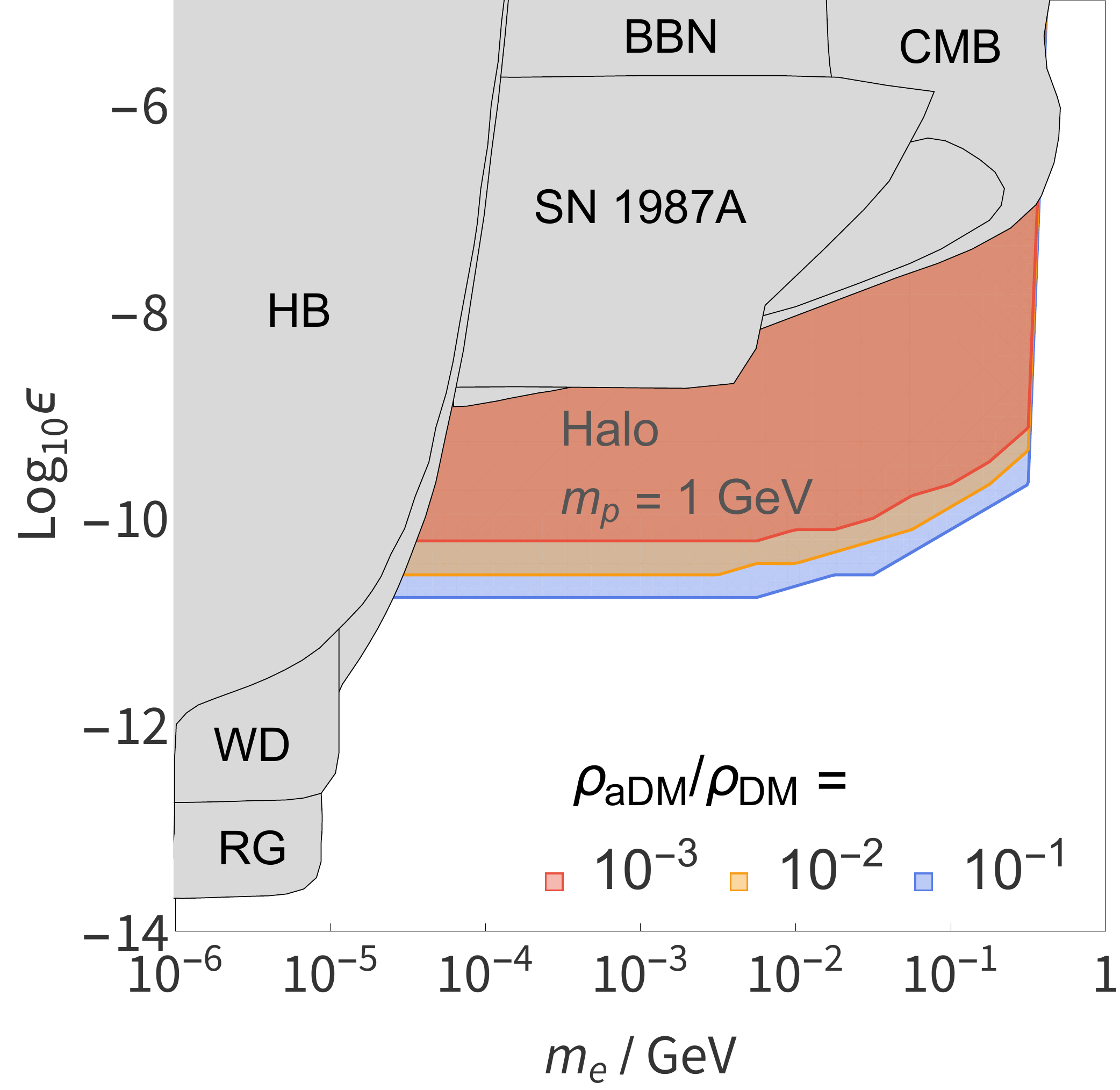}
        &
        \includegraphics[height=4.6cm]{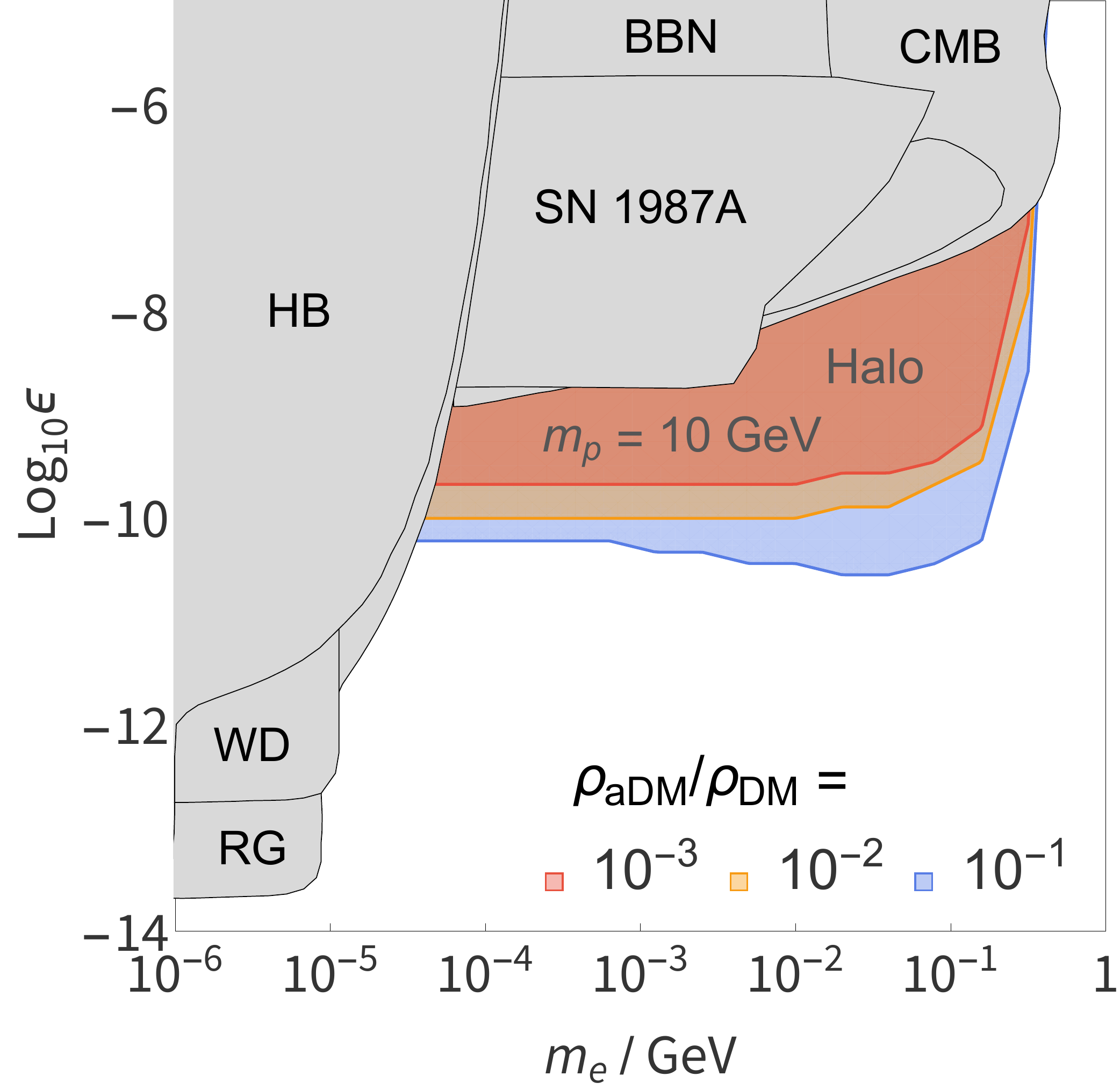}
        \\
        \includegraphics[height=4.6cm]{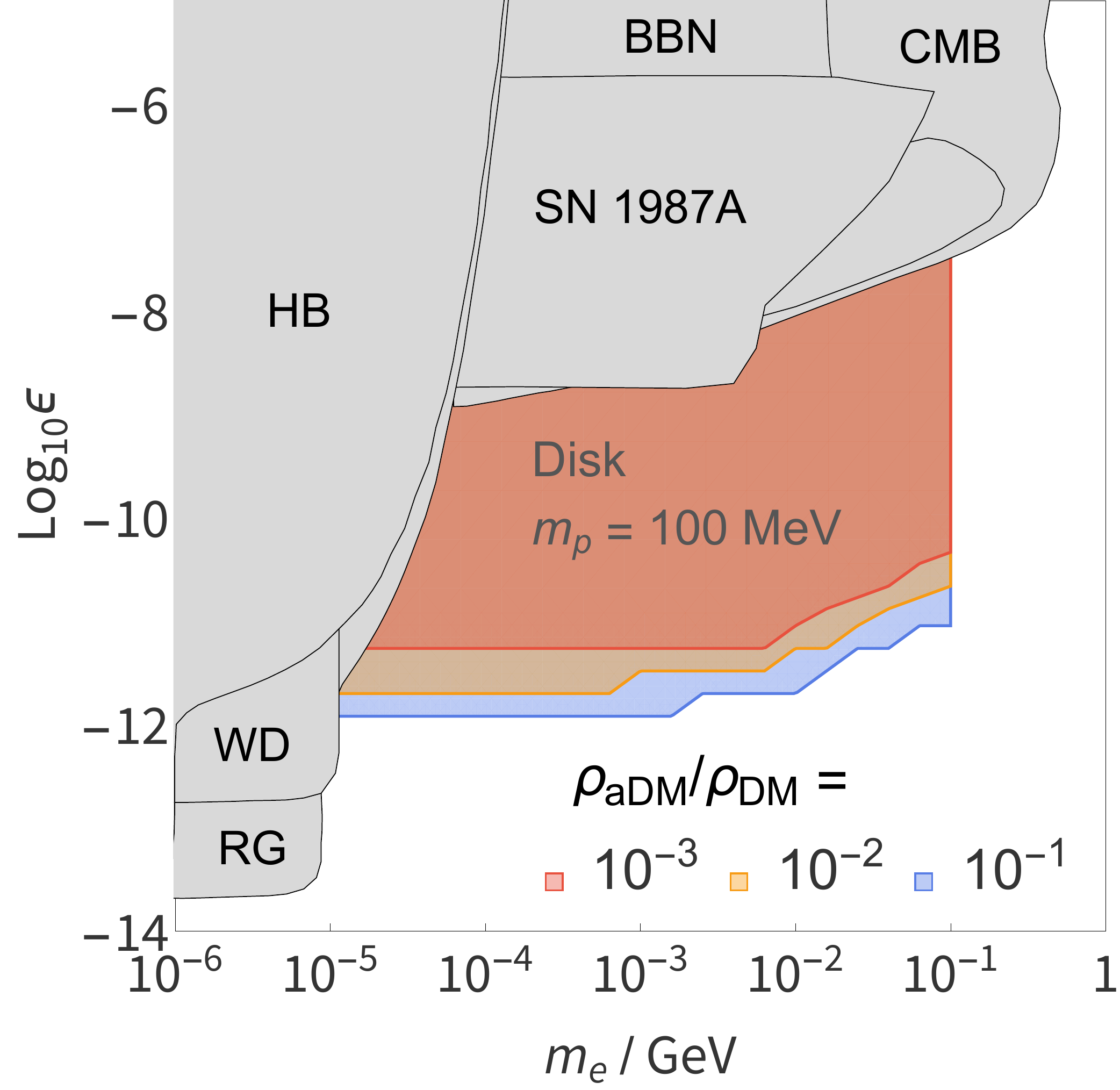}
        &
        \includegraphics[height=4.6cm]{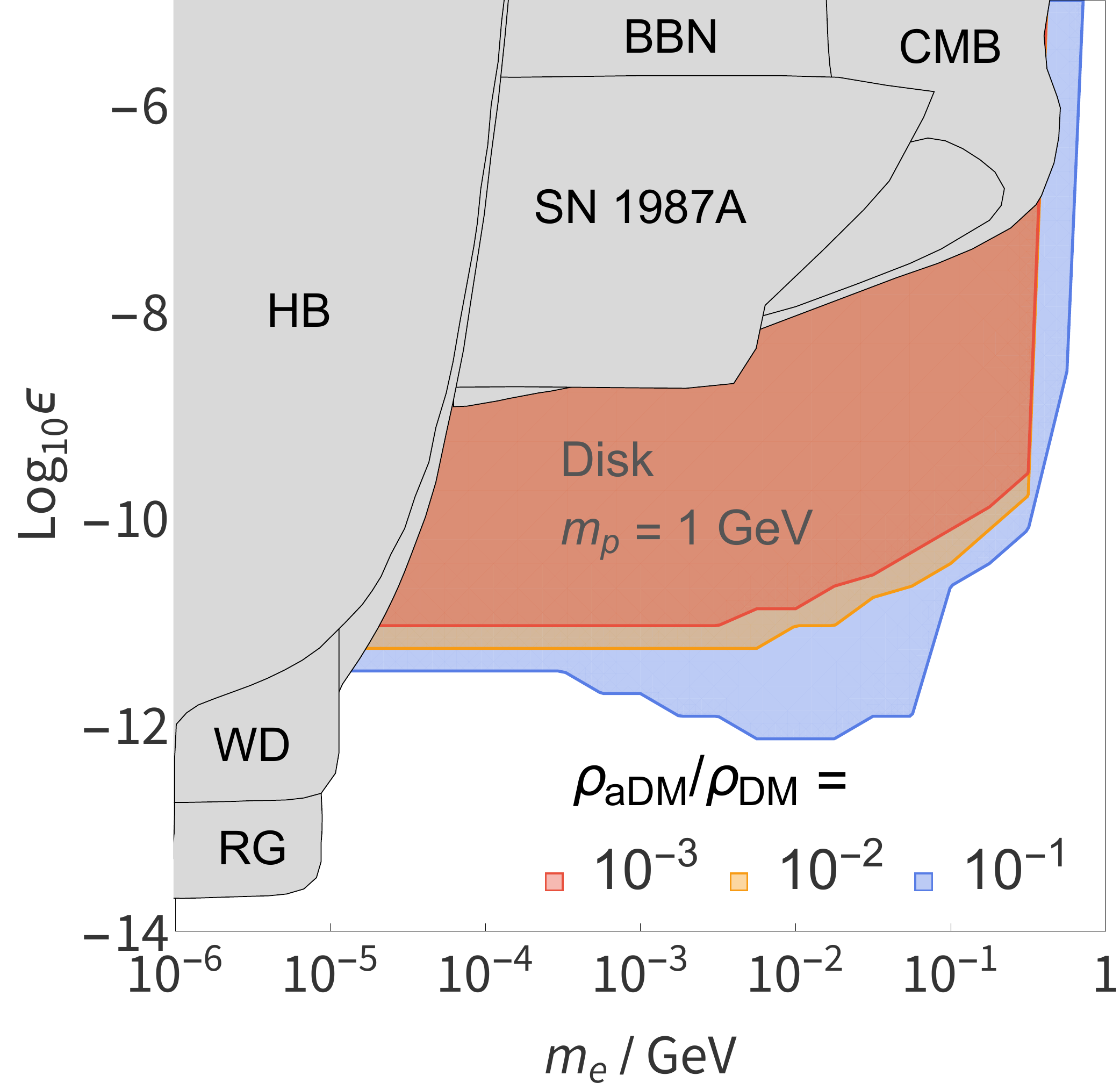}
        &
        \includegraphics[height=4.6cm]{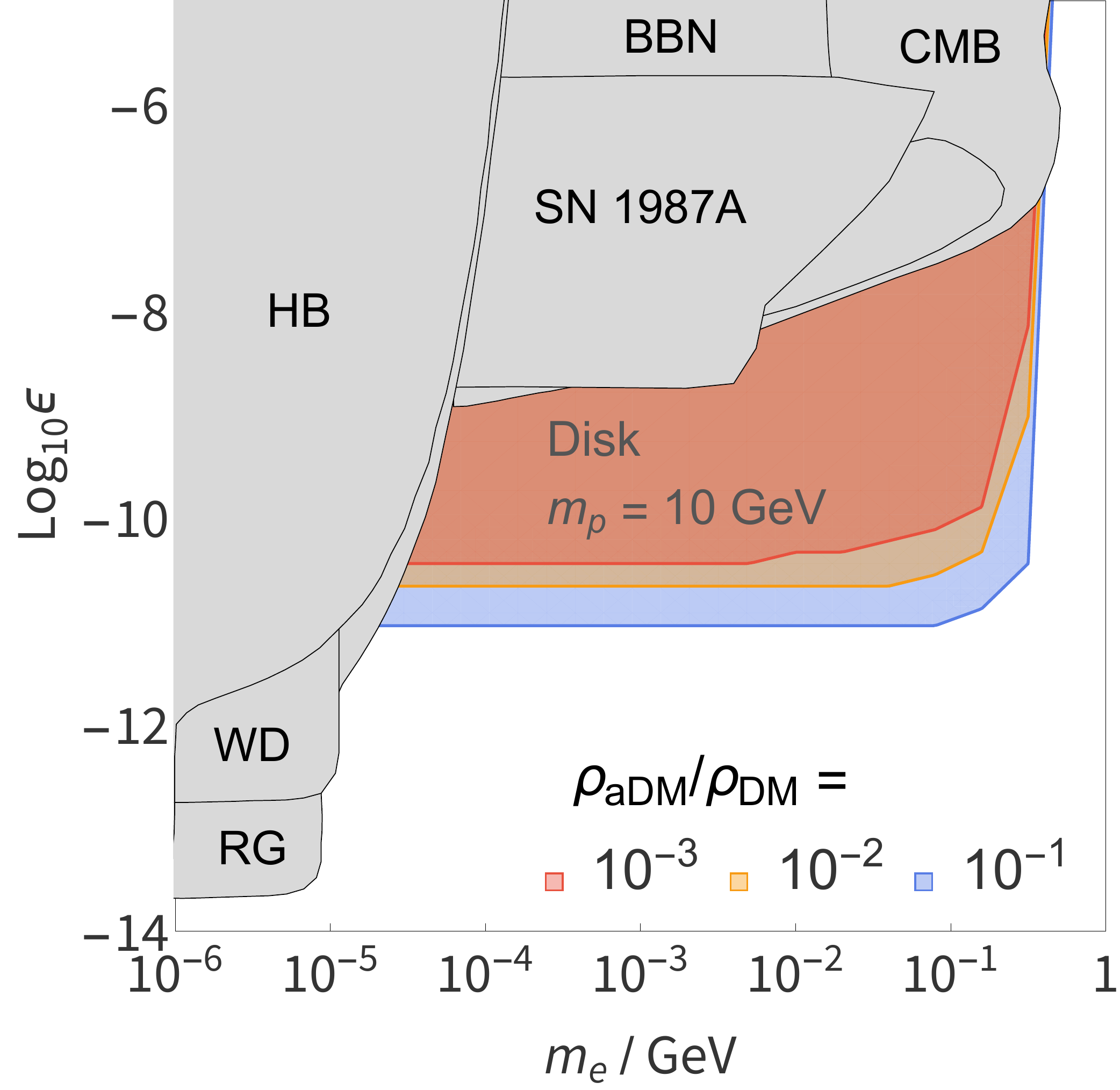}
    \end{tabular}
    \caption{
    Constraints on $\epsilon$ as a function of $m_{\tilde e}$, for three different values of the dark proton mass and $\alpha_D = \alpha_{em}$, assuming halo-like aDM velocity distribution (above) and the disk-like velocity distribution (below). The constraints shown here are from cooling of horizontal branch (HB) stars, white dwarfs (WD) and red giants (RG) \cite{Davidson:1991si,Davidson:1993sj,Raffelt:1996wa,Davidson:2000hf,Vogel:2013raa}, $\Delta N_\mathrm{eff}$ at BBN and CMB times \cite{Vogel:2013raa}, and from supernova 1987A \cite{Chang:2018rso}.
     Note that in the first column the constraints are cut off at $m_{\tilde e} = 100$ MeV, since the dark electron is defined to be always lighter than the dark proton. For heavier dark protons, the constraints die off due to inefficient bremsstrahlung for high electron masses and therefore low ionization fractions.}
    \label{fig:fixed_mp}
\end{figure}

\begin{figure}
    \centering
    \includegraphics[scale=0.4]{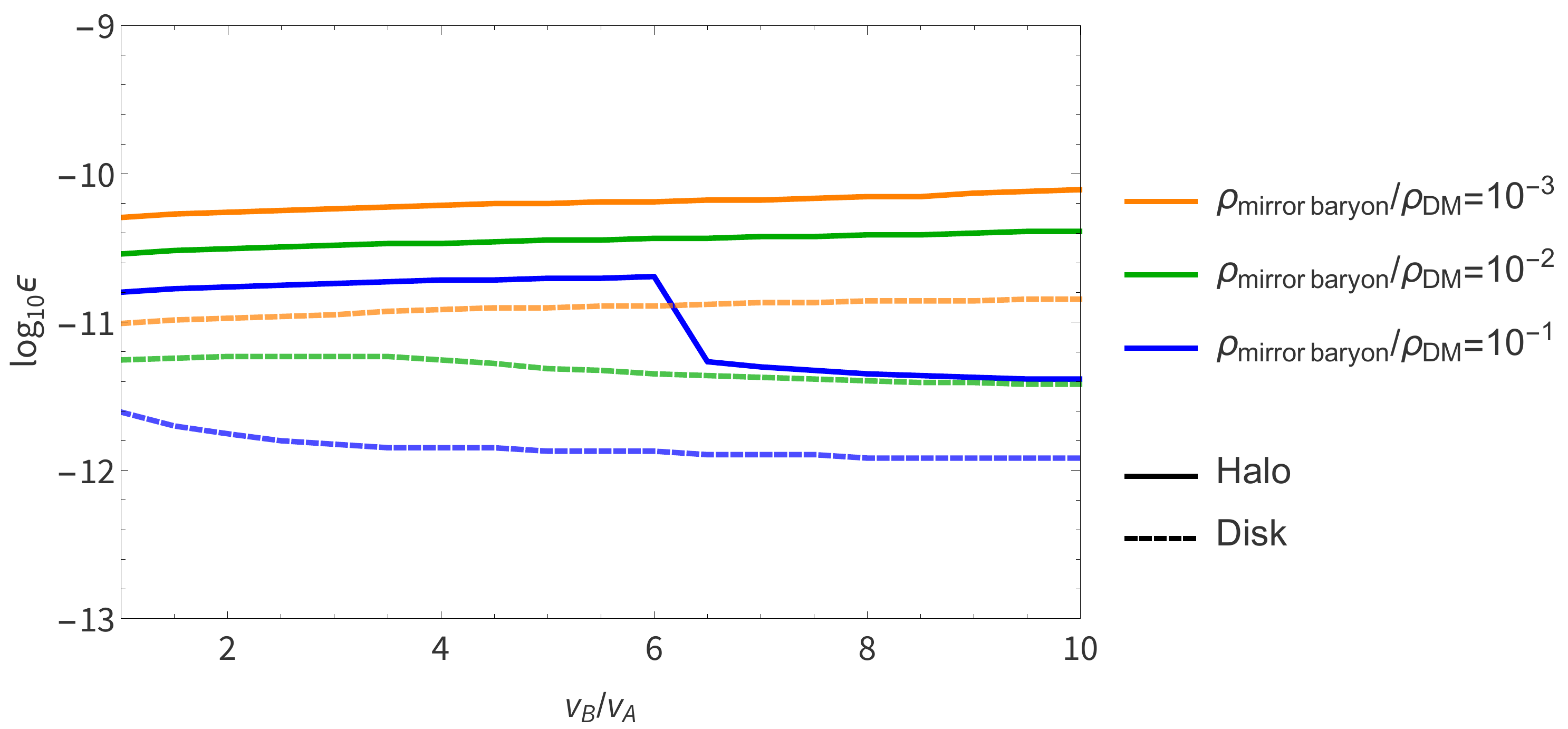}
    \caption{White dwarf cooling constraints on mirror photon kinetic mixing in the asymmetrically reheated Mirror Twin Higgs model as a function of the parameter $v_B / v_A$, the ratio of the Higgs VEVs in the hidden and visible sectors. The minimal Twin Higgs model gives a specific prediction for $m_{\tilde p}$ and $m_{\tilde e}$ as a function of this parameter (see Section~\ref{sec:models}). The solids lines are for $\vev{v} = 220$ km/s, while the dashed lines are for the disk-like $\vev{v} = 20$ km/s assumption.
    }
    \label{fig:fv_plot}
\end{figure}

\begin{figure}[t!h!]
    \begin{center}
    \begin{tabular}{ccc}
            \includegraphics[scale=0.15]{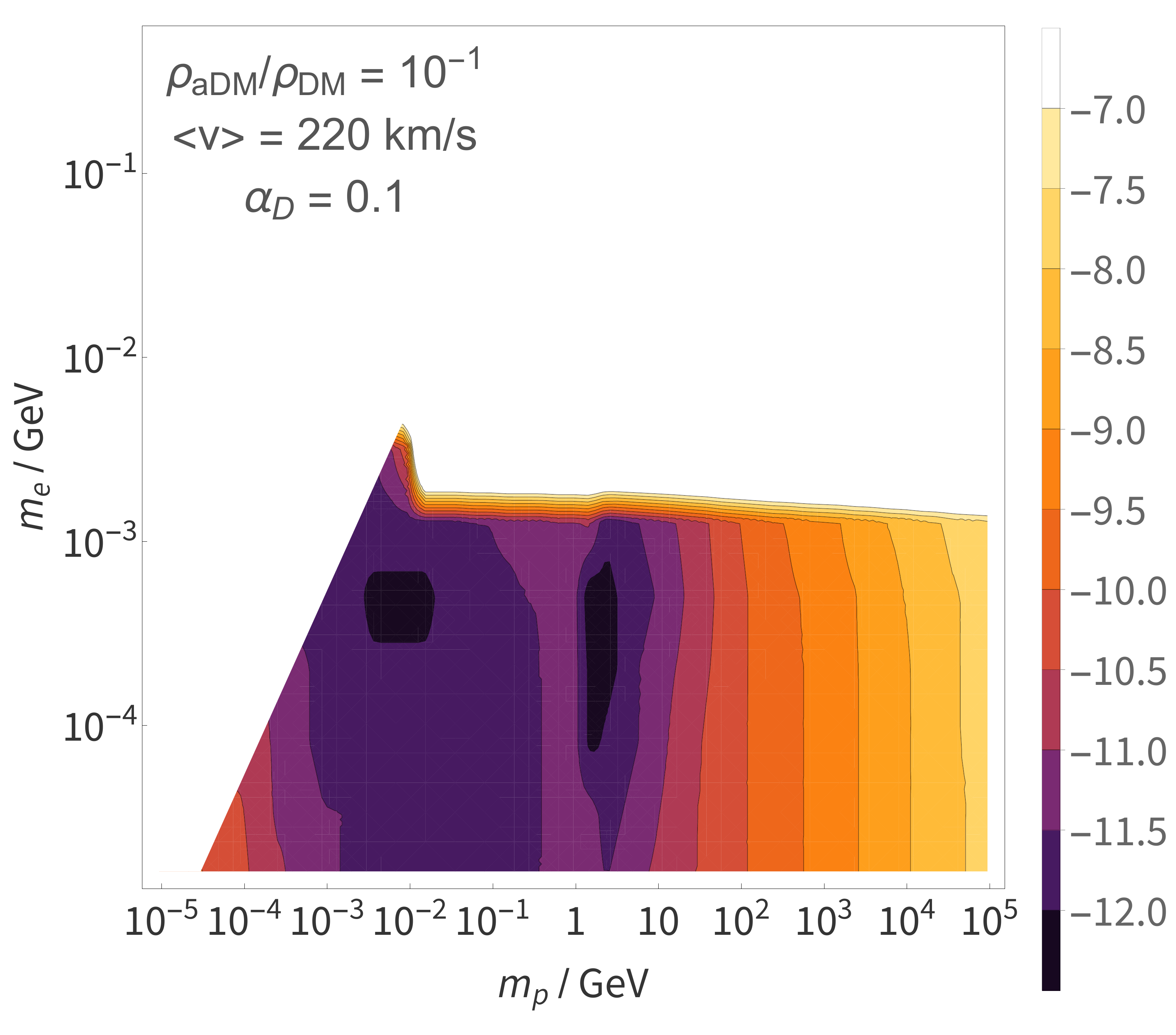}
   &
   
        \includegraphics[scale=0.15]{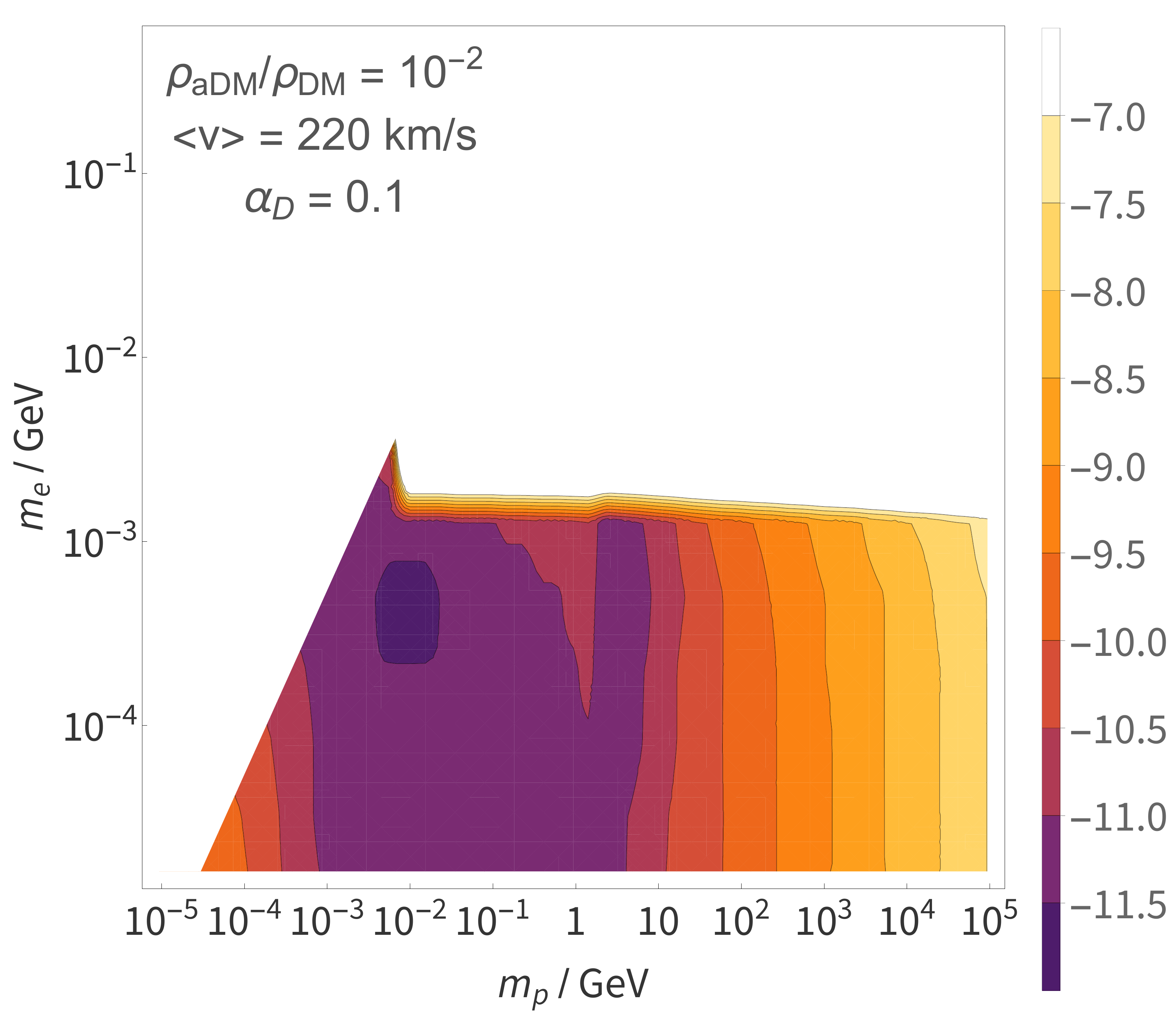}
   &
           \includegraphics[scale=0.15]{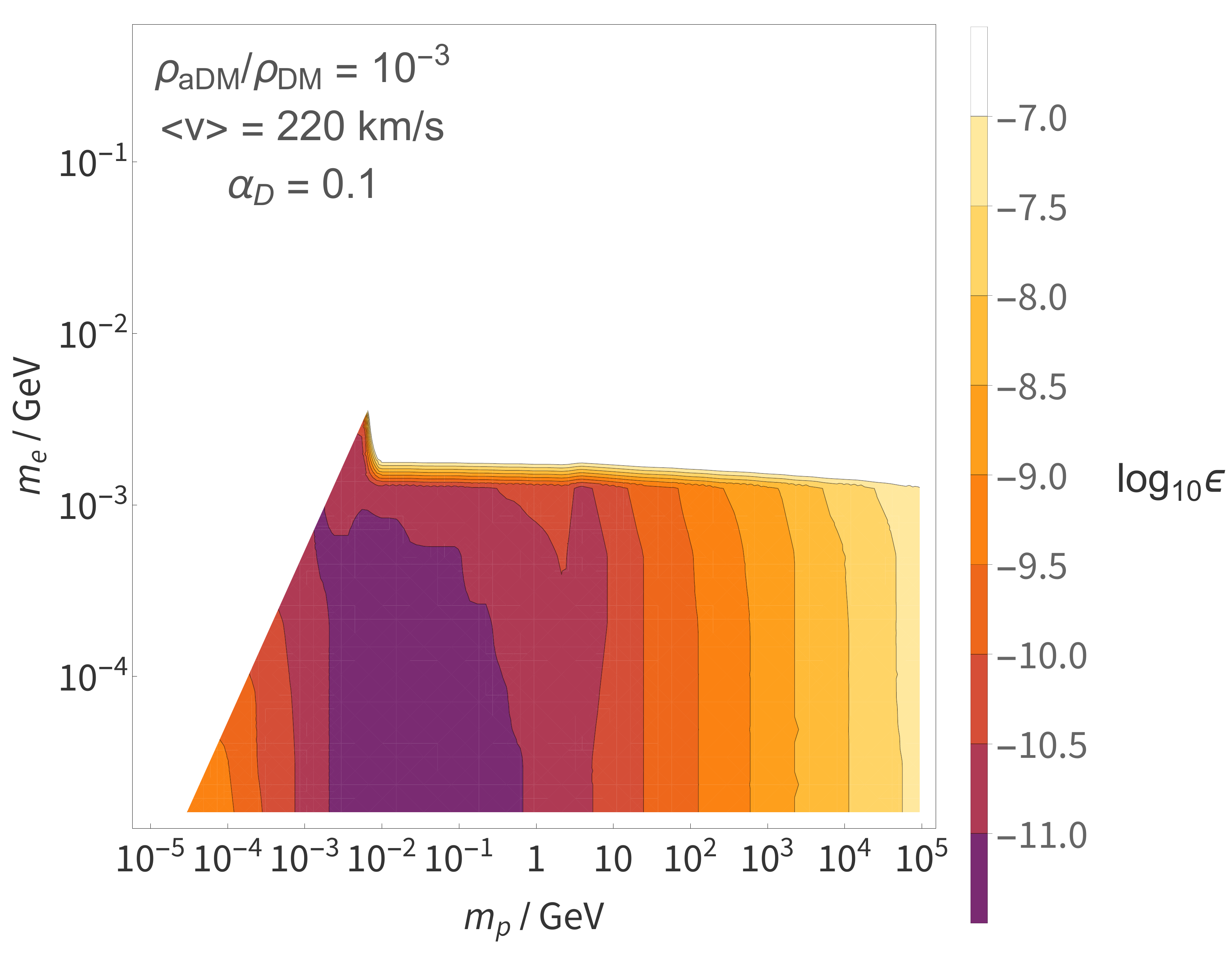}
           \\
        \includegraphics[scale=0.15]{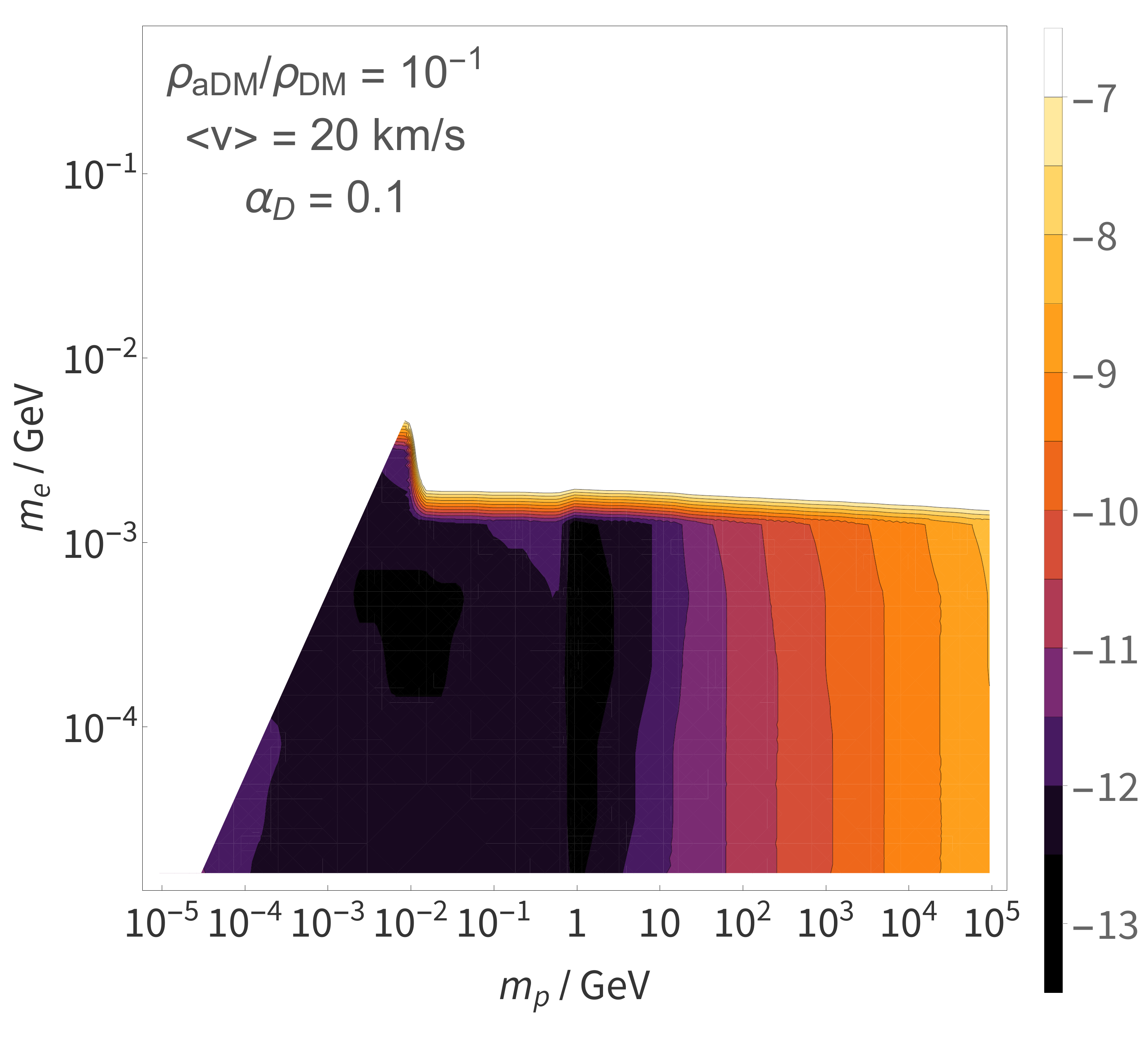}
&
        \includegraphics[scale=0.15]{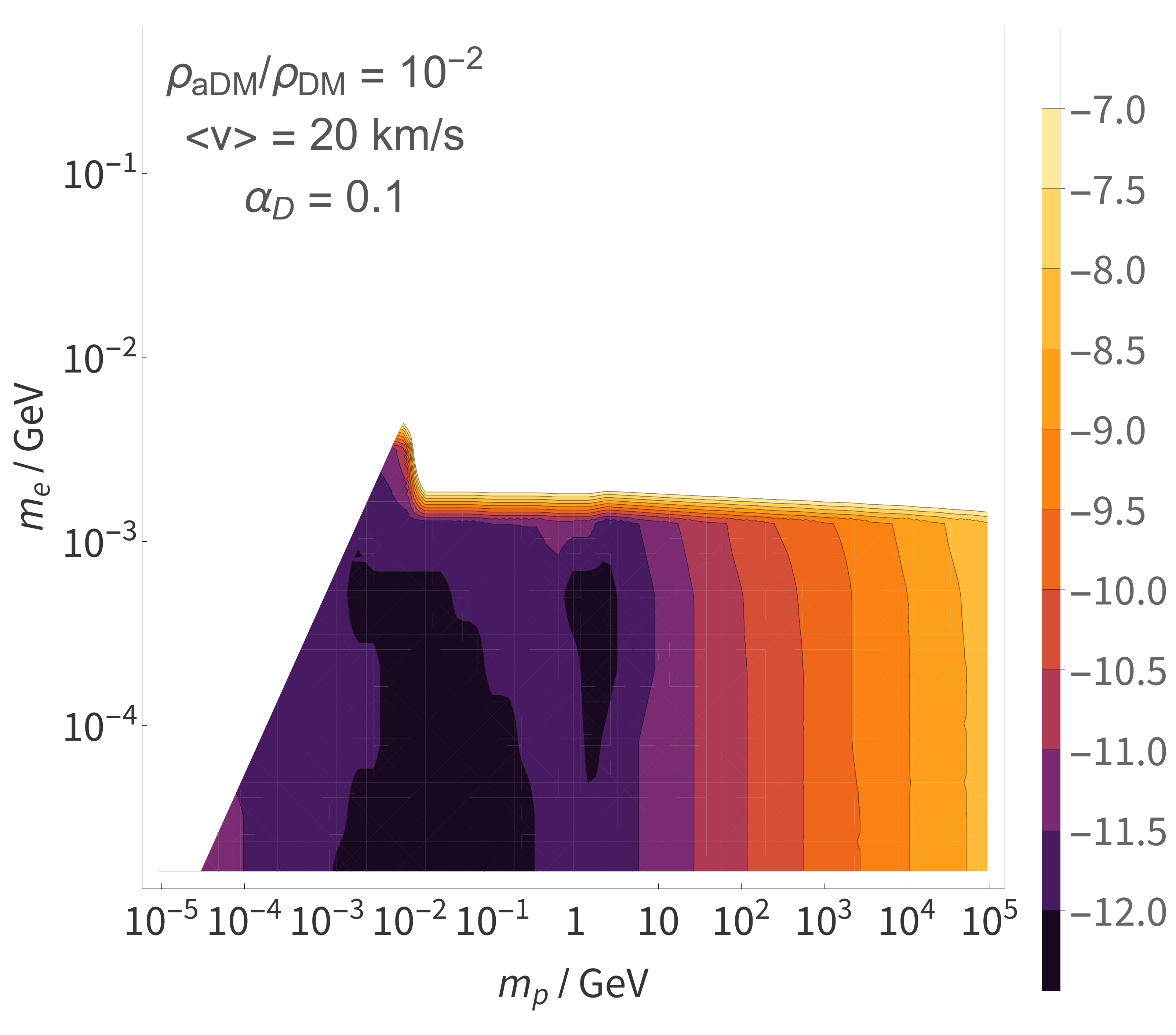}
&        \includegraphics[scale=0.15]{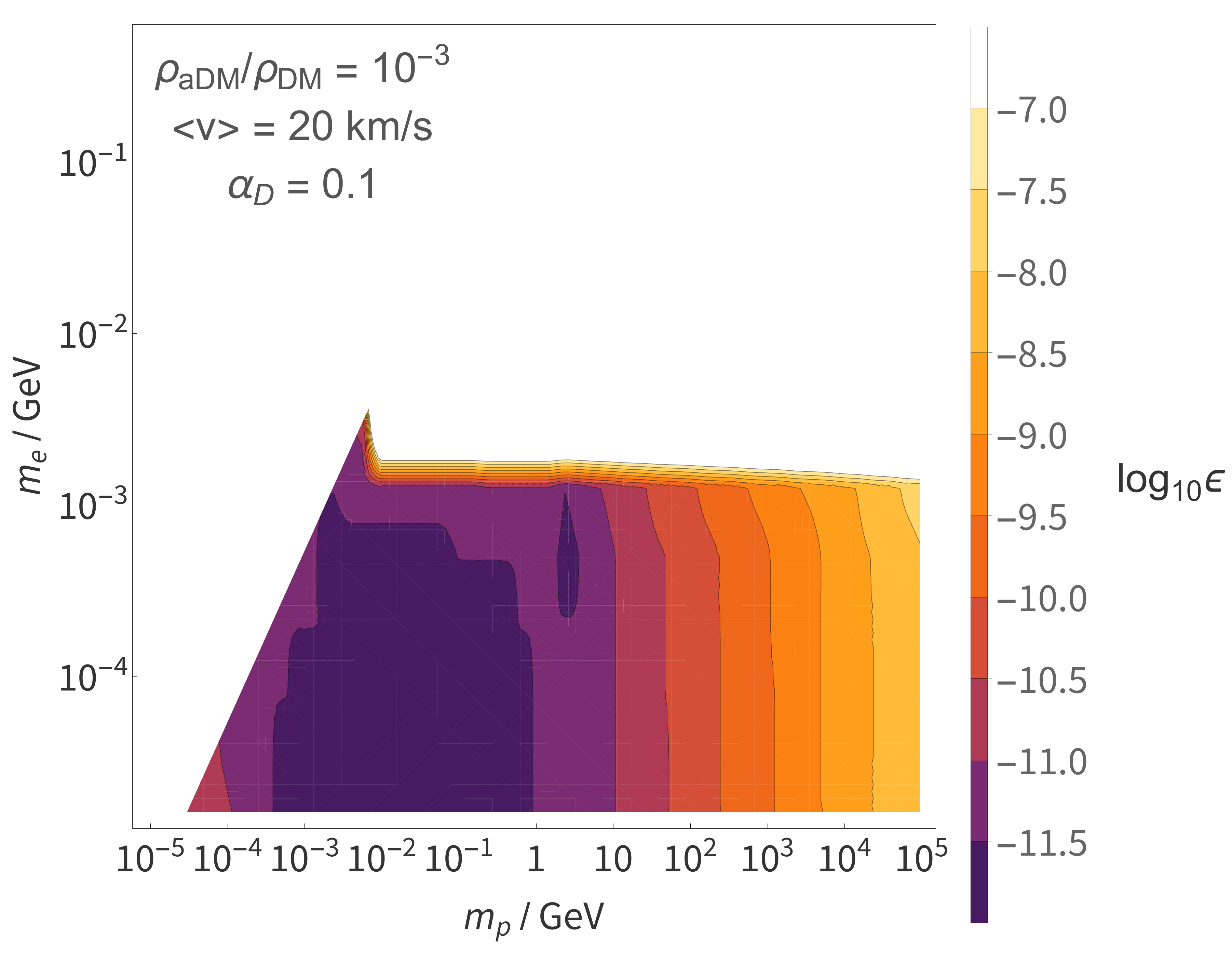}
  \end{tabular}
      \end{center}
    \caption{Same as Figure~\ref{fig:gridplots} but for $\alpha_D = 0.1$.}
    \label{fig:gridplots2}
\end{figure}

\begin{figure}[t!h!]
    \begin{center}
    \begin{tabular}{ccc}
      \includegraphics[scale=0.15]{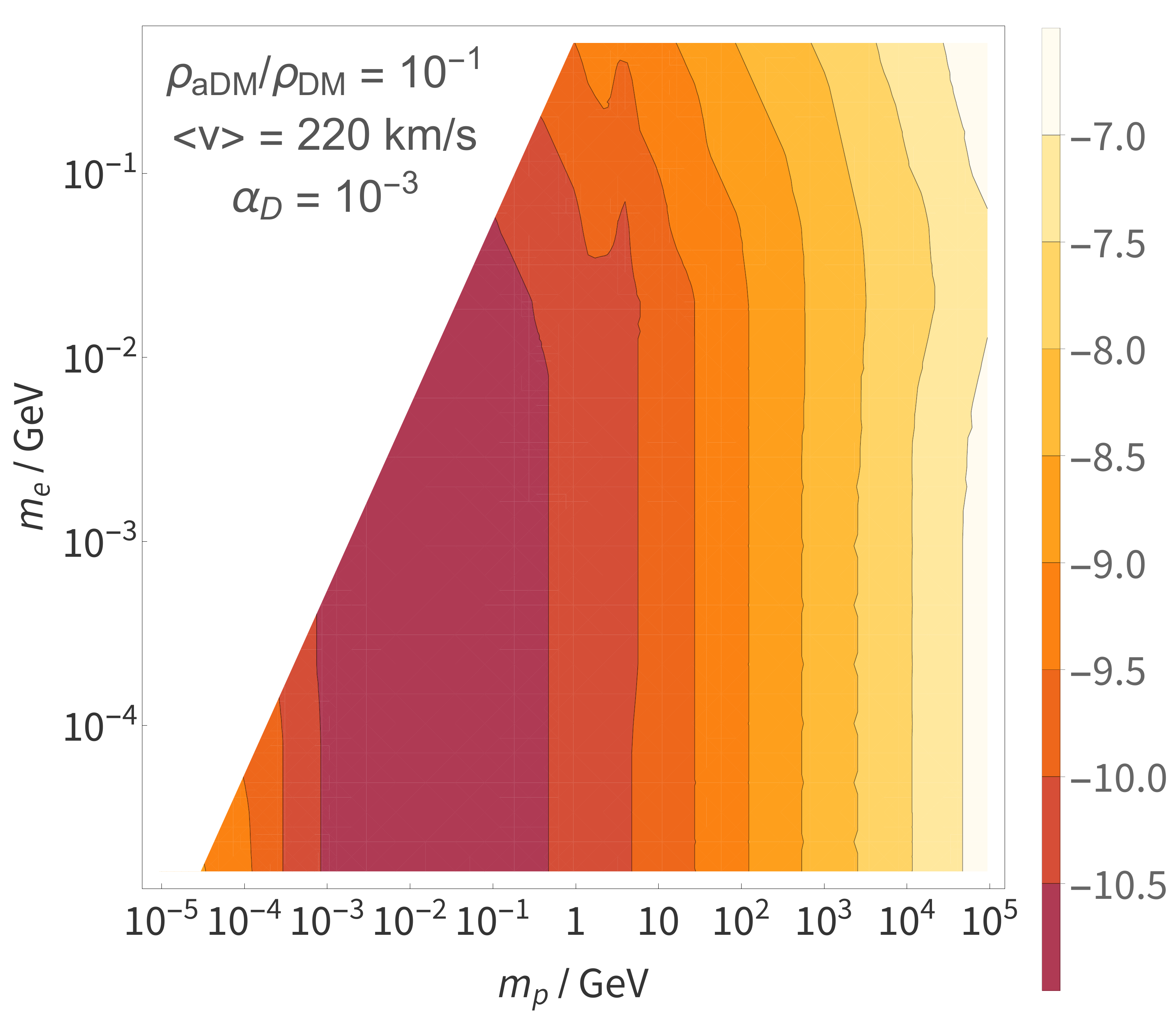}
   &
    \includegraphics[scale=0.15]{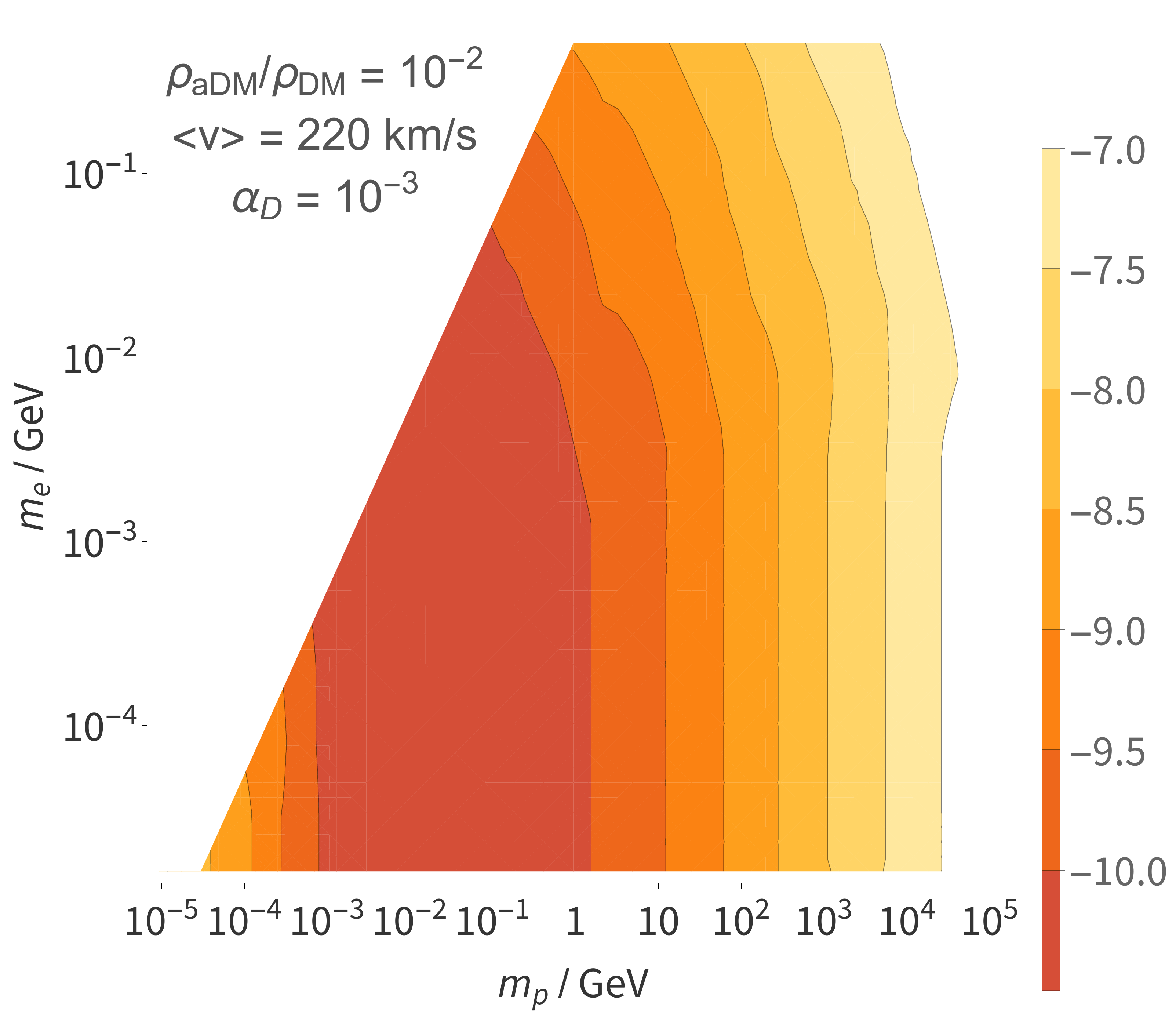}
  &
     \includegraphics[scale=0.15]{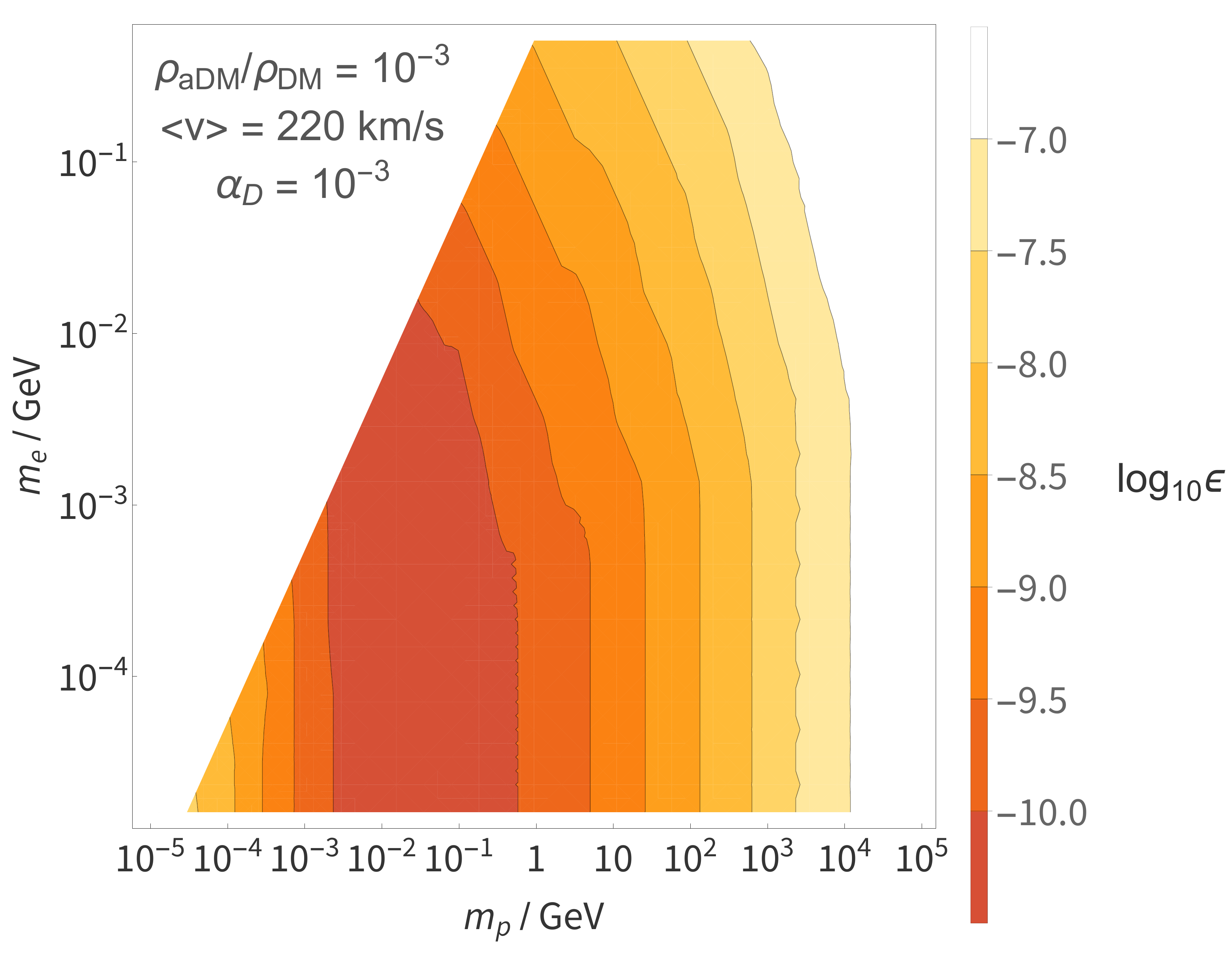}
  \\
    \includegraphics[scale=0.15]{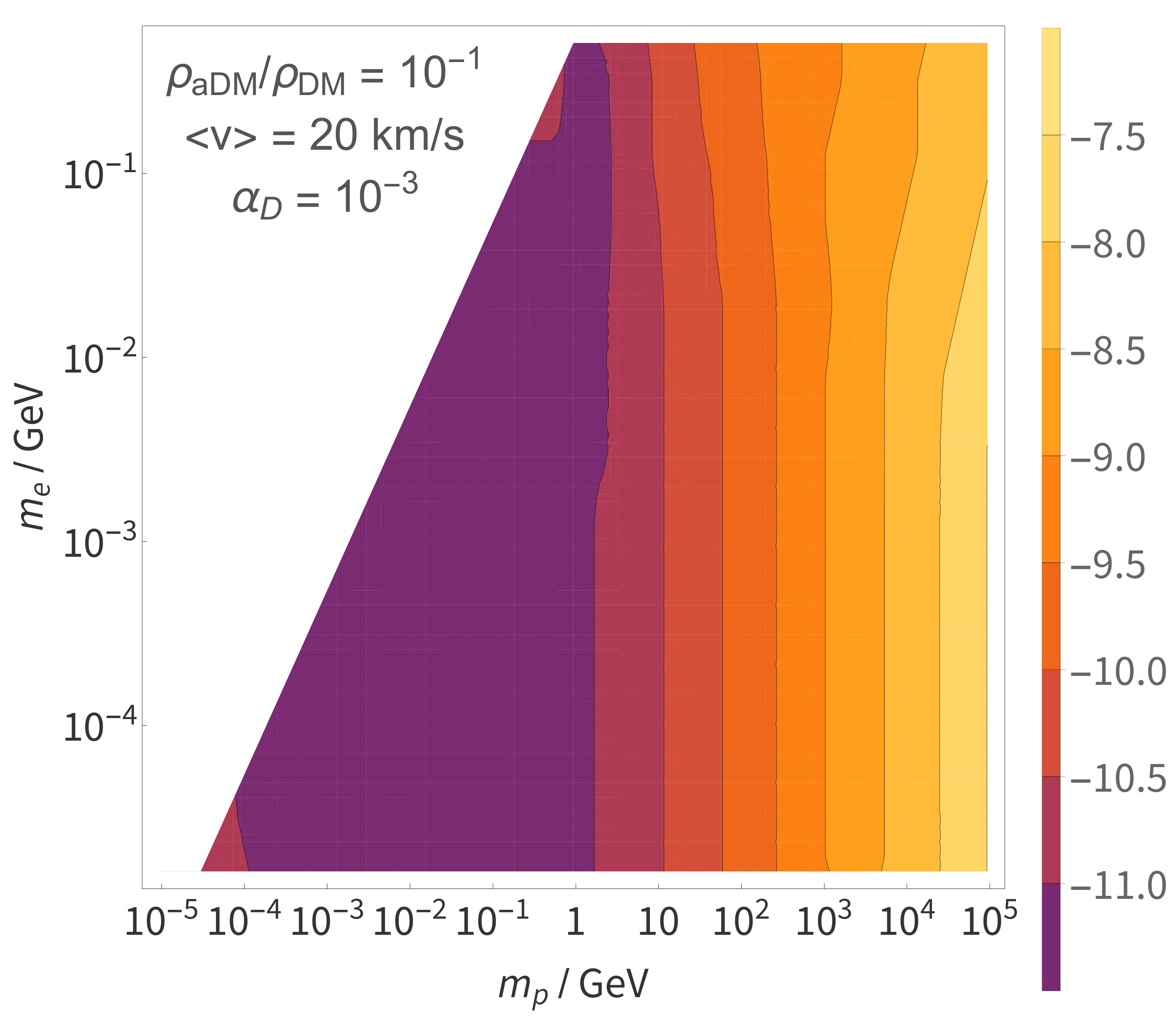}
 &
         \includegraphics[scale=0.15]{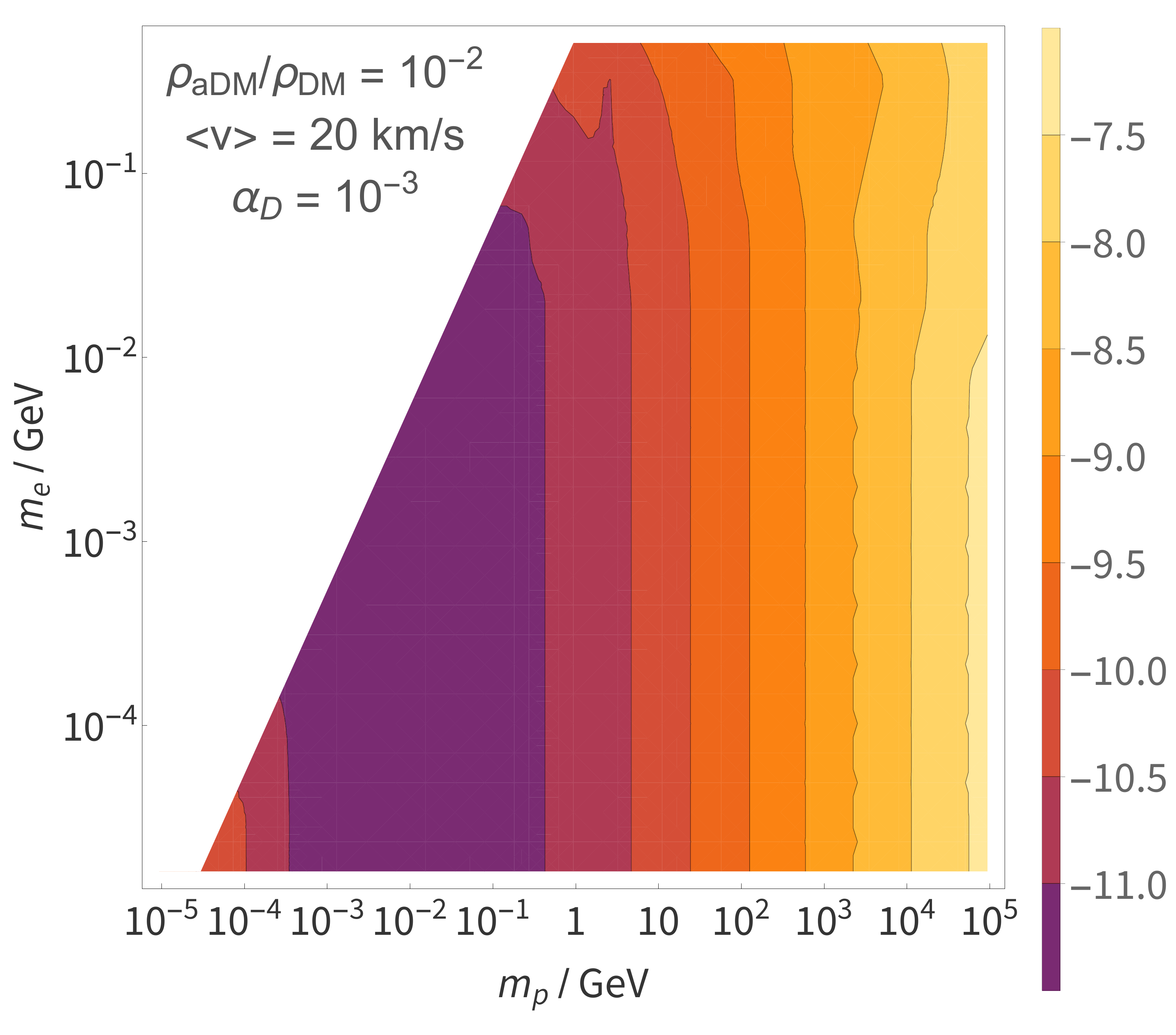}
  &
          \includegraphics[scale=0.15]{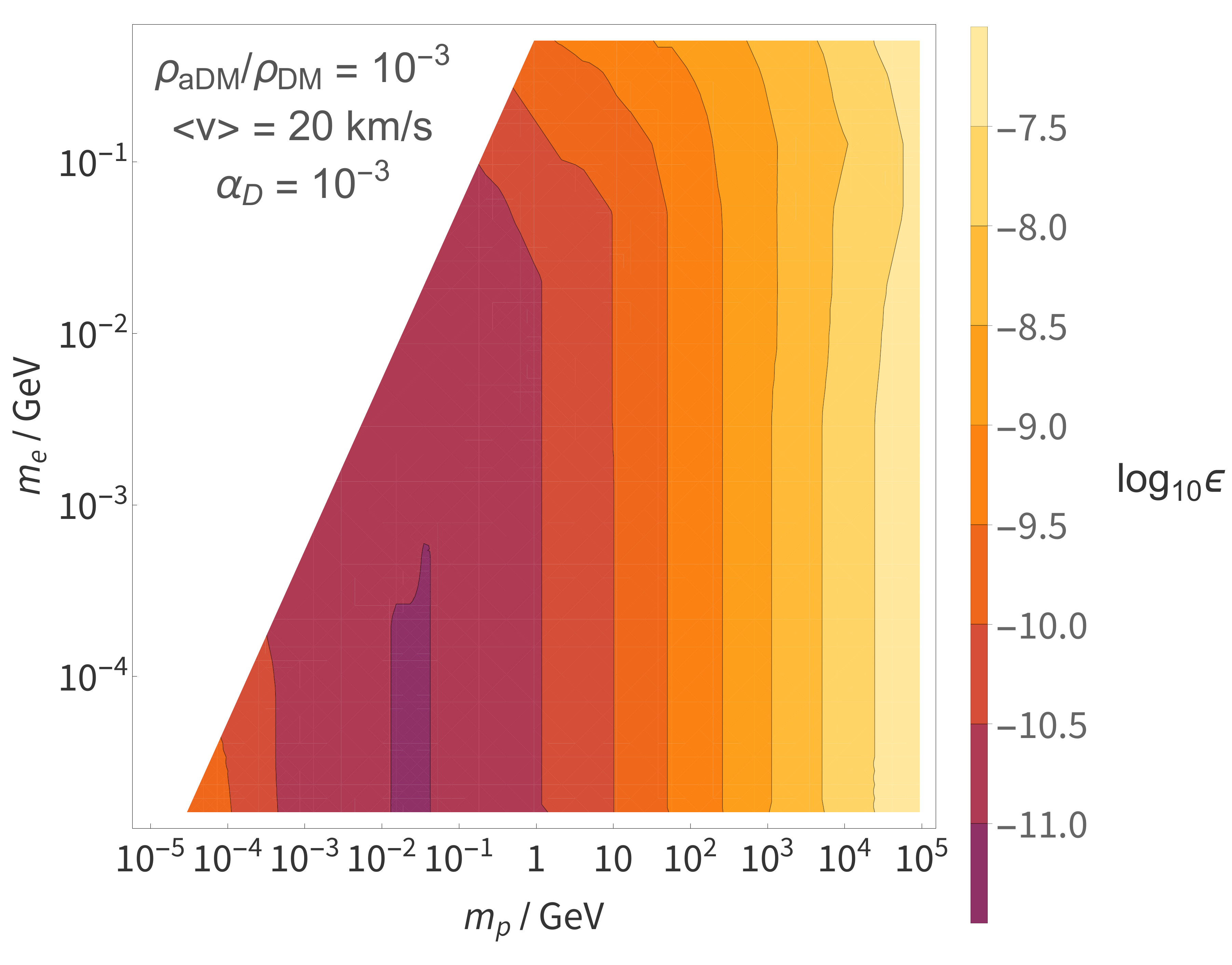}
  \end{tabular}
      \end{center}
    \caption{Same as Figure~\ref{fig:gridplots} but for $\alpha_D = 10^{-3}$ .}
    \label{fig:gridplots3}
\end{figure}

In most regions of parameter space where $m_{\tilde e} \gtrsim 10^{-5}$~GeV, white dwarf cooling presents the strongest bounds on aDM to date. This can be seen in Figure~\ref{fig:fixed_mp}, which shows our limits as a function of $m_{\tilde e}$ for three different fixed values of $m_{\tilde p}$, for $\alpha_{D} = \alpha_{em}$ and a halo-like aDM distribution.
Also shown are existing constraints from $\Delta N_\mathrm{eff}$ and stellar cooling. White dwarf cooling is more sensitive by several orders of magnitude.

In Section~\ref{sec:models} we discussed how the Mirror Twin Higgs model is a particularly motivated realization of aDM.
The green line in \fref{fig:gridplots} shows how the mirror electron and proton masses are correlated as $v_B/v_A$ is changed from 1 to 10. 
Figure~\ref{fig:fv_plot} shows the constraints on $\epsilon$ as a function of $v_B/v_A$. The jump in the constraint curve for $\rho_{aDM}/\rho_{DM} = 10^{-1}$ at $v_B/v_A \approx 6$ is physical and corresponds to a value of $\epsilon$ above which evaporation becomes important and begins to limits the accumulation at the end of the horizontal branch period.
These are the strongest direct detection limits to date, for either disk-like or halo-like mirror baryon distributions. 
They are even competitive with projected future sensitivities of electron-recoil experiments if the local aDM is atomic and/or disk-like, see~\cite{MTHastro}.

In Figs.~\ref{fig:gridplots2} and~\ref{fig:gridplots3} we show the constraints for different values of $\alpha_D$. As might be expected, the constraints are stronger (weaker) for larger (smaller) dark self-interactions, but the dark electron mass for which bremsstrahlung cooling loses efficiency decreases (increases) as well.

\section{Conclusion}
\label{sec:conclusion}

Atomic dark matter arises in many scenarios of dark complexity, and is an explicit prediction of mirror sector models, notably the Mirror Twin Higgs. It also gives rise to fascinating phenomenology that needs to be explored in detail.
A small aDM fraction of the total DM acts as a dissipative sub-component, radiating away dark photons and potentially collapsing into a dark disk, analogous to visible baryonic matter. 
Direct probes of this scenario are challenging due to the lower recoil in dark matter detectors from dark matter that is co-rotating with the Earth.

In this work we have derived the best constraints to date on aDM that couples to the SM  via a dark photon portal, using the well-measured cooling behaviour of white dwarf stars. 
White dwarfs are utilized as dark matter direct detection experiments, since in accumulating aDM throughout their lifetimes, they are sensitive to the same parameters of dark matter density and cross section with baryonic matter as terrestrial direct detection experiments. 
An important difference to terrestrial detectors is that the
 velocity dependence of the dark photon portal interaction implies \emph{increased} aDM capture in stars for \emph{lower} relative aDM velocities.
Stellar cooling bounds are therefore more sensitive for dark disks than for dark halos, complementing the reach of direct detection experiments in regions of dark sector parameter space where terrestrial experiments can lose sensitivity.
Depending on the dark proton and dark electron mass, white dwarf cooling constrains the kinetic mixing to be as low as $\epsilon \sim 10^{-12}$ for 
$\rho_{aDM}/\rho_{DM} = 10\%$.
In addition to being the most stringent constraints to date, our bounds are even competitive with future electron recoil direct detection experiments if the local aDM is atomic and/or disk-like, see \cite{MTHastro}.

We have focussed on a particularly simple scenario of atomic dark matter -- that is, models with a dark electron, proton and photon -- but it is likely that other dissipative dark matter models can be constrained using similar techniques. 
Furthermore, while our analysis provides a first conservative estimate of white dwarf cooling sensitivity to aDM, it is important to keep in mind that a more refined analysis could not only improve on our bounds, in the future it could even lead to a discovery if deviations from  robust SM predictions for white dwarf cooling are observed in the white dwarf luminosity function. 
Since the deviations in the WDLF for aDM are markedly different from those expected for other dark sector scenarios (most notably due to the fact that old cool dwarfs are more affected than young hot dwarfs), precision WDLF measurements could distinguish between aDM or dissipative DM scenarios and, for example, production of axions. Such a discovery would therefore also be the first step in characterizing the dark sector, demonstrating that explorations of dark complexity must proceed across many experimental, observational and theoretical frontiers.

\hspace{1cm}

\textbf{Acknowledgements:}
We thank Rouven Essig and Christopher Matzner for helpful conversations. The research of DC and JS was supported in part by a Discovery Grant from the Natural Sciences and Engineering Research Council of Canada, and by the
Canada Research Chair program. 

\bibliographystyle{JHEP}
\bibliography{References}

\end{document}